\documentclass{aastex62}
\usepackage[colorinlistoftodos]{todonotes}
\usepackage{CJK}
\submitjournal{ApJS}

\shorttitle{Exoplanets in the Antarctic sky. I.}
\shortauthors{Zhang et al.}

\begin{document}
\begin{CJK*}{UTF8}{gbsn}

\title{Exoplanets in the Antarctic Sky. I. The First Data Release of AST3-II (CHESPA) and New Found Variables within the Southern CVZ of \textit{TESS}}

\correspondingauthor{Hui Zhang}
\email{huizhang@nju.edu.cn}
\correspondingauthor{Ji-lin Zhou}
\email{zhoujl@nju.edu.cn}

\author{Hui Zhang (张辉)}
\affil{School of Astronomy and Space Science, Key Laboratory of Modern Astronomy and Astrophysics in Ministry of Education, \\
Nanjing University, Nanjing 210023, Jiangsu,China}

\author{Zhouyi Yu}
\affil{School of Astronomy and Space Science, Key Laboratory of Modern Astronomy and Astrophysics in Ministry of Education, \\
Nanjing University, Nanjing 210023, Jiangsu,China}

\author{Ensi Liang}
\affil{School of Astronomy and Space Science, Key Laboratory of Modern Astronomy and Astrophysics in Ministry of Education, \\
Nanjing University, Nanjing 210023, Jiangsu,China}

\author{Ming Yang}
\affil{School of Astronomy and Space Science, Key Laboratory of Modern Astronomy and Astrophysics in Ministry of Education, \\
Nanjing University, Nanjing 210023, Jiangsu,China}

\author{Michael C. B. Ashley}
\affil{School of Physics, University of New South Wales, NSW 2052, Australia}


\author{Xiangqun Cui}
\affil{Nanjing Institute of Astronomical Optics and Technology, Nanjing 210042, China}
\affil{Chinese Center for Antarctic Astronomy, Nanjing 210008, China}

\author{Fujia Du}
\affil{Nanjing Institute of Astronomical Optics and Technology, Nanjing 210042, China}
\affil{Chinese Center for Antarctic Astronomy, Nanjing 210008, China}

\author{Jianning Fu}
\affil{Department of Astronomy, Beijing Normal University, Beijing, 100875, China}

\author{Xuefei Gong}
\affil{Nanjing Institute of Astronomical Optics and Technology, Nanjing 210042, China}
\affil{Chinese Center for Antarctic Astronomy, Nanjing 210008, China}

\author{Bozhong Gu}
\affil{Nanjing Institute of Astronomical Optics and Technology, Nanjing 210042, China}
\affil{Chinese Center for Antarctic Astronomy, Nanjing 210008, China}

\author{Yi Hu}
\affil{National Astronomical Observatories, Chinese Academy of Sciences, Beijing 100012, China}
\affil{Chinese Center for Antarctic Astronomy, Nanjing 210008, China}

\author{Peng Jiang}
\affil{Polar Research Institute of China, Shanghai 200136, China}
\affil{Chinese Center for Antarctic Astronomy, 451 Jinqiao Rd, Nanjing 210008, China}

\author{Huigen Liu}
\affil{School of Astronomy and Space Science, Key Laboratory of Modern Astronomy and Astrophysics in Ministry of Education, \\
Nanjing University, Nanjing 210023, Jiangsu,China}

\author{Jon Lawrence}
\affil{Australian Astronomical Optics, Macquarie University, NSW 2109, Australia}

\author{Qiang Liu}
\affil{National Astronomical Observatories, Chinese Academy of Sciences, Beijing 100012, China}

\author{Xiaoyan Li}
\affil{Nanjing Institute of Astronomical Optics and Technology, Nanjing 210042, China}
\affil{Chinese Center for Antarctic Astronomy, Nanjing 210008, China}

\author{Zhengyang Li}
\affil{Nanjing Institute of Astronomical Optics and Technology, Nanjing 210042, China}
\affil{Chinese Center for Antarctic Astronomy, Nanjing 210008, China}

\author{Bin Ma}
\affil{National Astronomical Observatories, Chinese Academy of Sciences, Beijing 100012, China}
\affil{Chinese Center for Antarctic Astronomy, Nanjing 210008, China}
\affil{University of Chinese Academy of Sciences, Beijing 100049, China}

\author{Jeremy Mould}
\affil{Centre for Astrophysics and Supercomputing, Swinburne University of Technology, PO Box 218, Mail Number H29, Hawthorn, VIC 3122, Australia}
\affil{ARC Centre of Excellence for All-sky Astrophysics (CAASTRO)}

\author{Zhaohui Shang}
\affil{Tianjin Astrophysics Center, Tianjin Normal University, Tianjin 300387, China}
\affil{National Astronomical Observatories, Chinese Academy of Sciences, Beijing 100012, China}
\affil{Chinese Center for Antarctic Astronomy, Nanjing 210008, China}

\author{Nicholas B. Suntzeff}
\affil{George P. and Cynthia Woods Mitchell Institute for Fundamental Physics \& Astronomy, Texas A. \& M. University, \\
Department of Physics and Astronomy, 4242 TAMU, College Station, TX 77843, US}

\author{Charling Tao}
\affil{Aix Marseille Univ, CNRS/IN2P3, CPPM, Marseille, France}
\affil{Physics Department and Tsinghua Center for Astrophysics (THCA), Tsinghua University, Beijing, 100084, China}

\author{Qiguo Tian}
\affil{Polar Research Institute of China, 451 Jinqiao Rd, Shanghai 200136, China}

\author{C. G. Tinney}
\affil{Exoplanetar7y Science at UNSW, School of Physics, UNSW Sydney, NSW 2052, Australia}

\author{Syed A. Uddin}
\affil{Purple Mountain Observatory, Nanjing 210008, China}

\author{Lifan Wang}
\affil{George P. and Cynthia Woods Mitchell Institute for Fundamental Physics \& Astronomy, Texas A. \& M. University, \\
Department of Physics and Astronomy, 4242 TAMU, College Station, TX 77843, US}
\affil{Purple Mountain Observatory, Nanjing 210008, China}
\affil{Chinese Center for Antarctic Astronomy, Nanjing 210008, China}

\author{Songhu Wang}
\affil{Department of Astronomy, Yale University, New Haven, CT 06511, USA}

\author{Xiaofeng Wang}
\affil{Physics Department and Tsinghua Center for Astrophysics (THCA), Tsinghua University, Beijing, 100084, China}

\author{Peng Wei}
\affil{School of Astronomy and Space Science, Key Laboratory of Modern Astronomy and Astrophysics in Ministry of Education, \\
Nanjing University, Nanjing 210023, Jiangsu,China}

\author{Duncan Wright}
\affil{University of Southern Queensland, Computational Engineering and Science Research Centre, Toowoomba, Queensland 4350, Australia}

\author{Xuefeng Wu}
\affil{Purple Mountain Observatory, Nanjing 210008, China}
\affil{Chinese Center for Antarctic Astronomy, Nanjing 210008, China}

\author{Robert A. Wittenmyer}
\affil{University of Southern Queensland, Computational Engineering and Science Research Centre, Toowoomba, Queensland 4350, Australia}

\author{Lingzhe Xu}
\affil{Nanjing Institute of Astronomical Optics and Technology, Nanjing 210042, China}

\author{Shi-hai Yang}
\affil{Nanjing Institute of Astronomical Optics and Technology, Nanjing 210042, China}
\affil{Chinese Center for Antarctic Astronomy, Nanjing 210008, China}

\author{Ce Yu}
\affil{School of Computer Science and Technology, Tianjin University, Tianjin 300072, China}

\author{Xiangyan Yuan}
\affil{Nanjing Institute of Astronomical Optics and Technology, Nanjing 210042, China}
\affil{Chinese Center for Antarctic Astronomy, Nanjing 210008, China}

\author{Jessica Zheng}
\affil{Australian Astronomical Observatory, 105 Delhi Road, North Ryde, NSW 2113, Australia}

\author{Hongyan Zhou}
\affil{Polar Research Institute of China, 451 Jinqiao Rd, Shanghai 200136, China}

\author{Ji-lin Zhou}
\affil{School of Astronomy and Space Science, Key Laboratory of Modern Astronomy and Astrophysics in Ministry of Education, \\
Nanjing University, Nanjing 210023, Jiangsu,China}

\author{Zhenxi Zhu}
\affil{Purple Mountain Observatory, Nanjing 210008, China}
\affil{Chinese Center for Antarctic Astronomy, Nanjing 210008, China}


\begin{abstract}

Located at Dome A, the highest point of the Antarctic plateau, the Chinese Kunlun station is considered to be one of the best ground-based photometric sites because of its extremely cold, dry, and stable atmosphere\citep{Saunders09}. A target can be monitored from there for over 40 days without diurnal interruption during a polar winter. This makes Kunlun station a perfect site to search for short-period transiting exoplanets. Since 2008, an observatory has been built at Kunlun station and three telescopes are working there. Using these telescopes, the AST3 project has been carried out over the last six years with a search for transiting exoplanets as one of its key programs (CHESPA). In the austral winters of 2016 and 2017, a set of target fields in the Southern CVZ of \textit{TESS} \citep{Ricker09} were monitored by the AST3-II telescope. In this paper, we introduce the CHESPA and present the first data release containing photometry of 26,578 bright stars ($\textbf{\textit{m}}_\textit{i}\le15$). The best photometric precision at the optimum magnitude for the survey is around 2 mmag. To demonstrate the data quality, we also present a catalog of 221 variables with a brightness variation greater than 5 mmag from the 2016 data. Among these variables, 179 are newly identified periodic variables not listed in the AAVSO database\footnote{\url{https://www.aavso.org/}}, and 67 are listed in the Candidate Target List\citep{Stassun17}. These variables will require careful attention to avoid false-positive signals when searching for transiting exoplanets. Dozens of new transiting exoplanet candidates will be also released in a subsequent paper\citep{Zhang18b}.
\end{abstract}


\section{Introduction} \label{sec:intro}
Wide-field photometric surveys for transit signals have proved to be one of the most effective methods for finding exoplanets. To date, more than $80\%$ of known exoplanets have been discovered by photometric surveys using either ground- or space-based telescopes, with the Kepler project \citep{Borucki10}  has contributing the largest fraction of these transiting exoplanets. Besides just making new discoveries, photometric surveys provide key physical characteristics of exoplanets: the transit depth reveals the physical size of the transiting planet (relative to the host star), which is crucial in determining its nature (i.e. is it a gas giant or a super Earth?), while the shape of the transit event indicates the orbit's impact parameter, constraining the orbital inclination of the planet and providing insight into the dynamical evolution of the planetary system \citep{Seager03}. When coupled with dynamical masses determined from radial velocities of the host star, photometric measurements reveal the average density of the planet and hence constrain its composition and internal structure (see, e.g., \citealp{Seager07,Baraffe08}).   Ultra-high precision photometry can reveal secondary eclipses and phase curves. These provide a unique way to measure the reflected and thermal emission of an exoplanet, leading to estimates of its surface albedo and allowing modeling of the heat transport efficiency (see, e.g., \citealp{Deming05,Charbonneau05,Knutson07}). Combining all the information possible from photometry, we can establish a relatively detailed model of the exoplanet's atmosphere, and hence its possible habitability.

To search for and study exoplanets using photometry, a survey needs both a wide field-of-view (FOV), a high photometric precision and as complete time coverage of the obrial periods of interest as possible. Some pioneering ground-based projects have been very successful, e.g., WASP/SuperWASP \citep{Pollacco06}, HATNET \citep{Bakos04} , HATSouth \citep{Bakos13}, and KELT \citep{Pepper07}.
Hundreds of transiting exoplanets have been found from such efforts over the last two decades. These surveys have shown that stable instruments (e.g., low systematic errors, long-time stability), optimized operations (e.g., well organized duty cycle, precise auto-guiding) and superb observing conditions (e.g., good seeing, fewer interruptions from bad weather) are crucial for producing high-quality lightcurves, and thus reducing the number of false positives in lists of transiting exoplanet candidates.

On one hand, with experience gained and new technologies adopted, new generation wide-field transit surveys have overcome many obstacles in both hardware and software, e.g., NGTS \citep{Wheatley18} and Pan-Planets \citep{Obermeier16}.  On the other hand, a good observing site is still essential to guarantee a fruitful survey, and this is where the Antarctic Plateau has many advantages over traditional temperate latitude sites. The extremely cold atmosphere in Antarctica from the telescope up leads to very low and very stable water vapor content, which reduces photometric noise from varying water absorption. The decreased high-altitude turbulence above the plateau results in dramatically reduced scintillation noise \citep{Kenyon06}. This combination makes the Antarctic Plateau an ideal place to perform optical, infrared and THz observations. \citep{Lawrence04} reported a median seeing of $0.23^{\prime\prime}$ (average of $0.27^{\prime\prime}$) above a 30 m boundary layer at Dome C, drawing world-wide attention. Subsequently, many studies have been made of the astronomical conditions at various Antarctic sites; for a summary, see \cite{Storey05}, \cite{Burton07}, and \cite{Ashley13}. \cite{Saunders09} studied eight major factors, such as the boundary layer thickness, cloud coverage, auroral emission, airglow, atmospheric thermal backgrounds, precipitable water vapor, telescope thermal backgrounds, and the free-atmosphere seeing, at Dome A, B, C, and F, and also Ridge A and B. After a systematic comparison, they concluded that Dome A, the highest point of the Antarctic Plateau, was the best site overall. Besides the excellent photometric conditions, the polar night in Antarctica provides an opportunity to stare at a target field for over 40 days without interruption from the diurnal cycle. This is a great advantage to enhance the detectability of short-period transiting exoplanets.

Since the first visit to Dome A in 2005 by the 21st CHInese National Antarctic Research Expedition (CHINARE), Chinese astronomers and their international collaborators have performed a series of site-testing studies. The results from multiple experiments have shown that Dome A has a thin boundary layer with a 14 m median height \cite{Bonner10}, a strong temperature inversion above the snow surface, low wind speed \citep{Hu14}, low water vapor \citep{Shi16}, low sky brightness and a high clear-sky fraction \citep{Zou10,Yang17}. These results are consistent with previous predictions made largely from satellite observations by \cite{Saunders09}. The first generation telescope at Dome A (CSTAR, the Chinese Small Telescope ARray \citealp{Yuan08,Zhou10}) was installed in 2008. CSTAR produced more than 200,000 continuous images on a fixed FOV of $4.5^\circ \times 4.5^\circ $ centered at the South Celestial Pole in each austral winter of 2008, 2009 and 2010. After a series of refinements (such as corrections for the inhomogeneous effects of cloud, the treatment of ghost images and diurnal effects \citep{Wang12,Meng13,Wang14a}), a photometric precision of 4\,mmag was obtained for bright stars in CSTAR data. Hundreds of new variable stars were identified and studied from these data (e.g., \citealp{Wang11,Yang15,Zong15,Oelkers16,Liang16}), and the first 6 exoplanet candidates around the south celestial pole were identified \citep{Wang14b}. Based on this successful experience, and lessons learned from CSTAR project, a second generation of telescopes -- the Antarctic Survey Telescopes (AST3) -- were conceived for a wide-field, high-resolution photometric survey at Dome A. AST3 was envisaged as three 50\,cm-aperture telescopes, each with a different fixed filter. The AST3 telescopes not only have larger apertures than CSTAR, but they also have full tracking systems and higher angular resolutions ($\sim 1^{\prime\prime}$~pixel$^{-1}$), which provides wider sky coverage and better precision. The first and second AST3 telescopes -- AST3-I and AST3-II -- were installed at Dome A in 2012 and 2015 by the 28th and 31st CHINARE (respectively). Observing programs with AST3-I have been described by \cite{Liu18} and the corresponding data published by \cite{Ma18}. The third AST3 telescope is nearing completion in Nanjing, and will be equipped with a Kdark-band near-infrared camera \citep{Burton16}. Using the CSTAR and AST3 telescopes, we have started a long-term wide-field photometric survey searching for transiting exoplanets in the Antarctic sky -- the Chinese Exoplanet Searching Program from Antarctica (CHESPA).

In the austral winters of 2016 and 2017, we used  the AST3-II telescope to survey a group of selected fields near the southern ecliptic pole and within the Southern continuous viewing zone of \textit{TESS} \citep{Ricker09}. In this first paper, we describe the observations obtained and data reduction processes used on them, as well as presenting   some of the data products obtained in 2016. This data release includes reduced images, calibrated catalogs, and detrended lightcurves of 26578 bright ($7.5 \le \textbf{\textit{m}}_\textit{i}  \le 15.0$) stars near the southern ecliptic pole. We also present a catalog of variable stars found in the Southern continuous viewing zone of \textit{TESS}. Since wide-field surveys for transiting exoplanets often suffer from large pixel-scales and are notoriously plagued by false positives, high-resolution photometric follow-up observations are necessary to filter out objects such as eclipsing binaries. The precision of \textit{TESS} is expected to be much better than that of ground-based surveys, however, its pixel-scale is still relatively large, $\sim 21^{\prime\prime}$~pixel$^{-1}$, and a high false-alarm rate caused by blending events such as background eclipsing binaries is expected \citep{Collins18}. Our catalog of variables will be a good reference for identifying these blending events. In this work we only present obvious binaries and pulsating stars with regular period. A detailed classification and analysis of other variables, e.g. rotational spots modulation, semi-periodic/irregular variables and long-period variables, will be available soon in Fu et al.~(in preparation). The description of the lightcurve detrending, transit signal search and a detailed catalog of transiting exoplanet candidates will be presented by \cite{Zhang18b} Results for observations for specific targets, e.g., Proxima Centauri and $\beta$ Pictoris, will also be released in the near future.

This paper is organized as follows: we introduce the instruments and our survey strategy in Section \ref{sec:instru}, we describe the survey strategy and observations in Section \ref{sec:survey}; the detailed data reduction flow is described in Section \ref{sec:pipeline}; we present the survey results in Section \ref{sec:results} and summarize the paper in Section \ref{sec:summary}.

\section{The Instruments} \label{sec:instru}
\begin{figure}
\centering
\includegraphics[width =1.0 \textwidth]{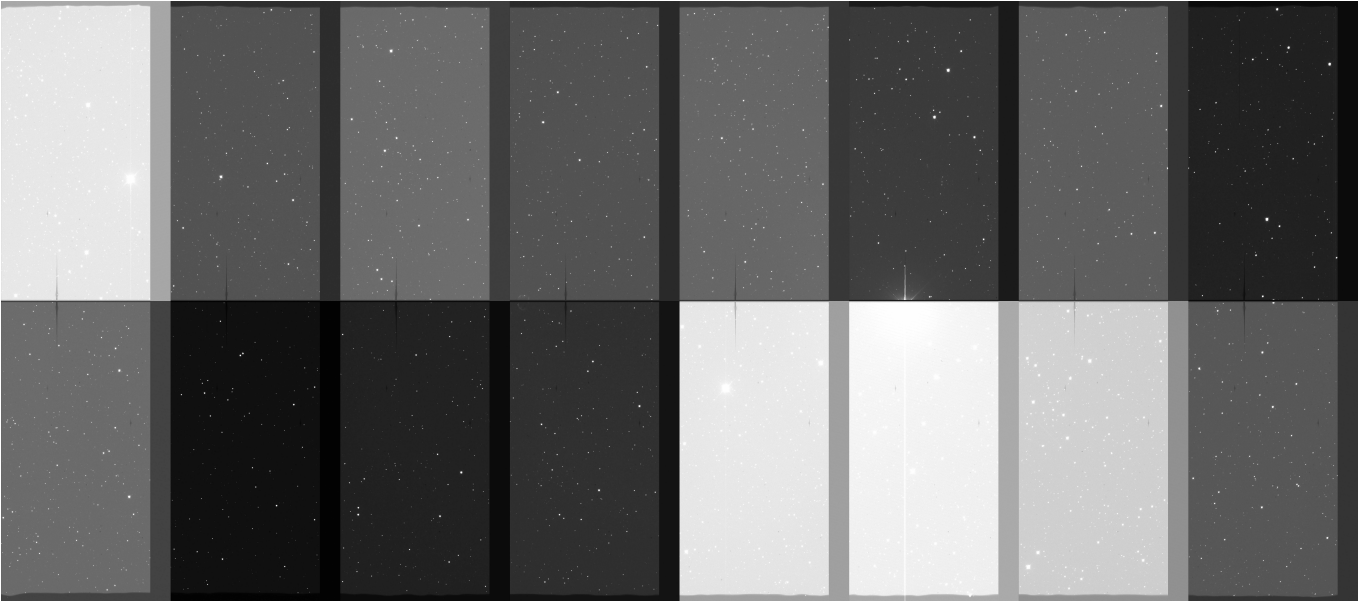}
\caption{Full frame of a raw image taken by AST3-II. There are 16 readout channels: 8 at the top and the other 8 at the bottom. Each channel has a physical resolution of $1500\times 2650$ pixels including overscan regions. Dark gaps between channels are those overscan regions which are extra pixels generated by the CCD electronics when the CCD is read out. They are not connected to real physical pixels on the CCD but can be used to estimate the bias of the image. Each gap will be modeled and subtracted from the readout channel to the left (see Section \ref{sec:overscan} for details). All these overscan regions are trimmed off during the data reduction processes.\label{fig:rawimage}}
\end{figure}

The AST3-I telescope was designed and built by the Nanjing Institute of Astronomical Optics and Technology (NIAOT) - see \cite{Cui08,Yuan14,Yuan15,Wang17} (and references therein) for more details. The AST3-II telescope (which was used to acquire all the data used in this work) is the second of the three planned AST3 telescopes. It is almost identical to AST3-I: they share the same modified Schmidt system design \citep{Yuan12}, the same entrance pupil aperture of 50\,cm, the same wide FOV of 4.3\,deg$^2$, the same Sloan $\textit{i}$ filter and the same CCD camera. Some improvements and innovations have been made in AST3-II in the area of focusing,  snow-proofing, and the defrosting system based on our experience from the earlier operations of AST3-I during austral winters since 2012. Thanks to these updates, AST3-II worked well during the extremely cold winters of 2016 and 2017, and has acquired over 30\,TB of high-quality images. A single raw AST3-II image  is $10\mbox{K} \times 5\mbox{K}$ pixels in size, using 122\,MB of storage. It is generated by a $10\mbox{K} \times10\mbox{K}$ frame transfer STA1600FT CCD camera with a pixel scale of $\sim 1^{\prime\prime}$~pixel$^{-1}$ over a FOV of $1.5 \times 2.9 \mbox{ deg}^2$ (RA $\times$ Dec). The average FWHM of stars obtained by focused AST3 telescopes is around $\sim 2.0$ pixels.  The CCD detector operates in frame transfer mode without a shutter and is divided into frame store regions (in its top and bottom quadrants) and an active imaging area in its central half. Although this configuration halved the maximum possible FOV, it eliminates the risk of shutter mechanism failure. Since the telescope has to be operated entirely remotely for 11 months with no possibility of repairs being carried out, any mechanism failure would be fatal to the whole project, and the shutter is one of the most fragile parts for these high-frequency observations. The image area of the CCD has 16 readout channels, each of $1500 \times 2650$ pixels (including overscan regions). A full-frame raw image is shown in Figure \ref{fig:rawimage} and more details on the AST3 CCD performance can be found in \cite{Ma12} and \cite{Shang12}.

During the austral winter observing season, the whole AST3 system is operated remotely and the scheduled observations are executed in a fully-automatic mode. Consequently, safety, reliability and stability are key issues for the AST3 design. The core systems, consisting of the main control computer, the storage disk array and the pipeline computer, were intensively customized to handle almost any conceivable hardware, software or network failure. Each component in the system had two identical copies for redundancy, minimising the risk to the system from single point failures. As an example, the fiber optic data link from the CCD camera was split into two fibers so that two separate computers could be used to control the CCD. The hardware and software for the operation, control and data (COD) system were developed by the National Astronomical Observatories, Chinese Academy of Sciences (NAOC) \citep{Shang12,Hu16}. The electrical power supply and internet communication were provided by a similarly reliable on-site observatory platform, PLATO-A, which is an improved version of UNSW's automated PLATO observatory platform for CSTAR and other earlier instruments. PLATO-A was designed to provide a continuous 1 kW power source for the AST3 telescopes \citep{Lawrence09,Ashley10} for at least one year without servicing.

\section{The Exoplanet Survey Program} \label{sec:survey}
The Kepler project has dramatically broadened our horizon in exoplanet science over the last five years. Although it has surveyed only a small part of the sky, its unprecedented photometric precision and continuous observing capability have been superbly successful at discovering planets down to Earth sizes. However, most candidates found by Kepler are just too faint to be followed up with dynamical mass measurements from the ground. Kepler's successor mission, \textit{TESS}, will be similarly revolutionary, albeit in a different way -- by surveying the whole sky for short-period transiting exoplanets orbiting bright and/or nearby stars that {\em can} be followed up with dynamical masses from the ground. \textit{TESS} launched on April 18th, 2018 started surveying all bright stars in the southern hemisphere in June 2018. Thousands of exoplanet candidates are expected to be detected over the next two years. However, to achieve the whole-sky coverage, \textit{TESS} must use a large pixel-scale ($\sim 21^{\prime\prime}$~pixel$^{-1}$), which may lead to a higher false alarm rate than Kepler. Many candidates detected by \textit{TESS} may be astrophysical mimics, e.g., blended eclipsing binaries. Therefore, high-resolution photometric follow-up observations are necessary before undertaking time-consuming and competitive radial velocity observations to measure dynamical masses. The AST3 telescopes were designed to perform wide-field (FOV $\sim4.3 \mbox{ deg}^2$) and high-resolution (pixel-scale $\sim1^{\prime\prime}$~pixel$^{-1}$) time-domain photometric surveys and are well suited to this task. The observational conditions and uninterrupted polar nights at Dome A deliver additional benefits for this science.

To maximize the scientific value of our program, we select 48 fields within \textit{TESS}' Southern Continuous Viewing Zone (which is centered at the southern ecliptic pole  at $\mbox{RA} =06^{\mbox{h}}\,00^{\mbox{m}}\,00^{\mbox{s}} , \mbox{Dec} =-66^\circ\,33^{\prime}\,00^{\prime\prime}$). These fields are scientifically important because (in addition to their southerly declinations meaning they can be observed from Dome A at low airmasses), they will be observed by \textit{TESS} uninterruptedly for 12 months. This long observing baseline will greatly increase \textit{TESS}' sensitivity for short-period exoplanets, as well as allowing the detection of long-period planets in single-star and binary systems \citep{Stassun17}.  Additionally, the continuous viewing zone of JWST will also be located in this region, so any targets of interest will be studied by much more intensive detailed follow-up in the future.

\begin{deluxetable}{c|c|c|c}
\tablecaption{Center coordinates of 10 AST3-II target fields surveyed in 2016. \label{tab:2016fields}}
\tablehead{
\colhead{Field Name} & \multicolumn{2}{c}{Field Center}  &\colhead{Valid Images} \\
\colhead{} & \colhead{RA(J2000)} & \colhead{Dec(J2000)} &\colhead{}
}
\startdata
AST3II004 &   93.000  &  $-73.000$   & 3179 \\
AST3II005 &   98.500  &  $-73.000$   & 3080 \\
AST3II006 &  104.000  &  $-73.000$   & 3049 \\
AST3II007 &  109.500  &  $-73.000$   & 3103 \\
AST3II008 &  115.000  &  $-73.000$   & 3248 \\
AST3II009 &   93.000  &  $-70.000$   & 3090 \\
AST3II010 &   97.750  &  $-70.000$   & 3000 \\
AST3II011 &  102.500  &  $-70.000$   & 2991 \\
AST3II012 &  107.250  &  $-70.000$   & 3021 \\
AST3II013 &  112.000  &  $-70.000$   & 3128 \\
\enddata
\end{deluxetable}

\begin{figure}
\centering
\includegraphics[width =1.0 \textwidth]{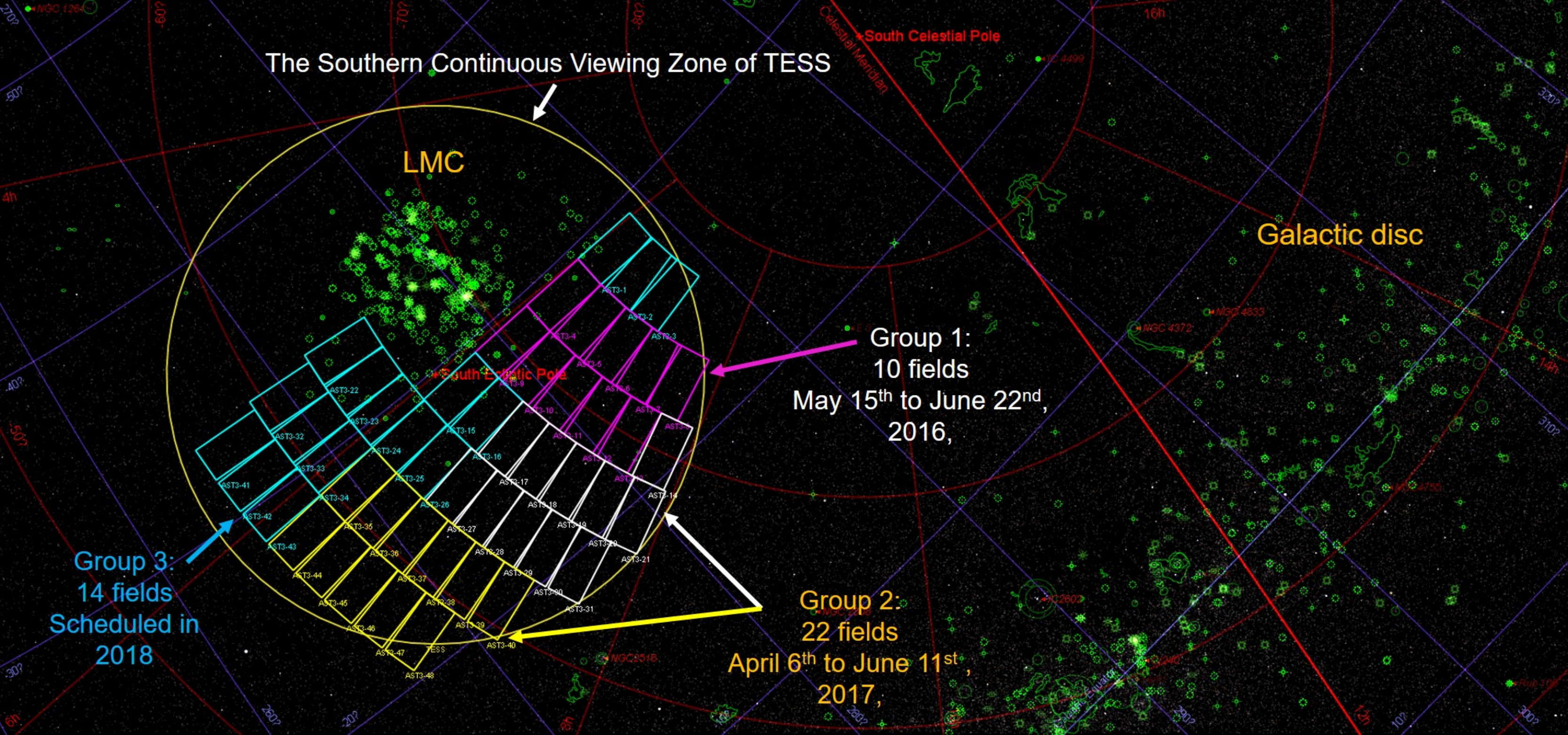}
\caption{Planned survey of 48 target fields in 2016, 2017 and 2018. Each field has a sky coverage of $\sim4.3\mbox{ deg}^2$. Fields close to the LMC are excluded to avoid crowded fields of giant stars. Group 1 (10 fields) and group 2 (22 fields) were scanned in 2016 and 2017, respectively.\label{fig:48fields}}
\end{figure}

\begin{figure}
\centering
\includegraphics[width =1.0 \textwidth]{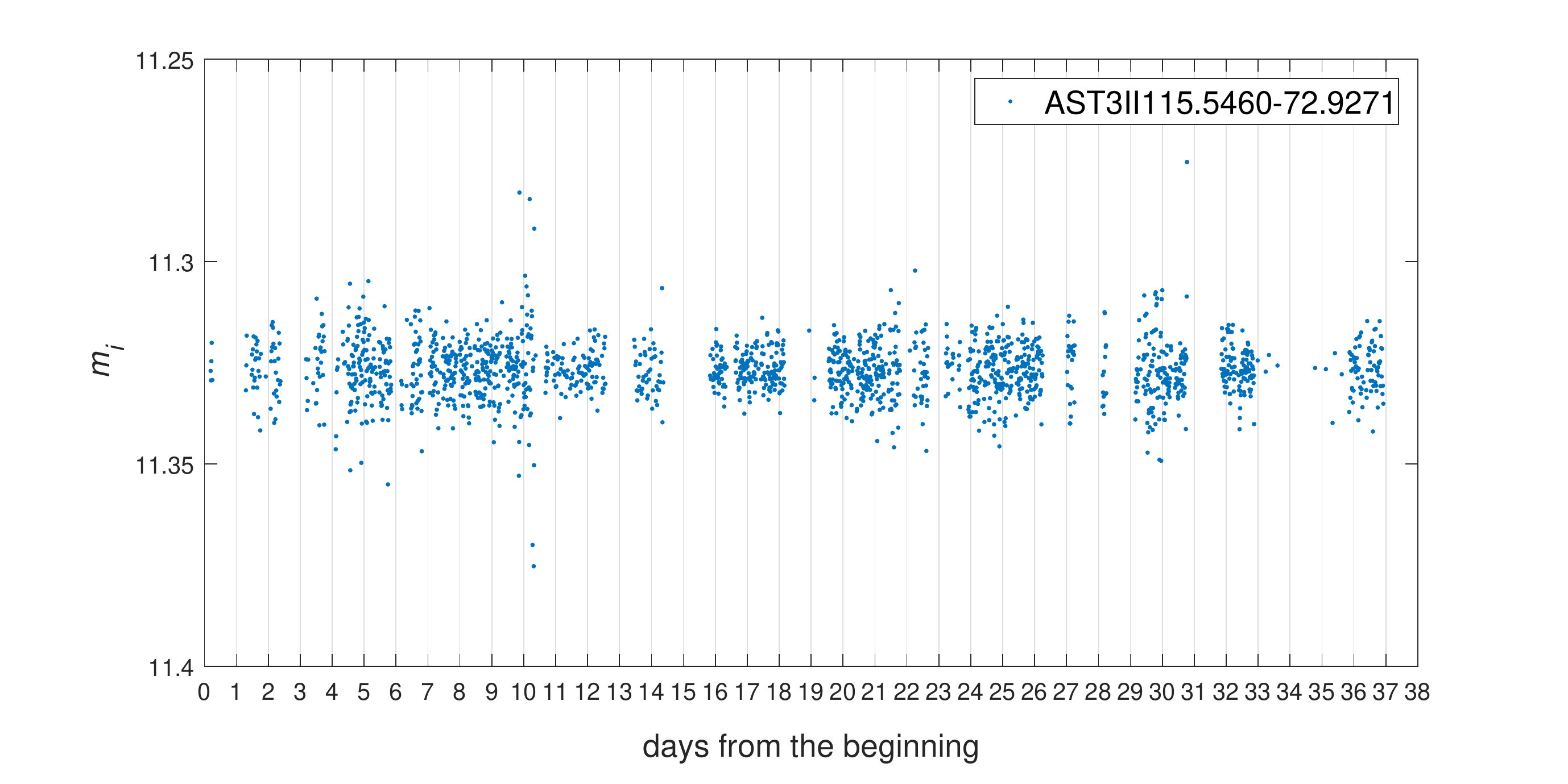}
\caption{\textbf{Lightcurve of target ``AST3II115.5460-72.9271'' which is a new found eclipsing binary. The sampling cadence is 12 minutes. A detrended, binned and phase-folded lightcurve of this target can be found in Figure \ref{fig:binaries2}.} The x-axis shows the days from beginning of the survey. Thanks to the polar nights, our observation may span several days without interruption by daylight. The gaps between continuous observations are caused by bad weathers, instrument maintenance and other \textbf{surveys}. \label{fig:days_coverage}}
\end{figure}

Our selected fields were numbered and divided into three groups, each of these three group being scheduled to be surveyed for a whole austral winter in 2016, 2017 and 2018, respectively (see Figure \ref{fig:48fields}). Fields close to the Large Magellanic Cloud (LMC) were excluded, because they are too crowded and because most bright stars in the LMC will be giants, making transiting exoplanet signals exceedingly weak. From May 16th to June 22nd in 2016, we scanned the first group of 10 adjacent fields (AST3II004 -- AST3II013, see Table \ref{tab:2016fields}). The available dark time increased day by day during the first half of our observation campaign and it finally became 24 hours a day at the end of June. A part of the very dark nights were spent on the other programs searching for transient targets, e.g. \textbf{supernovae}. And after excluding interruptions caused by bad weather and regular instrument maintenance, we observed for more than 350 hours spanning 37 nights (Figure \ref{fig:days_coverage}). The overall operation coverage is around $\sim 40\%$ which is less than the coverage we expected: $\sim 75\%$. In 2017, we spent most of the \textbf{nights where weather and instrumentation were suitable} on CHESPA and its working coverage reached $\sim 80\%$. To avoid saturation by bright stars, and to reduce the electromagnetic interference in our CCD images (see Section \ref{sec:EIcorrection}), we adopted a short-exposure-stacking strategy. In each observing night, we started from field AST3II004, and took three consecutive 10\,s exposures, then moved to field AST3II005 and repeated, then moved to AST3II006, and so on. When the last field -- AST3II013 -- completed, we jumped back to the first field -- AST3II004 -- and started a new loop. The dead time between each field was about 72\,s, including three 16\,s readout intervals and a 24\,s slewing operation. Every three adjacent images of the same field were median combined to achieve a high signal-noise ratio. Thus, the overall sampling cadence was about 12\,minutes for each field.

Twilight sky frames were taken at each dawn and dusk during the periods when the Sun was still rising each day at Dome A. These sky frames were combined to produce a master flat-field image (see Section \ref{sec:flatfield}). However, fluctuations in the super-sky flat-field are a major source of systematic noise for wide-field surveys. To reduce this, we first adopted a moderate defocusing of the telescope. The average seeing on the ground of Dome A is estimated to be around $\sim 1^{\prime\prime}$ or better. Convolved with the optics of a telescope, the actual size of PSF (point-spread-functions) of stars is usually larger than the seeing. So the AST3-II pixel-scale of $1^{\prime\prime}$~pixel$^{-1}$ is designed to be better than the average FWHM (full-width-half-maximum) of stars acquired by the focused AST3-II at Dome A, which is around $\sim 2.0$ pixels. During the exoplanet survey, we adjusted the focus to produce star images with PSFs between $\mbox{FWHM}=3$ pixels and $\mbox{FWHM}=5$ pixels. The shapes and sizes of the star images vary with the environment temperature, airmass, and etc. Sometimes the PSF deviates from the standard Gaussian profile and forms a flat top. But it does not cause much of a problem, since we adopt aperture photometry. For each star, the variations in the FWHM, the Elongation\footnote{\textbf{This is a ratio between the semi-major axis and the semi-minor axis of the star image. Large elongation usually means the telescope is not stable during exposure, either it is moving or shaking.}} and the Fraction-of-light radius\footnote{\textbf{This is the radius that encloses $90\%$ of the light from this star}} are recorded. In the latter EPD (external parameter detrending) process, these variations will be detrended from the lightcurve. Second, a semi-autoguiding method was adopted. For each field, a template image with a well-determined astrometric solution is constructed before the survey begins. Every time the telescope points to a new field, the first image of this field is cross-matched with its template and offsets are fed back to the pointing model and the two following images acquired after this pointing correction. The interval between two pointing corrections is the same as the dead time between moving fields, i.e., every $\sim 72$\,s. Within this short time, the position variation of a star is still less than the average size of a PSF, $\mbox{FWHM} \sim 4\mbox{ pixels}$. Even so, it is still very difficult to fix stars to the same pixels during the entire observation campaign. Finally, we detrend the brightness variation of each star against its inter- and intra-pixel position changes and variances to further reduce any systematic error caused by position shifting of stars.

The telescopes working at Dome A are entirely unattended during the year, apart from about 3 weeks over summer when servicing is performed on-site. The internet connection to Dome A relies on the Iridium satellite system, which has a typical latency of about 2\,s and can occasionally drop out and take up to a minute to reconnect. The operation of the telescope and CCD is therefore controlled automatically from scripts running on-site at Dome A. The whole system can be shifted to manual control from China if required (e.g. if the telescope optics and/or gear mechanisms need defrosting). Images taken by AST3-II are stored on a disk-array on-site. The bandwidth of the Iridium connection makes it impractical to download even a single whole image, so we only perform relatively simple data reduction processes on-site -- e.g. to check data quality and set up alarms for specific events. The detailed reduction of the entire data-set waits until the hard drives are physically returned by the expedition team in April or May of the following year.

\begin{figure}
\centering
\includegraphics[width =1.0 \textwidth]{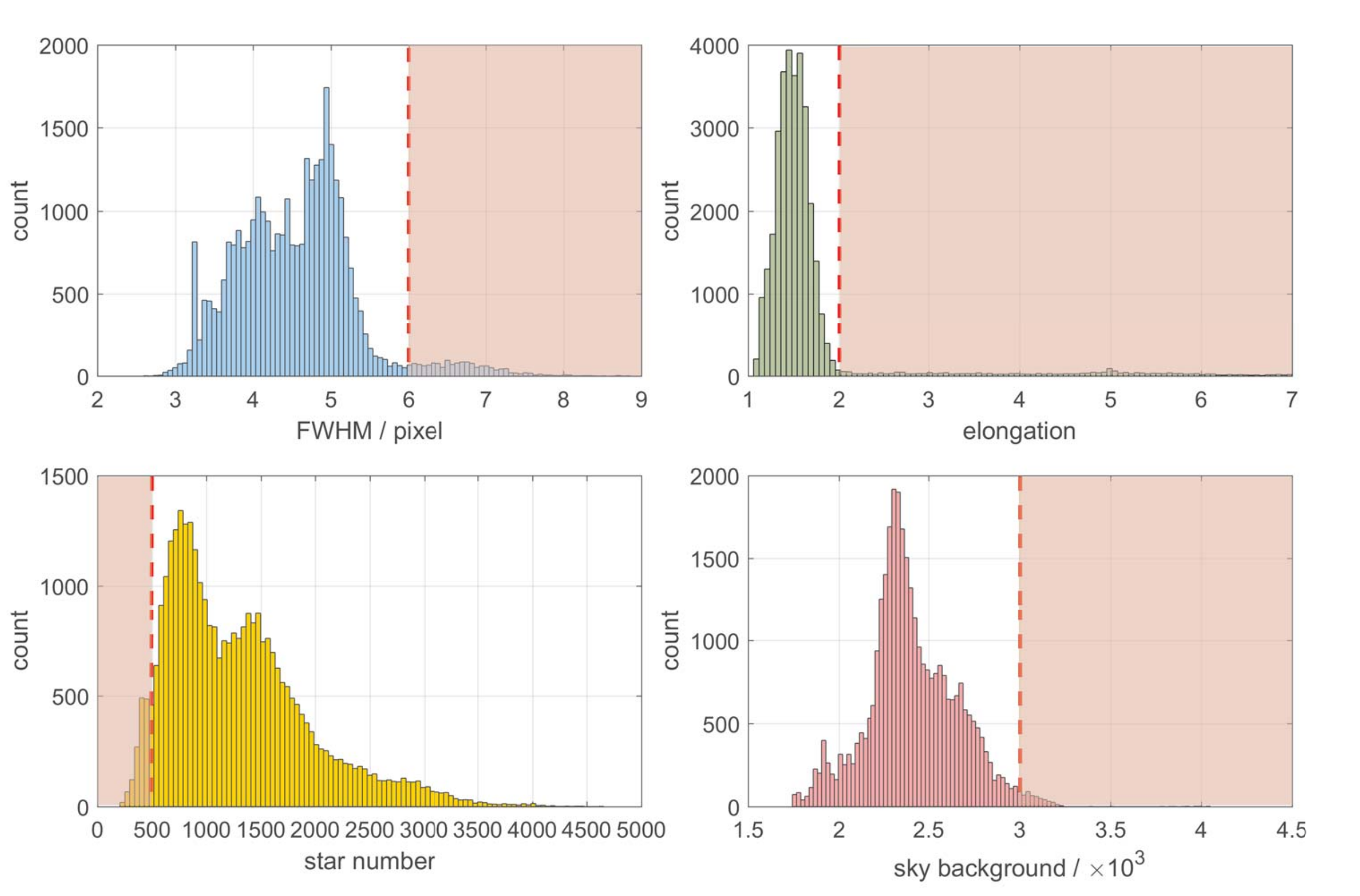}
\caption{Statistics of image qualities. Images within the shadowed region were dropped to maintain high data quality. \label{fig:imagestatitics}}
\end{figure}

At the end of the exoplanet survey program in 2016, more than 35000 science frames were taken by the AST3-II telescope. Some statistics on the data quality are shown in Figure \ref{fig:imagestatitics}. We removed images with flaws including high sky backgrounds, a large fraction of saturated area, large average FWHM ($ > 6$\,pix), large average elongations ($> 2$) and small numbers of detectable stars ($< 500$). After this quality filtering, 30889 high-quality images remained. They were further reduced and coadded to produce 18729 science images. catalogs of these images were matched with a master catalog generated from the APASS database \citep{Henden15,Henden16}.  Finally, lightcurves for of 26578 stars brighter than Sloan $\textbf{\textit{m}}_\textit{i}$ = 15\,mag were extracted.

\section{Data Reduction Pipeline} \label{sec:pipeline}
The on-site power budget and limited computational resources mean we run only a simple on-site data reduction pipeline including only standard image processes such as overscan subtraction, flat-field correction and simple aperture photometry using Sextractor \citep{Bertin96}. Reduced catalogs are then downloaded through the Iridium satellite connection while observing. However, the limited connection bandwidth and monthly quota mean the catalogs of only  selected fields are downloaded. We perform further astrometric solutions and flux calibration on these catalogs, and extract lightcurves of selected targets. Although the precision of this on-site data reduction is not optimal, it does return RMS$\sim$1\%lightcurves which are useful for inspecting the daily data quality and triggering alarms of high valuable targets (e.g., potential transit events of $\beta$ Pictoris and Proxima Centauri).
As described above, full data processing awaits the return of the data on hard drives from Dome A.

\begin{figure}
\centering
\includegraphics[width =1.0 \textwidth]{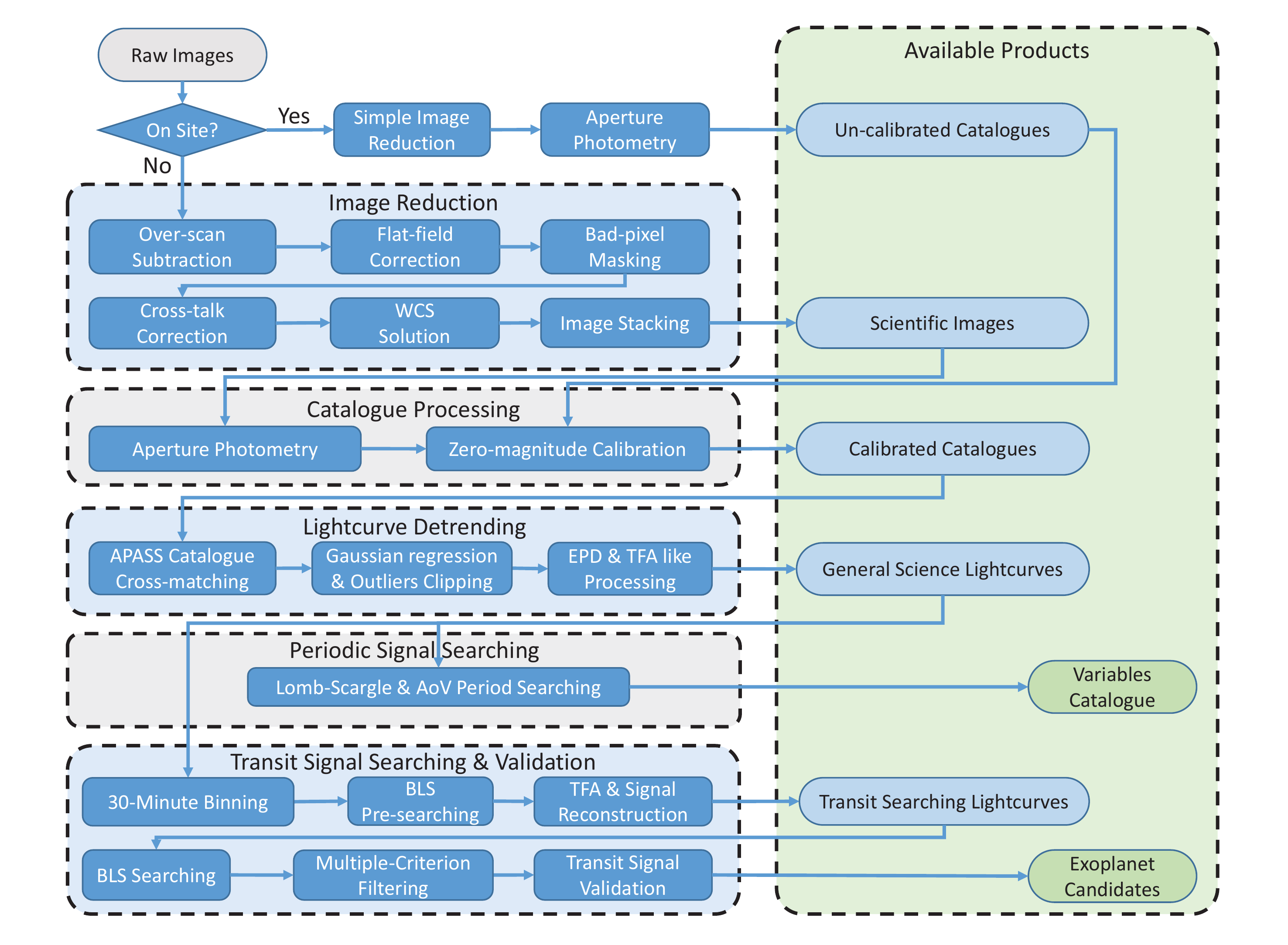}
\caption{Data reduction flow chart of the exoplanet searching project using the AST3 telescopes.\label{fig:flowchart}}
\end{figure}

A detailed data reduction flowchart for this full processing is shown in Figure \ref{fig:flowchart}. The pipeline consists of five major components, which are described below: 1. Image Reduction, which produces cleaned, WCS solved and coadded science images; 2. catalog Processing, which produces flux-calibrated catalogs; 3. Lightcurve Detrending, which produces lightcurves for general scientific usages; 4. Periodic Signal Searching, which searches for periodic variables and generates the variables catalog that is published in this work; 5. Transit Signal Searching, which polishes the lightcurves further, and searches for transit-like signals and validates exoplanet candidates. The pipeline operates in the $\emph{\text{MATLAB}}$ environment and has been fully parallelized. Some functions from well-tested open-source packages are used, including Sextractor \citep{Bertin96}, Swarp \citep{Bertin02}, VARTOOLS  \citep{Hartman16} and \emph{Astrometry.net} \citep{Lang10}.

\subsection{Image Reduction}
The image reduction component includes some standard steps (e.g., overscan subtraction, dark current subtraction and flat field correction), as well as some special steps to correct specific features within our data (e.g., cross-talk effects and electromagnetic fringes).

\subsubsection{Overscan Subtraction} \label{sec:overscan}
\begin{figure}
\centering
\includegraphics[width =1.0 \textwidth]{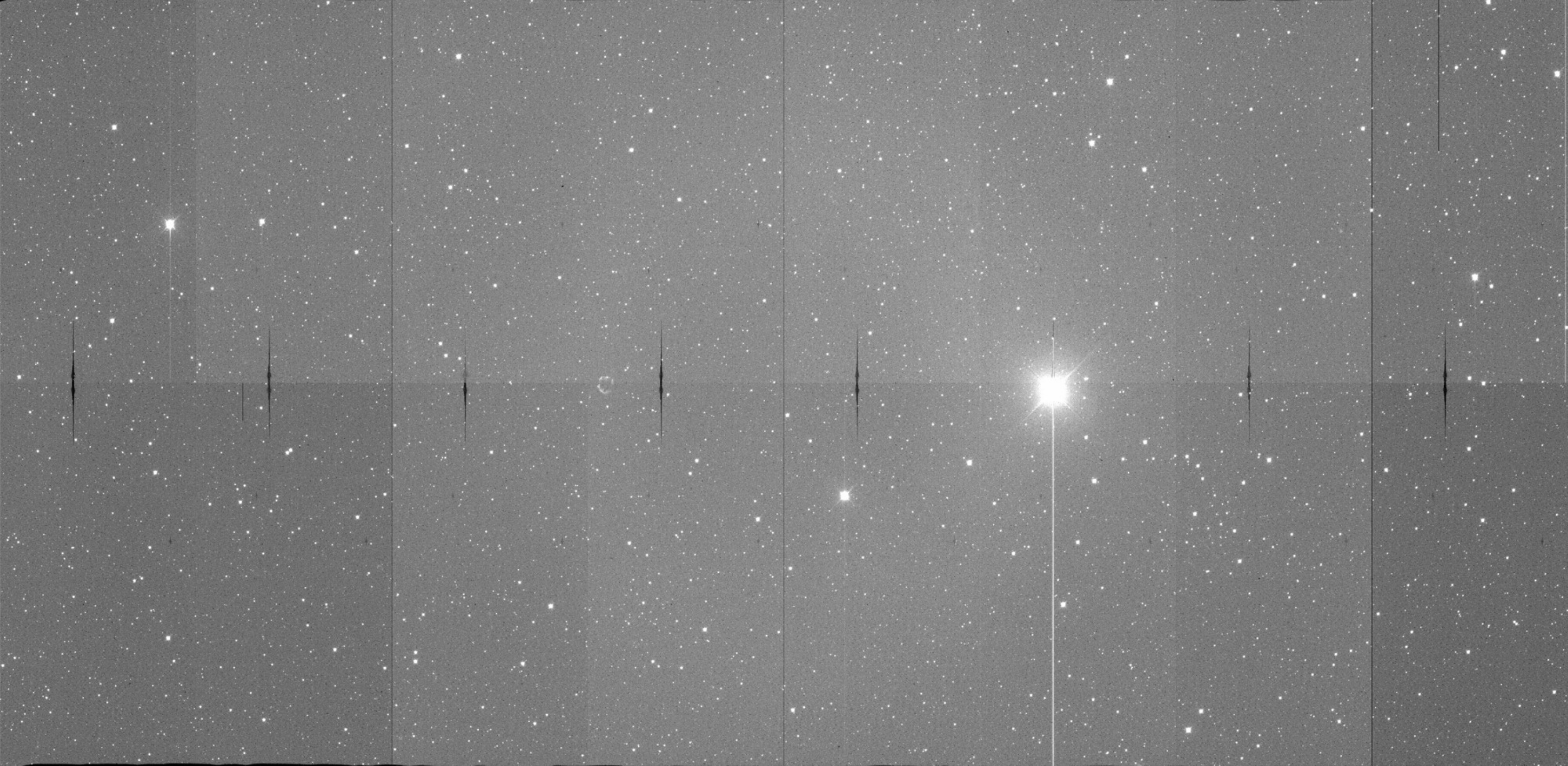}
\caption{Overscan subtracted image. The dark vertically-extended spots are caused by cross-talk from the highly-saturated star.\label{fig:cross-talk}}
\end{figure}

As shown in Figure \ref{fig:rawimage}, the $10\mbox{K}\times10\mbox{K}$ STA1600FT CCD camera installed on AST3-II works in a frame-transfer mode and produces an image size of $12000 \times 5300$, including overscan regions. At both the top and the bottom boundaries, there are 60 rows that are insensitive to light---these are trimmed off before further reductions. The rest of the image is divided into 16 readout channels with 8 at the top and the other 8 at the bottom. The vertical readout direction of channels at the top or bottom of an image is from the center to the top or bottom, respectively. The horizontal readout direction is always from the left to right within each channel.  Within a readout channel, the first 10 rows and the last 180 columns compose a reverse ``L''-shaped overscan region surrounding the photosensitive area, which has a size of $1320\times 2580$ pixels. Instead of simple bias subtraction, we perform a 2-D overscan subtraction for each channel: horizontal subtraction first and then vertical subtraction. Taking a top channel as an example, its horizontal overscan region is located at its bottom. This region, which is composed of 10 rows and 1500 columns, is median-combined to a single row, and this combined row is fitted using a second-order polynomial function and subtracted from the other 1320 rows (including the vertical overscan region). Then the vertical overscan region at the right, which is now 1320 rows by 180 columns, is median-combined, 2nd-order polynomial fitted and subtracted from each column of the final photosensitive area. The channels at the bottom are then processed in a vertically-flipped direction. After the 2-D overscan subtraction, the photosensitive areas of all the channels are fitted together to form a new image (see Figure \ref{fig:cross-talk}).

\subsubsection{Cross-talk Correction}
\begin{figure}
\centering
\includegraphics[width =0.8 \textwidth, height =0.9 \textwidth]{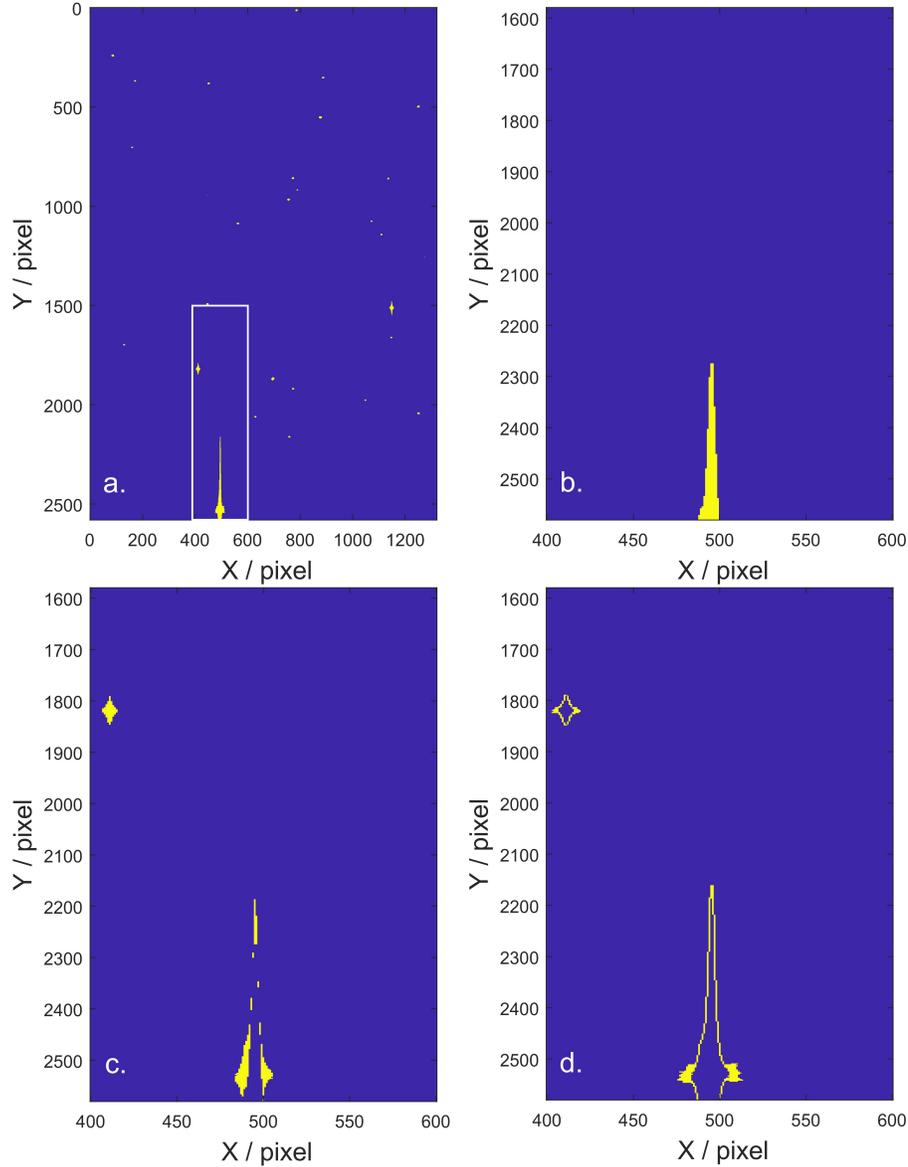}
\caption{Panel (a). The template channel of cross-talk effects. This is a binary map with cross-talk affected pixels marked as bright yellow and other pixels marked as dark blue. Each star-like spot in this template is produced by a saturated star lying in 1 of the 16 readout channels. And it shows the local positions of the darkened spot (which is also called the ghost of a saturated star) in each of the other 15 channels. This template channel is then copied 16 times and grouped as the formation of the original image ($8 \mbox{ Column} \times2 \mbox{ Row}$) to generate a completed map which marks all the pixels affected by the cross-talk effects. With this map we can simply mask all these unreliable pixels, or try to reduce the cross-talk effects down to a level less than the variation of the local sky background. The \textbf{latter} choice requires a detailed structure of the darkened area. And this is done by decomposing this template into three layers. For better illustration, we zoom in around two ghosts (enclosed by the white box) and show the layers of the overlapped region, core region and halo region in panels (b)–(d), respectively.  Panel (b). The bright yellow pixels show the overlapped region of two ghosts caused by two saturated stars lying in the similar positions of different channels. Note that the overlapped region does not necessarily exist in every ghost---the one on the top left has no such overlapped region. Panel (c). \textbf{The bright yellow region shows only the non-overlapped parts of the core regions of the two ghosts.} Note that only non-overlapping saturated pixels are included in the core region. Panel (d). The halo region of each ghost is shown in bright yellow. These pixels are not saturated but are affected by the charge-bleed issues. \label{fig:cross-talklayers}}
\end{figure}

\begin{figure}
\centering
\includegraphics[width =1.0 \textwidth]{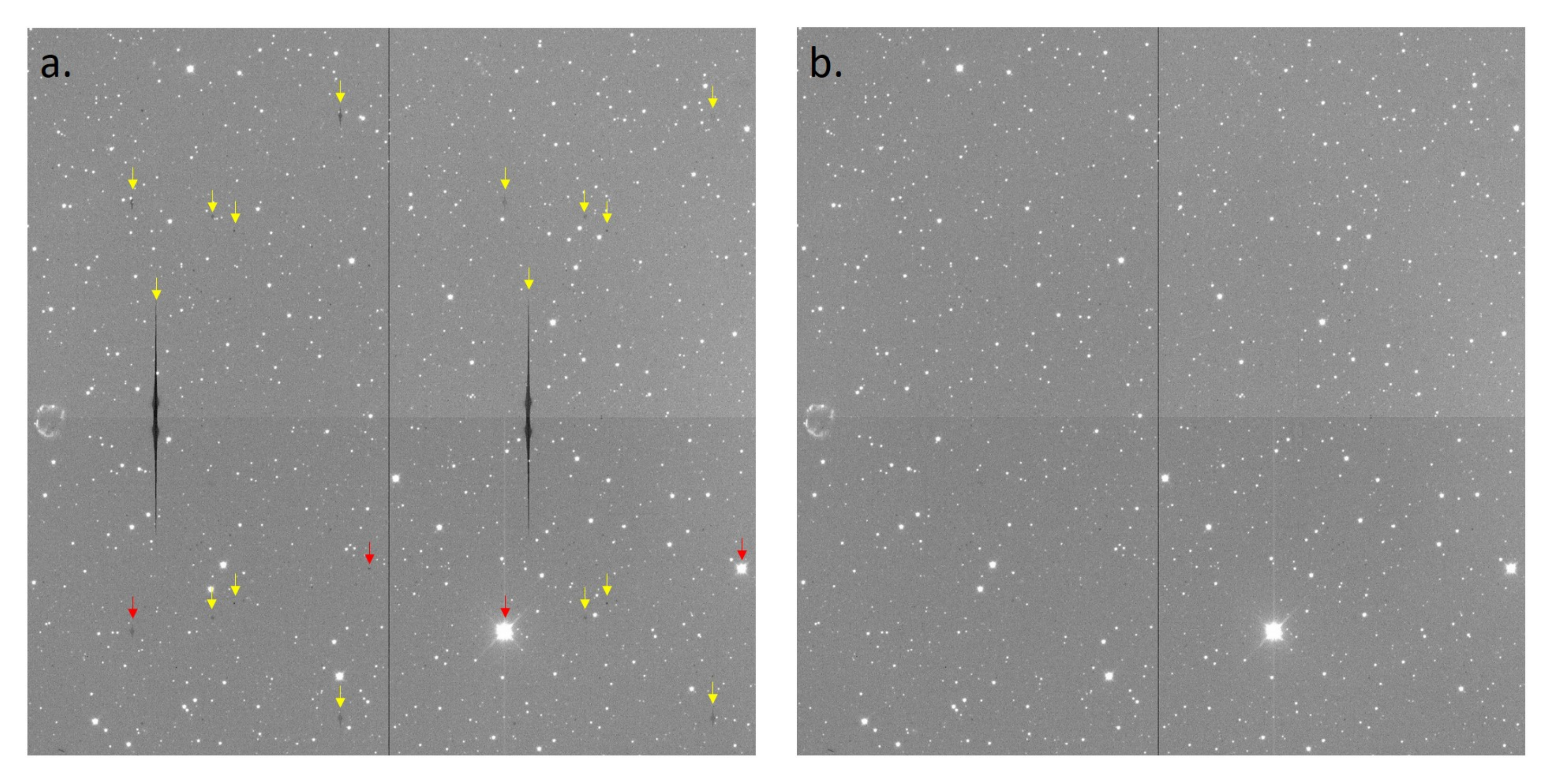}
\caption{Result of our cross-talk correction. Panel (a). This is the central part of an image, which consists of four corners from four channels. The dark spots/ghosts marked by yellow arrows are caused by the cross-talk effects. Note that their positions are the same in different channels. Red arrows mark two ghosts and their sources. Panel (b). The same region after cross-talk correction. The darkening caused by the cross-talk effect has been reduced down below the variation of the local sky background.\label{fig:cross-talkcorrection}}
\end{figure}

In the overscan subtracted image, serious dark spots can be seen at the same positions along the readout direction in each channel (see Figure \ref{fig:cross-talk}). These are the cross-talk effects caused by saturated stars in other channels. When one amplifier is reading out a saturated pixel, the video signal can contaminate the pixel streams that are being read out simultaneously from the other amplifiers. In the AST3-II images, the cross-talk effects appear as negative ghosts of their saturated source stars. The depth of such a ghost is several tens of ADUs below the sky background and varies from channel to channel. Since there are 16 readout amplifiers in our CCD camera, these cross-talk effects are significant. Each saturated star results in 15 negative ghosts in the other 15 readout channels and any real stars at the positions of these ghosts are made to appear artificially faint. A detailed discussion of the causes of this cross-talk is beyond the scope of this work. Here we only describe how we eliminate its influence from our data.  The basic idea can be summarized into the following steps:
\begin{enumerate}
\item Mark all saturated pixels (by setting the value of unsaturated pixels as 0) in every readout channel and combine all channels into a single template channel (flipping the top channels vertically) which reflects the positions of all the saturated stars/pixels in the entire images.
\item Redistribute this template channel into all 16 channels (flipping the top channels vertically) to produce a copy of the original image showing the locations of all the ghosts.
\item Find the difference between the bottom of each ghost spot and the local sky background, then add this back to the pixels inside the spot. Repeat this process for every readout channel.
\end{enumerate}
However, there are some complicating factors. First, some ghosts may overlap with each other and the overlapped region is darkened twice. Second, a saturated star always generates a halo that also causes cross-talk, even if the halo itself is not saturated. Thus the corresponding ghost may have a halo, and this halo region may escape from the very first marking step (which marks only saturated pixels). As a result, every ghost has its own structure and the depth within a ghost is not constant. Finally, the gains of amplifiers are different from channel to channel. So the depths of the ghosts caused by the same saturated star are different in different channels. And the depths of ghosts within the same channel, but caused by sources in different channels, are different as well. To solve these problems, we decompose each ghost into three parts: the core region which is caused by the saturated star itself, the halo region which is caused by charge bleed from the saturated star and the overlapped region of two or more saturated stars. The pixels within the halo region are not saturated, but contain extra charge due to charge bleed issues caused by the neighboring saturated pixels. The template channel is also decomposed into three corresponding layers (see the captions of panel a., b., c. and d. in Figure \ref{fig:cross-talklayers} for more details.). Within each layer, we identify all connected pixels and group them as a `sub-ghost'. This is similar to identifying stars within a image when performing photometry. We then redistribute each layer into all 16 channels to show the locations of all `sub-ghosts' in the original image. Finally, we examine every identified `sub-ghost', fit its median depth and add it back to the pixels affected by this `sub-ghost' in the original image. This process is repeated for all readout channels. Figure \ref{fig:cross-talkcorrection} shows the result of our cross-talk correction. Darkening caused by the cross-talk effects is reduced and less than the variation of the local sky background in the corrected image.

\subsubsection{Flat-field Correction}\label{sec:flatfield}
During twilight periods before the polar winter in 2016, hundreds of twilight sky images were taken. These twilight images are combined to produce the master flat field. Owing to the relatively large FOV of 4.3 deg$^2$, a sky brightness gradient of $\sim$1-10\% remains in individual twilight images after preprocessing for cross-talk, overscan and dark current (for a detailed treatment of dark current see \citealp{Ma14}). Two hundred twilight images were  selected to model this sky brightness gradient. For each of the selected images, the brightness gradient was first fitted with an empirical function based on the Sun's altitude and position angle. Then the fitted gradient was divided from each twilight image, and the resulting 200 twilight images were median-combined to produce a master flat-field. The final flat-field error is well below $1\%$. More details of the brightness gradient model and uncertainties in the flat-field  can be found in \cite{Wei14}.

\subsubsection{Electromagnetic Interference Correction}\label{sec:EIcorrection}
\begin{figure}
\centering
\includegraphics[width =1.0 \textwidth]{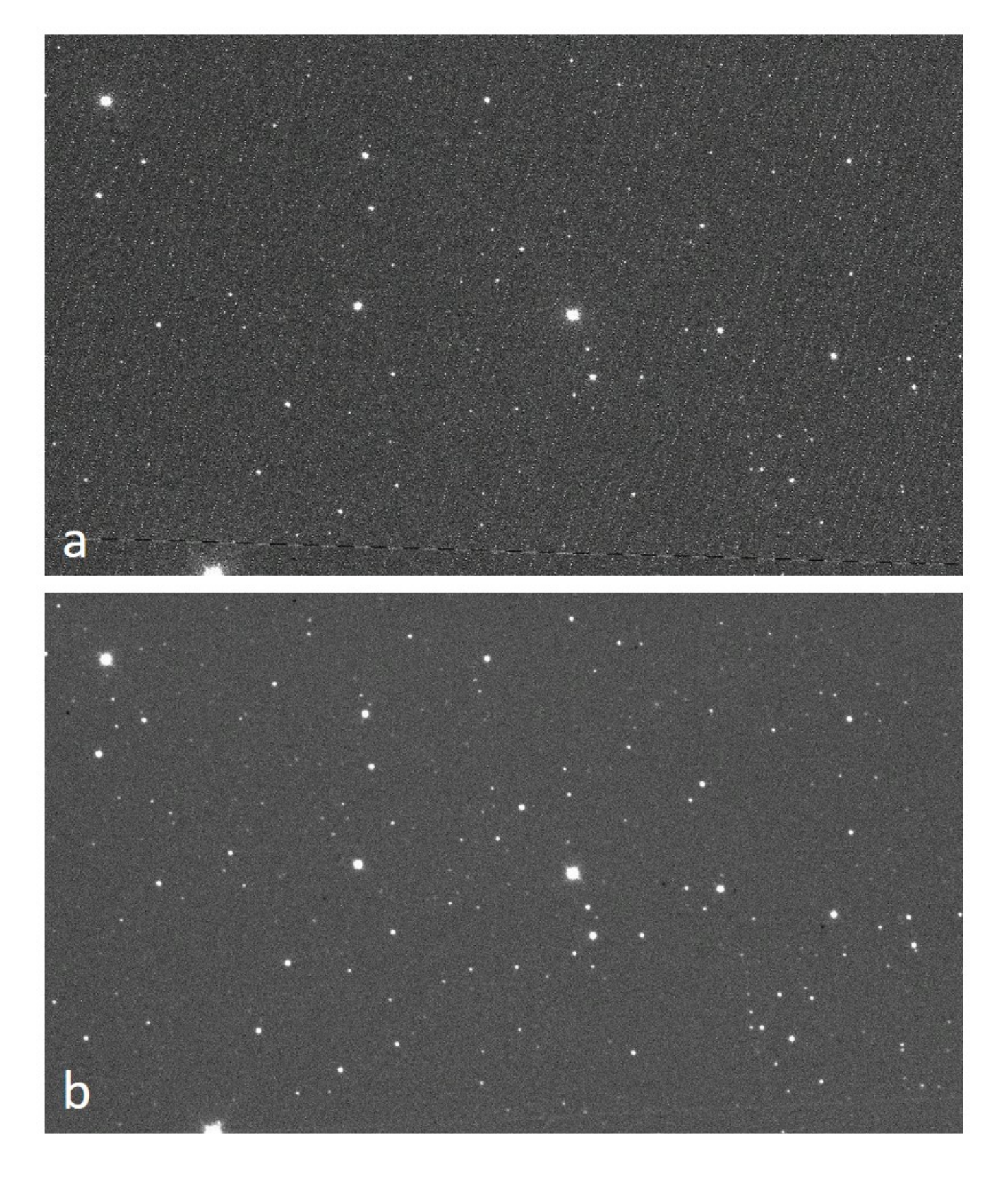}
\caption{Electromagnetic interference correction. Panel a. Strong fringes can be seen across the whole image. Panel b. After co-added stacking of three adjacent frames, the relative intensity of fringes is reduced. \label{fig:interferencecorrection}}
\end{figure}

Another feature of our CCD images is electromagnetic interference. Straight and inclined fringes can be easily identified  in each channel (see the panel a. in Figure \ref{fig:interferencecorrection}). The inclined angles are different in the top  and bottom halves of each image, thus a series of ``$<$''-shape fringes are formed and spread across the entire image. Pixels affected by these fringes will be brightened or darkened by about 20-50\,ADU, which has to be corrected before we can achieve millimagnitude precision. These interference fringes are thought to be caused by noise from the camera's power supply being electrically and/or magnetically coupled into the CCD readout signal chain (though this has not yet be confirmed). The best way to remove this noise would be at the source by improving the grounding and shielding of the entire power supply system.
In the absence of such a solution, we model the positions and influences of each noise fringe and correct them in software. However, this has turned out not to be easy. While the pattern of the fringes remains very similar, they move across the image with time. So, the affected pixels and impact on them vary from frame to frame. Even within a single fringe, only a fraction of the pixels are affected and the amplitudes of the affected pixels also vary. It is very difficult to identify affected pixels, especially those that are also overlapped by stars. Our adopted solution is not yet ideal, but it is simple.
We take three 10-second exposures for the same field, resample and match the last two frames to the first one to guarantee pixel alignments. \textbf{Finally}, all three pixel-aligned images are median combined by the \emph{Swarp} code \citep{Bertin02} to produce a new image with lower background variation \citep{Zackay17}. Since the fringe pattern is moving, a pixel is usually affected in only a single frame of the three. Therefore, the median combination may have a good chance to reject the polluted value as an outlier and reveal the real sky background. This simple process helps to reduce the influence of the fringe noise to an acceptable level. A comparison of images before and after stacking is presented in Figure \ref{fig:interferencecorrection}.

\subsection{Photometry and Calibration }
\begin{figure}
\centering
\includegraphics[width=\textwidth]{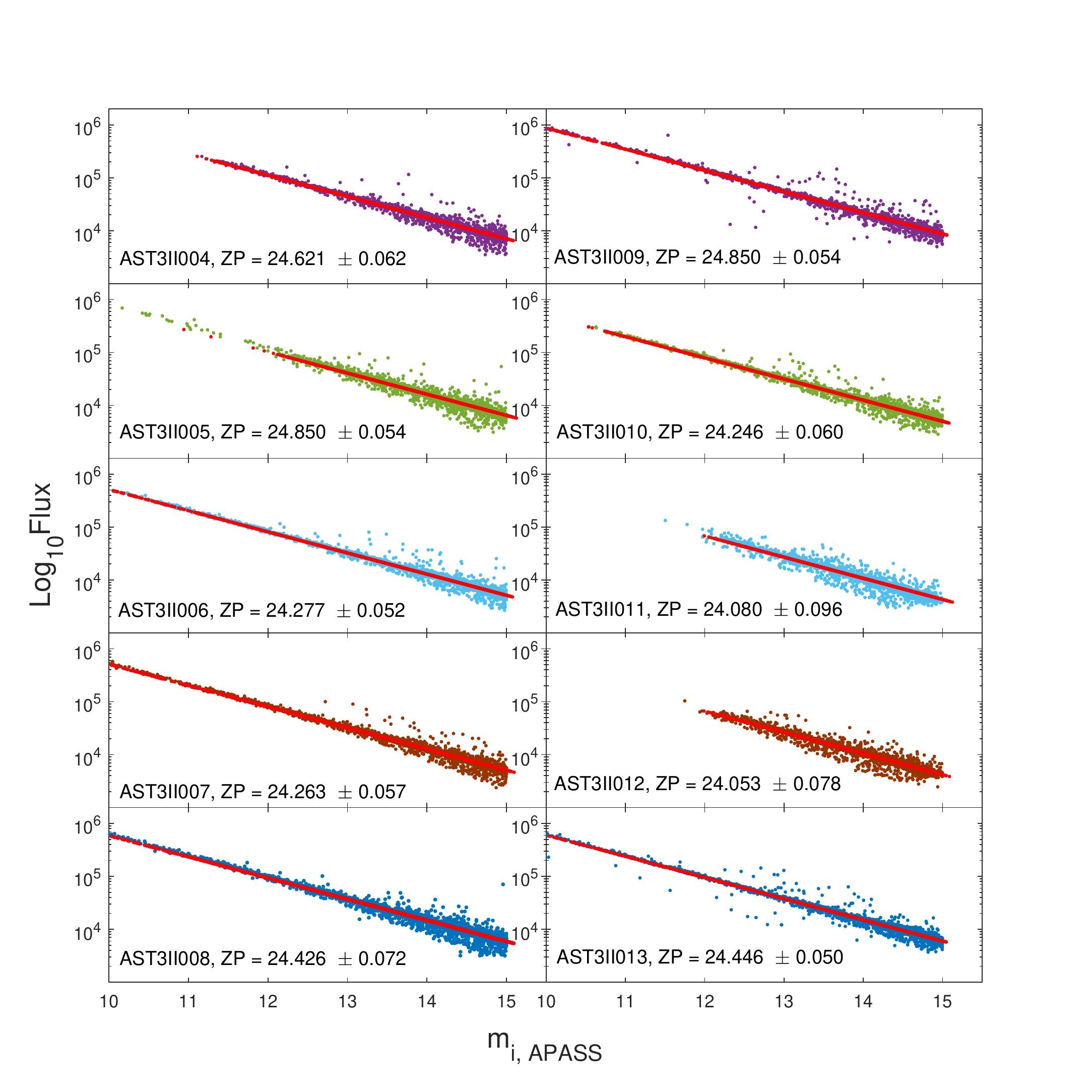}
\caption{The fitted zero-point magnitude ($\textit{i}_{\mbox{zp}}$) in Sloan $\textit{i}$-band for each target field. The y-axis is the logarithmic flux ($\log \mbox{F}$) measured by the AST3-II and the x-axis is the $\textit{i}$-band magnitude ($\textit{i}_{\mbox{apass}}$) in APASS. The solid red line shows the fitting result of this relation: $\textit{i}_{\mbox{apass}} = \textit{i}_{\mbox{zp}} - 2.5\log \mbox{F}$. \label{fig:zeropoint}}
\end{figure}

\begin{figure}
\centering
\includegraphics[width=\textwidth]{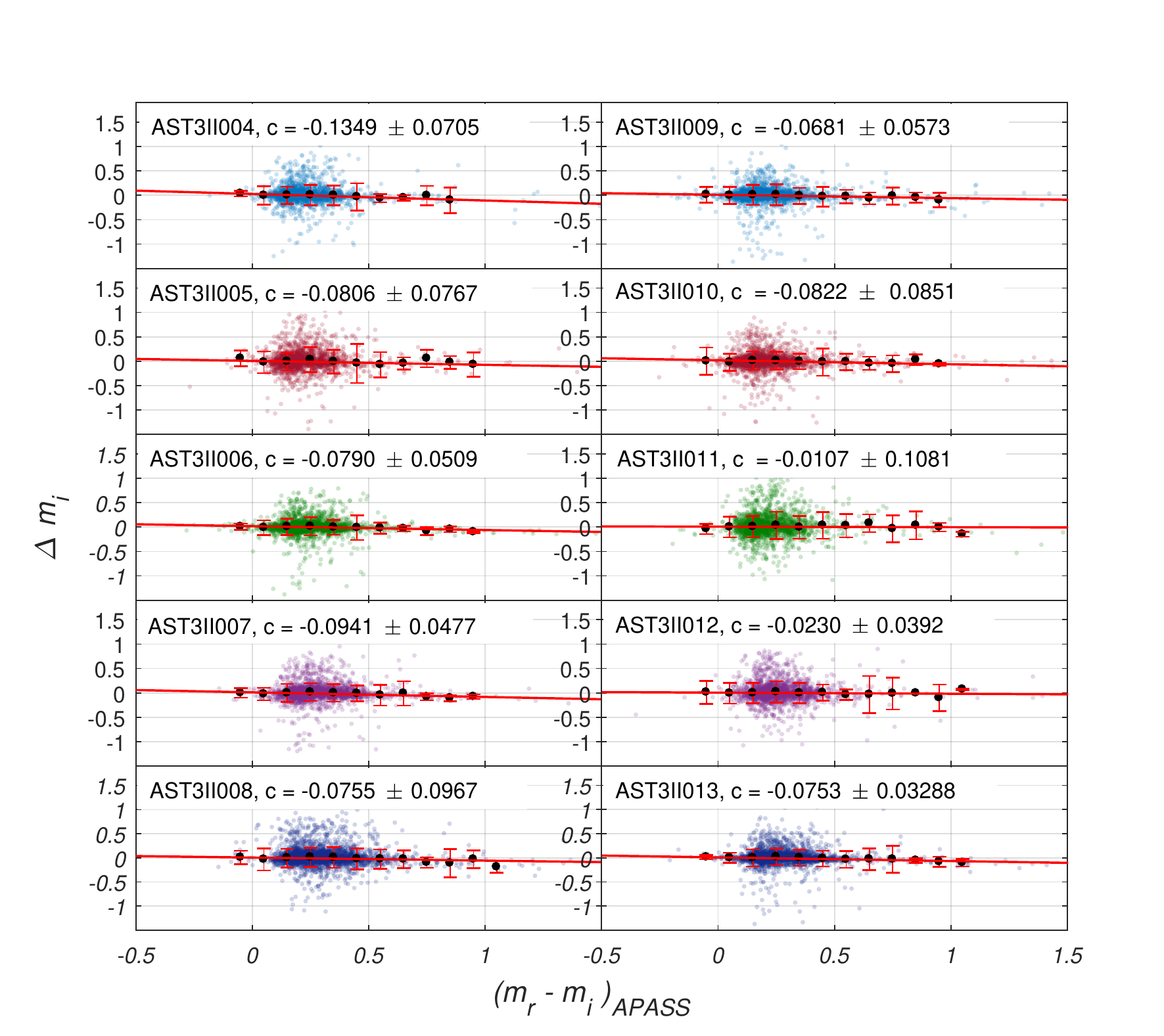}
\caption{Color dependence of magnitude calibration in each target field. Here we demonstrate the relation between $\Delta \textbf{m}_\textit{i}$, the difference between the calibrated $\textit{i}$ magnitude and the corresponding the $\textit{i}$ magnitude in the APASS catalog, and the color term in APASS ($textbf{m}_\textit{r} - textbf{m}_\textit{i}$). The solid black dots are median-binned $\Delta textbf{m}_\textit{i}$ with a color interval of $0.1$. Only those stars brighter than $\textbf{\textit{m}}_\textit{i}=12$ are adopted to produce median bins. The error bar of each bin shows the RMS of measurements within each interval. The color dependence is linearly fitted to these bins and showed by a solid red line in each field. The fitted slope of the color dependence of each field is also labelled in each panel.\label{fig:colorissue}}
\end{figure}

The cleaned images from the preceding steps are then registered and coadded by \emph{Astrometry.net} \citep{Lang10} and \emph{Swarp} \citep{Bertin02}, respectively. As we mentioned above, the cadence of stacked images is about 12 minutes with an effective exposure time of $\sim30$\,s. These images are labeled as `scientific images', and are a major component of our data products. Aperture photometry is then performed on these scientific images and star catalogs are generated using Sextractor \citep{Bertin96}. For each image, the total photometry process is divided into two steps. First, we output a simple catalog that contains only the positions and fluxes using optimal apertures (corresponding to the keyword: FLUX\_AUTO). This simple catalog is then cross-matched with the APASS catalog (the AAVSO Photometric All-Sky Survey which is conducted in five filters: Johnson $B$ and $V$, plus Sloan $\textit{g}$, $\textit{r}$, $\textit{i}$ bands)  \citep{Henden15}. The $\textit{i}$-band magnitudes and fluxes of all matched stars are extracted from the APASS and AST3 catalogs, respectively. The zero-point magnitude is then fitted from the power-law relation of magnitude and flux Figure \ref{fig:zeropoint}. Second, we adopt the fitted zero-point magnitude and run Sextractor again to produce the final calibrated catalog. Although we use a Sloan $\textit{i}$-band filter, there may be some systematic offsets or color dependences between our results and that from the APASS database. The major sources may include the manufacturing difference between our filter and the one adopted by APASS, the absence of strong water feature in the $\textit{i}$-band due to the extremely low water vapor in Antarctica and the differences in the magnitude cutoff and colors of the different ensembles of reference stars. To demonstrate these issues, we fit a simple relation, $\Delta \textbf{\textit{m}}_\textit{i} = \textbf{\textit{m}}_{\textit{i},ast3}-\textbf{\textit{m}}_{\textit{i},apass} = c(\textbf{\textit{m}}_\textit{r}-\textbf{\textit{m}}_\textit{i})_{apass}+b$, in each field to show the color dependence of our magnitude calibration. $c\equiv d\Delta \textbf{\textit{m}}_\textit{i} / d(\textbf{\textit{m}}_\textit{r}-\textbf{\textit{m}}_\textit{i}) $ is the slope of the color dependence and $b$ is an magnitude offset. Results from our 10 target fields are shown in Figure \ref{fig:colorissue}. The fitted slope and uncertainty of the color dependence in each field is also labeled in each panel. And we find a median value of the slope, $c_{median} = -0.0771$ , among all the 10 target fields. The slopes of color term are close to zero in all target fields, which indicates that the color issue is insignificant within our result. However it \textbf{is still worth noting} that this color term hasn't been corrected in this data release and it will be considered in the future data releases.

In the calibrated catalog, we output image positions, J2000 equatorial coordinates and their variances, photometry results of optimal aperture and three fixed apertures (8, 10 and 12 pixels, which are, according to our plate scale, $8^{\prime\prime}$, $10^{\prime\prime}$ and $12^{\prime\prime}$, respectively.), parameters listing observation conditions, and parameters from image statistics. All these pieces of information are inherited by the lightcurves and are used as external trend parameters in the following lightcurve detrending process. An example of a catalog header is shown in Table \ref{tab:cataloghead}.

\begin{table*}[h!]
\caption{An example catalog header of an image taken on `2016-05-03T16:34:44.34'. Position and flux related measurements (keywords) are generated by Sextractor\citep{Bertin96}\label{tab:cataloghead}}
\begin{center}
\begin{tabular}{lll}
\hline
\hline
Name                        & Example/description                                 &   Units       \\
\hline
Observation start time      & $2016-05-03T16:34:44.34  $                          &   UT          \\
Exposure time               & $30                      $                          &   sec      \\
CCD temperature             & $217.92                  $                          &   K           \\
Airmass                     & $1.07                    $                          &   mag         \\
Solar altitude              & $-23.87                  $                          &   deg      \\
Moon distance               & $90.82                   $                          &   deg      \\
Zero-point mag              & $21.93                   $                          &   mag         \\
 \#  1  XWIN\_IMAGE\tablenotemark{a}   &   image position in x axis                                 &   pixel       \\
 \#  2  YWIN\_IMAGE                    &   image position in y axis                                 &   pixel       \\
 \#  3  X2WIN\_IMAGE                   &   position variance in x axis                              &   pixel$^2$   \\
 \#  4  Y2WIN\_IMAGE                   &   position variance in y axis                              &   pixel$^2$   \\
 \#  5  XYWIN\_IMAGE                   &   position covariance between x and y axis                 &   pixel$^2$   \\
 \#  6  ERRX2WIN\_IMAGE                &   error of position variance in x axis                     &   pixel$^2$   \\
 \#  7  ERRY2WIN\_IMAGE                &   error of position variance in x axis                     &   pixel$^2$   \\
 \#  8  ERRXYWIN\_IMAGE                &   error of position covariance between x axis and y axis   &   pixel$^2$   \\
 \#  9  CXXWIN\_IMAGE\tablenotemark{b} &   the first ellipse parameter of object's isophotal shape  &   pixel$^{-2}$\\
 \# 10  CYYWIN\_IMAGE                  &   the second ellipse parameter of object's isophotal shape &   pixel$^{-2}$\\
 \# 11  CXYWIN\_IMAGE                  &   the third ellipse parameter of object's isophotal shape  &   pixel$^{-2}$\\
 \# 12  XWIN\_WORLD                    &   Right Ascension (Ra) in J2000                            &   deg         \\
 \# 13  YWIN\_WORLD                    &   Declination (Dec) in J2000                               &   deg         \\
 \# 14  ERRX2WIN\_WORLD                &   error of position variance in Ra                         &   deg$^2$     \\
 \# 15  ERRY2WIN\_WORLD                &   error of position variance in Dec                        &   deg$^2$     \\
 \# 16  ERRXYWIN\_WORLD                &   error of position covariance between Ra and Dec          &   deg$^2$     \\
 \# 17  MAG\_AUTO                      &   Kron-like elliptical aperture magnitude                  &   mag         \\
 \# 18  MAGERR\_AUTO                   &   RMS error for AUTO magnitude                             &   mag         \\
 \# 19  FLUX\_AUTO                     &   flux within a Kron-like elliptical aperture              &   count       \\
 \# 20  FLUXERR\_AUTO                  &   RMS error for AUTO flux                                  &   count       \\
 \# 21  MAG\_APER                      &   fixed aperture magnitude vector                          &   mag         \\
 \# 24  MAGERR\_APER                   &   RMS error vector for fixed aperture mag.                 &   mag         \\
 \# 27  FLUX\_APER                     &   flux vector within fixed circular aperture(s)            &   count       \\
 \# 30  FLUXERR\_APER                  &   RMS error vector for aperture flux(es)                   &   count       \\
 \# 33  FWHM\_IMAGE                    &   FWHM assuming a Gaussian core                            &   pixel       \\
 \# 34  ELONGATION    &   ellipticity of object's isophotal shape                  &      -        \\
 \# 35  BACKGROUND                     &   background at centroid position                          &   count       \\
 \# 36  THRESHOLD                      &   detection threshold above background                     &   count       \\
 \# 37  FLUX\_RADIUS  &   fraction-of-light radius                                 &   pixel       \\
 \# 38  FLAGS\tablenotemark{c}         &   extraction flags                                         &      -        \\
\hline
\end{tabular}
\end{center}
\tablenotetext{a}{All the position related measurements, e.g. XWIN\_IMAGE, XWIN\_WORLD and etc., are performed through a Gaussian window funcion, i.e. \textbf{the centroid coordinates are calculated through an iterative process updated by a Gaussian-weighted integration of coordinates of isophotal pixels. The start point is set to be the first order moment of the source. More details can be found in \cite{Bertin96}.} All `Windowed' measurements are denoted with a `WIN'. `\_IMAGE' means in pixel coordinates and `\_WORLD' means in equatorial coordiantes. }
\tablenotetext{b}{CXXWIN\_IMAGE, CYYWIN\_IMAGE and CXYWIN\_IMAGE are three ellipse parameters describing the shape of object's isophotal ellipse. 
$\mbox{CXXWIN\_IMAGE}(x - \bar{x})^2 + \mbox{CYYWIN\_IMAGE}(y -  \bar{y})^2 + \mbox{CXYWIN\_IMAGE}(x -  \bar{x})(y - \bar{y}) = R^2$, where $x$ and $y$ are XWIN\_IMAGE and YWIN\_IMAGE, $\bar{x}$ and $\bar{y}$ are mean values of XWIN\_IMAGE and YWIN\_IMAGE, respectively. The last constant $R=3$.}

\tablenotetext{c}{FLAGS contains, coded in decimal, all the extraction flags as a sum of powers of 2: 
1: The object has neighbours, bright and close enough to signifcantly bias the MAG\_AUTO photometry, or bad pixels (more than $10\%$ of the integrated area affected).
2: The object was originally blended with another one. 
4: At least one pixel of the object is saturated (or very close to).
8: The object is truncated (too close to an image boundary).
16: Object’s aperture data are incomplete or corrupted.
For example, an object close to an image border may have FLAGS = 16, and perhaps FLAGS = 2+4+8+16 = 30. 
}
\end{table*}

\subsection{Lightcurve Detrending}
The general concept behind our lightcurve ``polishing'' process is to perform blind detrending that removes the common variations among all targets. Two such methods have been commonly employed in data reduction for wide-field photometric surveys: SYSREM (SYStematic effects REMove, \citealp{Tamuz05}) and TFA (Trend Filtering Algorithm, \citealp{Kovacs05}). We implemented a TFA-like algorithm with some minor adjustments. First,  the reference stars of a target lightcurve were selected according to their distances to the target star (e.g., they must be in the same readout channel as the target star). Second, we built an individual trend matrix (whose columns are magnitudes of reference stars and external parameters) for each target lightcurve and updated it dynamically according to the correlation coefficients between the brightness variations of the target and its reference stars. Those stars with large variations were then clipped from the reference star list after iteration. Third, the external parameters of the target star, e.g., its airmass, distance to the moon, FWHM, local background variation, and etc. (see Table \ref{tab:lightcurvehead} for details), were recorded simultaneously with the brightness measurements and are now appended to its trend matrix as additional columns. The variations present in these external parameters will be used to detrend the target lightcurve. This is usually called the EPD (external parameter detrending) process. Here we combine the EPD- and TFA-like processes together and detrend the target lightcurve using its external parameters and brightness of reference stars simultaneously. Finally,  instead of the Least-Square Fit to the linear combination of the reference lightcurves in the standard version of TFA, the detrending process in our pipeline is achieved by a Multi-variable Linear Regression with each column in the trend matrix marked as a `variable'. We noticed that some outliers, caused by miss-matches of stars or bad weather, caused serious problems in the detrending processes, and occasionally they may crash the BLS (Box Least Square fitting) searching function that follows. To eliminate these outliers and retain large physical variations at the same time, we performed a Gaussian Regression to model the time-dependent variations within each lightcurve, and we then clipped all measurements that are more than 3-$\sigma$ away from the model before using them in the detrending process.

\begin{figure}
\centering
\includegraphics[width=\textwidth]{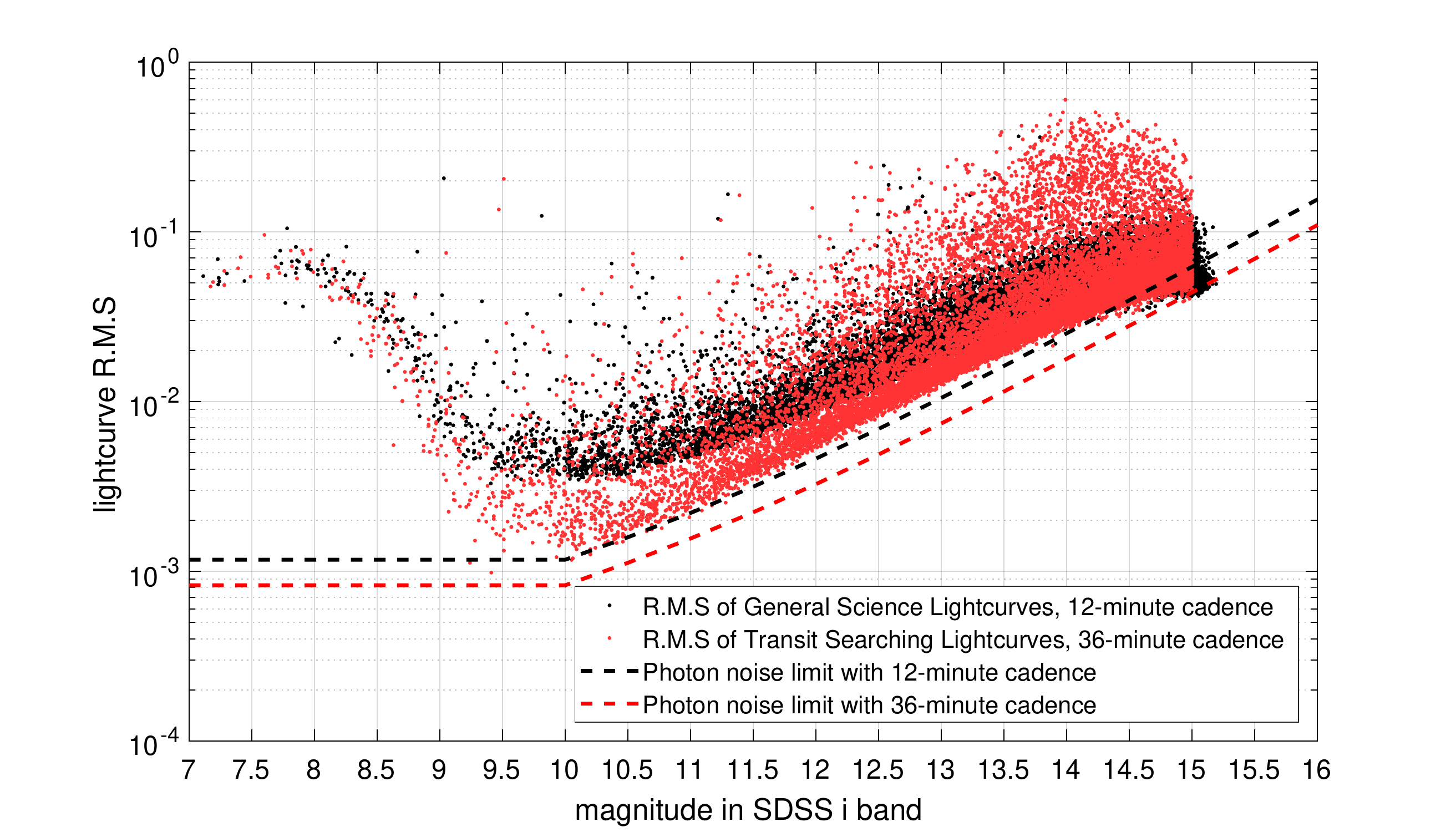}
\caption{Lightcurves RMS VS. Magnitude of 26578 stars. \textbf{Each point represents the overall RMS of a detrended lightcurve with time spanning the whole observation campaign.} The black points are lightcurves with a cadence of $\sim12$ minutes and the red ones are lightcurves binned to 36 minutes. The expected photon limits (black dash for \textbf{12-minute} cadence and red dash for \textbf{36-minute} cadence) against magnitude are also overploted. Note that when we calculate the photon noises, sky background noise, readout noise, saturation limit and averge number of measurements falling in the binned interval are also considered. So the photon limit of \textbf{36-minute} cadence is not simply improved by a factor of $\sqrt{3}$, but a factor around $\sim \sqrt{2}$, to that of \textbf{12-minute} cadence.  Stars brighter than $\textbf{\textit{m}}_\textit{i}=10$ are likely to be saturated and suffer large variations. However, we still found some obvious variables in this magnitude range (see Section \ref{sec:results}). \label{fig:rms}}
\end{figure}

In this data release we present two versions of lightcurves for general science usage: the raw lightcurves that contains all the original information (see Table \ref{tab:lightcurvehead}), and the detrended lightcurves that contain only time, magnitude and the error in the magnitude. The cadence of these lightcurves is $\sim12$ min and the best RMS after detrending is $\sim4$ mmag at the bright end, $\textbf{\textit{m}}_\textit{i}=10$. This precision is good enough to search binaries or pulsating variables, however, to search for transiting exoplanets we need higher precision. To achieve this the lightcurves are then further ``polished'' and binned to 36 min, after which they achieve a best RMS of 2 mmag at $\textbf{\textit{m}}_\textit{i}=10$ (see Figure \ref{fig:rms}). The curves of expected photon limited RMS against magnitude are also overploted in Figure \ref{fig:rms}. Note that when we calculate the photon noises, sky background noise, readout noise, saturation limit and averge number of measurements falling in the binned intervals are also considered. As a result, the photon limit of lightcurves with \textbf{36-minute} cadence is not simply improved by a factor of $\sqrt{3}$, but a factor around $\sim sqrt{2}$, to that of \textbf{12-minute} cadence. The RMS of binned lightcurves is improved a lot at the bright end, however there are still hidden systematic errors that prevent us from reaching the photon limited precision. The detrending process and further polishing determine the final precision of our lightcurves and directly influence the detection rate of transit signals. The detailed algorithm and parameter settings, e.g., the number of reference stars, the selection of external parameters, will be presented in a following paper \citep{Zhang18b} which concentrates on the transiting exoplanet searching.

\begin{deluxetable*}{c|lll}
\tablecaption{Columns in raw lightcurves. Keywords have the same meaning as those in \textbf{Talbe \ref{tab:cataloghead}.} \label{tab:lightcurvehead}}
\tablehead{
\colhead{Column No.} & \colhead{Name} & \colhead{Explanation} & \colhead{Units}
}
\startdata
 1      &  time                &   AST3 observation time \textbf{in UT} (JD --2456000)               &   day         \\
 2      &  XWIN\_IMAGE         &   image position in x axis                                 &   pixel       \\
 3      &  YWIN\_IMAGE         &   image position in y axis                                 &   pixel       \\
 4      &  X2WIN\_IMAGE        &   position variance in x axis                              &   pixel$^2$   \\
 5      &  Y2WIN\_IMAGE        &   position variance in y axis                              &   pixel$^2$   \\
 6      &  XYWIN\_IMAGE        &   position covariance between x and y axis                 &   pixel$^2$   \\
 7      &  ERRX2WIN\_IMAGE     &   error of position variance in x axis                     &   pixel$^2$   \\
 8      &  ERRY2WIN\_IMAGE     &   error of position variance in x axis                     &   pixel$^2$   \\
 9      &  ERRXYWIN\_IMAGE     &   error of position covariance between x axis and y axis   &   pixel$^2$   \\
 10     &  CXXWIN\_IMAGE       &   the first ellipse parameter of object's isophotal shape  &   pixel$^{-2}$\\
 11     &  CYYWIN\_IMAGE       &   the second ellipse parameter of object's isophotal shape &   pixel$^{-2}$\\
 12     &  CXYWIN\_IMAGE       &   the third ellipse parameter of object's isophotal shape  &   pixel$^{-2}$\\
 13     &  XWIN\_WORLD         &   Right Ascension (Ra) in J2000                            &   deg         \\
 14     &  YWIN\_WORLD         &   Declination (Dec) in J2000                               &   deg         \\
 15     &  ERRX2WIN\_WORLD     &   error of position variance in Ra                         &   deg$^2$     \\
 16     &  ERRY2WIN\_WORLD     &   error of position variance in Dec                        &   deg$^2$     \\
 17     &  ERRXYWIN\_WORLD     &   error of position covariance between Ra and Dec          &   deg$^2$     \\
 18     &  MAG\_AUTO           &   Kron-like elliptical aperture magnitude                  &   mag         \\
 19     &  MAGERR\_AUTO        &   RMS error for AUTO magnitude                             &   mag         \\
 20     &  FLUX\_AUTO          &   flux within a Kron-like elliptical aperture              &   count       \\
 21     &  FLUXERR\_AUTO       &   RMS error for AUTO flux                                  &   count       \\
 22--24 &  MAG\_APER           &   fixed aperture magnitude vector                          &   mag         \\
 25--27 &  MAGERR\_APER        &   RMS error vector for fixed aperture mag.                 &   mag         \\
 28--30 &  FLUX\_APER          &   flux vector within fixed circular aperture(s)            &   count       \\
 31--33 &  FLUXERR\_APER       &   RMS error vector for aperture flux(es)                   &   count       \\
 34     &  FWHM\_IMAGE         &   FWHM assuming a Gaussian core                            &   pixel       \\
 35     &  ELONGATION          &   ellipticity of object's isophotal shape                  &      -        \\
 36     &  BACKGROUND          &   background at centroid position                          &   count       \\
 37     &  THRESHOLD           &   detection threshold above background                     &   count       \\
 38     &  FLUX\_RADIUS        &   fraction-of-light radius                                 &   pixel       \\
 39     &  FLAGS               &   extraction flags                                         &    -          \\
 40     &  exposure\_time      &   exposure time                                            &   sec         \\
 41     &  CCD\_temperature    &   temperature of CCD chip                                  &   K           \\
 42     &  airmass             &   airmass                                                  &    -          \\
 43     &  sun\_altitude       &   altitude of the Sun                                      &   deg         \\
 44     &  moon\_distance      &   angular distance to the Moon                             &   deg         \\
 45     &  zeropoint           &   zero-point magnitude                                     &   mag         \\
 \enddata
 \end{deluxetable*}

\subsection{Periodic Signal Searching}
\begin{figure}
\centering
\includegraphics[width=\textwidth]{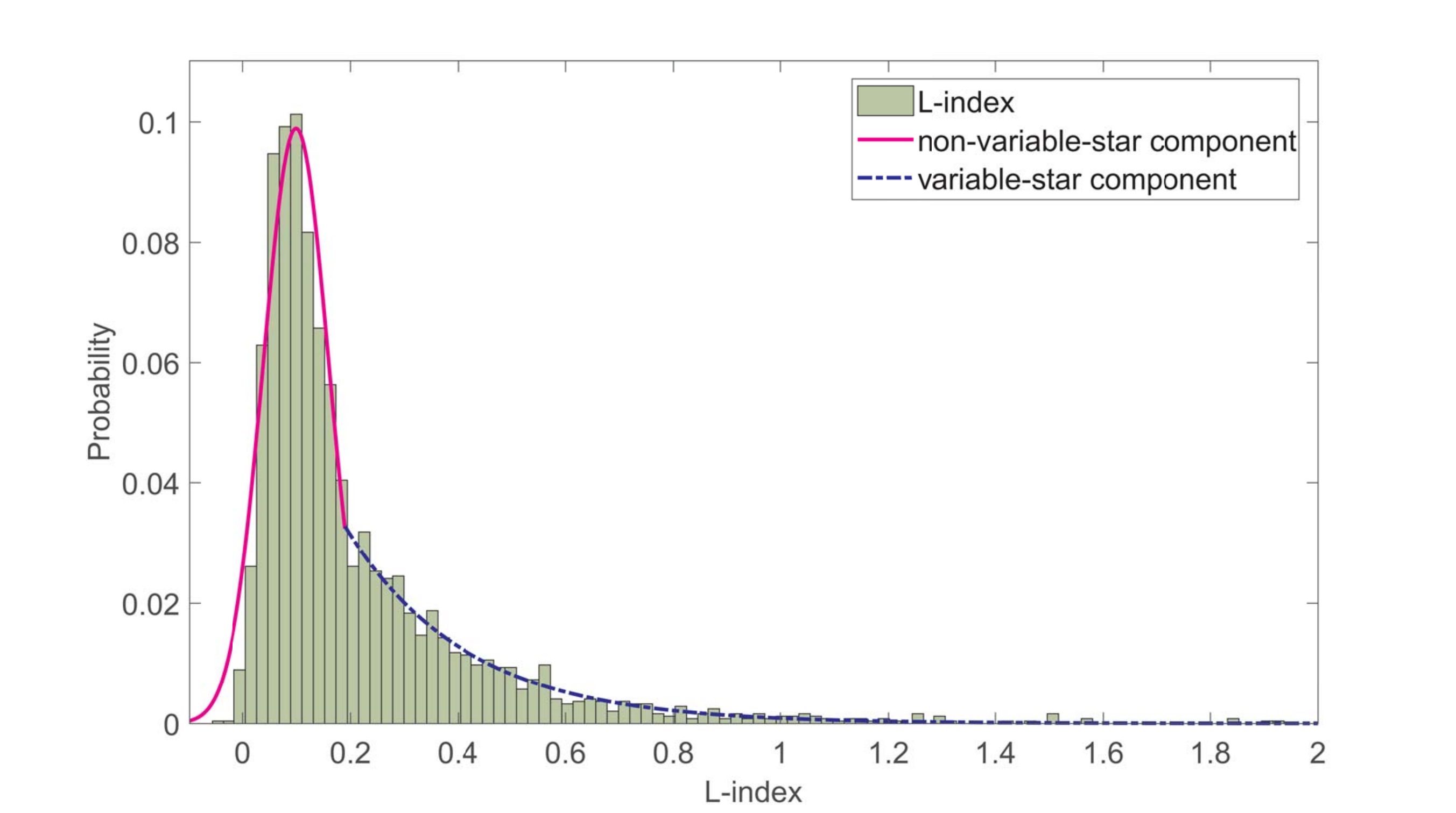}
\caption{Distribution of the Welch-Stetson variability statistic L for the 26578 bright stars in the AST3-II data-set. Two components can be fitted from this distribution: an exponential profile corresponds to variable stars and a Gaussian profile which represents stars with insignificant variations.\label{fig:Lindex}}
\end{figure}

The `L' variability index \citep{Welch93,Stetson96} is used to select potential variable candidates. For each lightcurve we calculated its L-index using the $-Jstet$ command of VARTOOLS \citep{Hartman16}. The overall distribution of L-index in our sample is shown in Figure \ref{fig:Lindex}. Two components can be fitted from this distribution, one of which shows a Gaussian profile and corresponds to stars with insignificant variation and the other of which shows an exponential tail and is likely to correspond to variable stars.  The intersection of these two profiles is at $L\approx0.19$. Stars with $L\geq0.19$ are selected as variable star candidates and delivered to the ``Periodic Signal Searching'' procedure. Periodic and sinusoidal variations are then identified using both the Generalized Lomb-Scargle \citep{Zechmeister09,Press92} and the AoV \citep{Schwarzenberg89,Devor05} methods. In practice, we integrate the $-LS$ and $-aov$ commands of VARTOOLS  into our pipeline. This module produces two data-sets from the LS and AoV processes. Periodic signals with SNR above 1.5-$\sigma$ in both data-sets, and those with SNR above 3-$\sigma$ in single data-sets, then proceed to visual inspection. Some targets detected by the BLS method \citep{Kovacs02} in the latter Transit Signal Searching module, most of which are eclipsing binaries, are also appended to the results of this module. Only obvious and regular periodic signals that could be classified as binaries or pulsating stars are selected and published in this work. Signals that classified as rotating spots, rotating ellipsoidal variables and other semi-periodic/irregular variables will be studied and released in future works (e.g. Fu et al., in preparation).

To determine the period of a variable that has multiple candidate periods, we fold its lightcurve with the period found from each of the LS, AoV and BLS methods, and select the one that results in the smallest RMS. Then we re-run the corresponding method at a range of $\pm0.1$ days around the selected period with a time step of 0.0001 days. We further fold the lightcurve according to the updated period and inspect the plot visually. Confusions are usually caused by some binaries whose primary and secondary eclipses are similar. In this case, the half period often mimics the true period. To distinguish the tiny difference between eclipses, we double the period when we fold the lightcurve. Therefore, the plot shows the phase from 0.0 to 2.0 and the second phase is not identical to the first one. This ensures that the plot shows two or four deep eclipses, no more no less (if it has only one or more than 4 deep eclipses, then we know the original period is wrong, see panels (b1), (b2) and (c1), (c2) in Figure 14), and makes it easier to compare the odd and even eclipses. In most cases, we will find a slight but distinguishable difference between the depth of the odd and even eclipses (e.g. AST3II096.5462-68.5901 in panel (b1) of Figure 14) or a tiny secondary eclipse between two adjacent deep eclipses (e.g. AST3II106.3396-69.1610 in panel (c1) of Figure 14). For contact eclipsing binaries, If the odd and even eclipses are just identical, then we assume the primary and secondary eclipses are both present but indistinguishable and select the period which makes 4 eclipses from phase 0.0 to 2.0 (e.g. AST3II108.2786-71.7907 in Figure 16). However, for detached eclipsing binaries, if there is no obvious secondary eclipse found, we assume the secondary eclipse is present but undetectable at our precision and select the period that makes two eclipses from phase 0.0 to 2.0 (e.g. AST3II098.1909-70.8216 in Figure 17).

\section{Results} \label{sec:results}
\begin{figure}
\centering
\includegraphics[width=\textwidth]{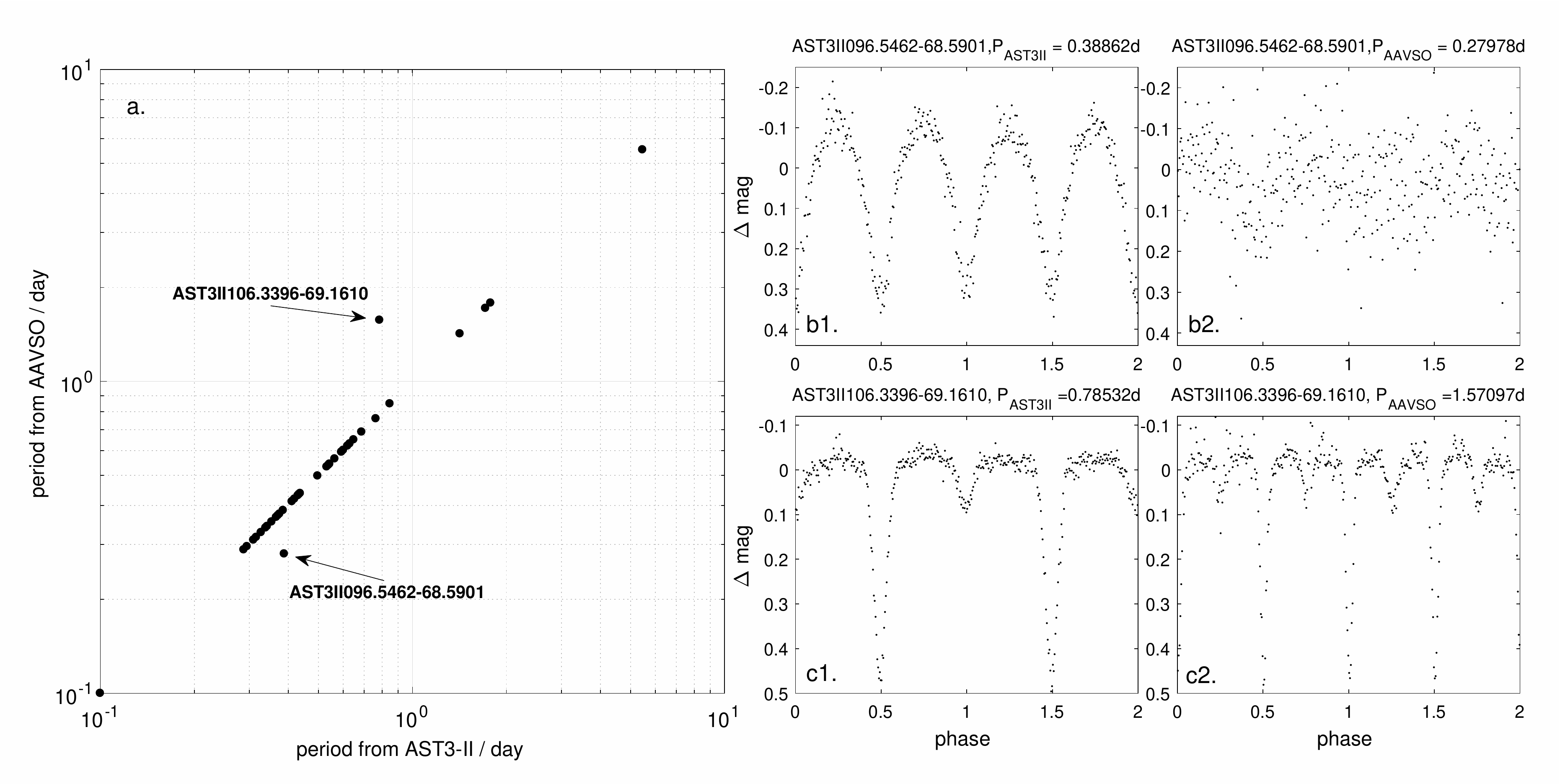}
\caption{Comparison between the periods of 42 common variables found by AST3-II and present in the AAVSO database. Panel a. Periods from AAVSO VS. Periods given in AAVSO database. There are only two common variables whose periods are not consistent. Panel b1. Phase-folded Lightcurve of target AST3II096.5462-68.5901 according to the period found by AST3-II $P_{\mbox{\scriptsize AST3II}} = 0.38862$ days. Panel b2. Phase-folded Lightcurve of target AST3II096.5462-68.5901 according to the period given in AAVSO database $P_{\mbox{\scriptsize AAVSO}} = 0.27978$ days. Panel c1. Phase-folded Lightcurve of target AST3II106.3396-69.1610 according to the period found by AST3-II $P_{\mbox{\scriptsize AST3II}} = 0.78532$ days. Panel c2. Phase-folded Lightcurve of target AST3II106.3396-69.1610 according to the period given in AAVSO database $P_{\mbox{\scriptsize AAVSO}} = 1.57097$ days. For both variables, the AST3-II period produces a better result. \label{fig:cross-match}}
\end{figure}

\begin{figure}
\centering
\includegraphics[width=\textwidth]{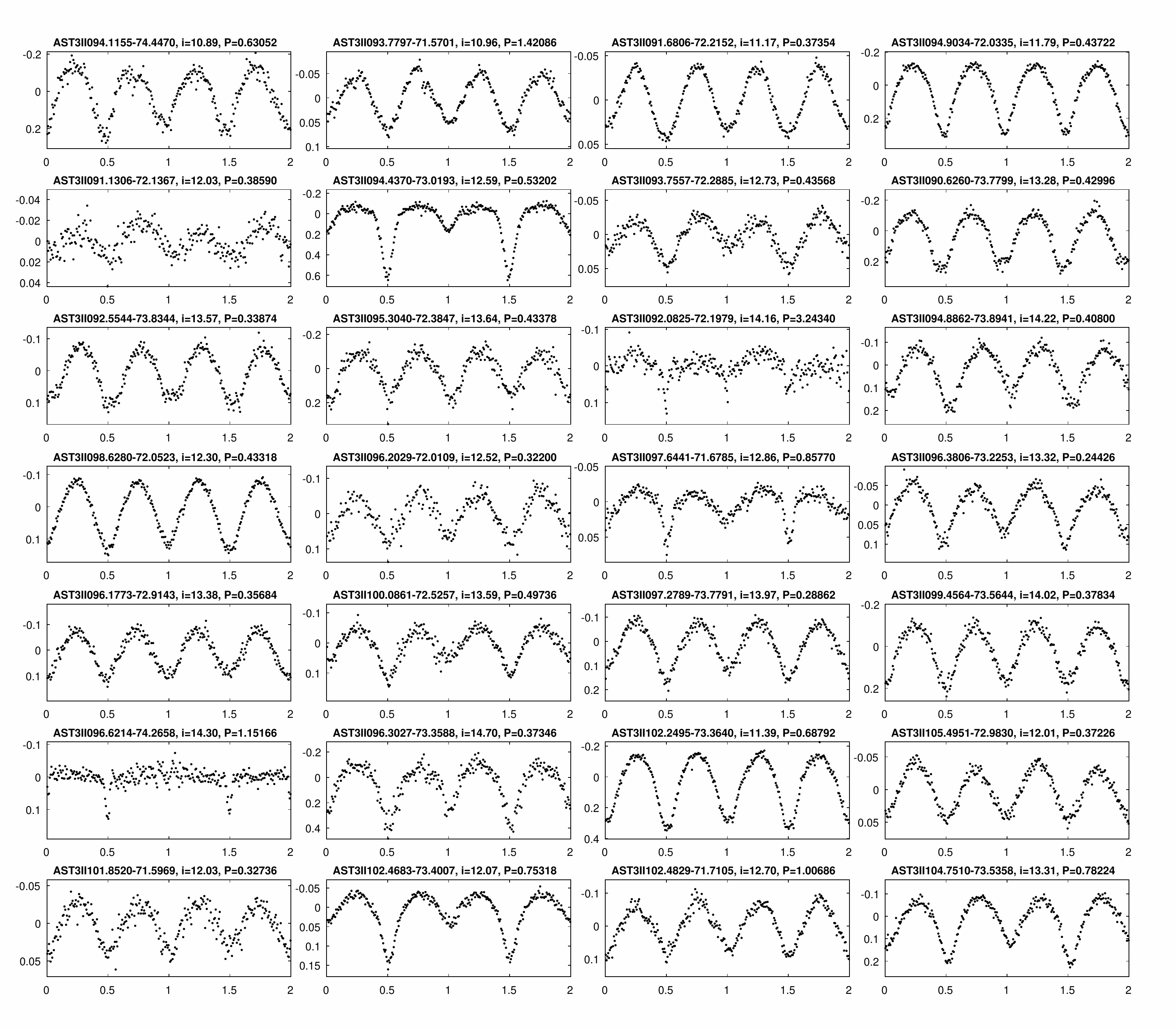}
\caption{Eclipsing binaries found within the data obtained in 2016 by AST3-II. The label above each panel contains the $\textit{i}$-band median magnitude (which has been subtracted) for the curves, and the period in days. The x-axis and the y-axis of each panel are the phases [0,2] (Note that each lightcurve is folded to 2 times of its period, so the measurements around phase 1.5 is not identical to that around phase 0.5.) and the vartiation in magnitude $\Delta \textbf{\textit{m}}_\textit{i}$, respectively.\label{fig:binaries1}}
\end{figure}

\begin{figure}
\centering
\includegraphics[width=\textwidth]{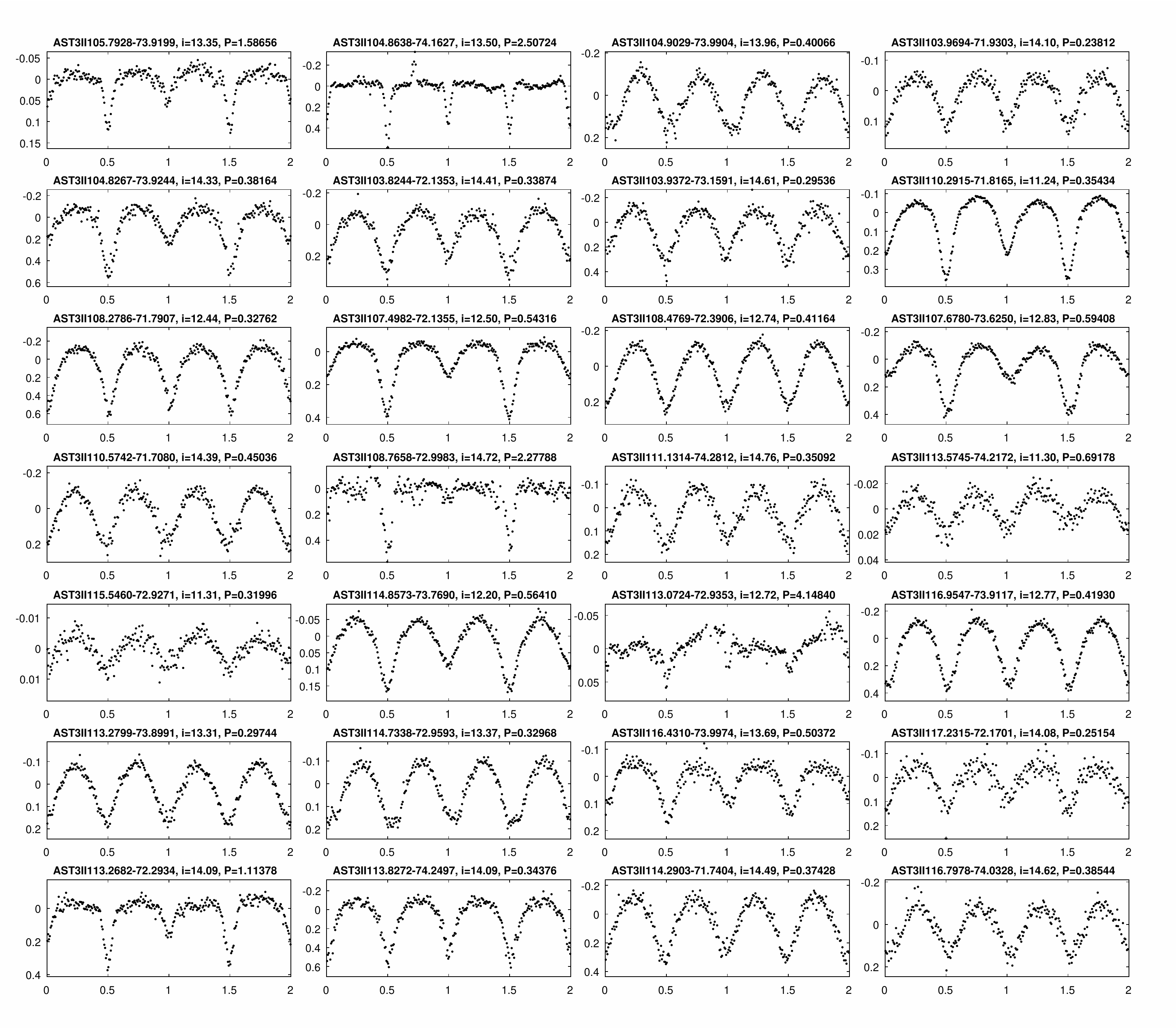}
\caption{Eclipsing binaries found within the data obtained in 2016 by AST3-II, continued...\label{fig:binaries2}}
\end{figure}

\begin{figure}
\centering
\includegraphics[width=\textwidth]{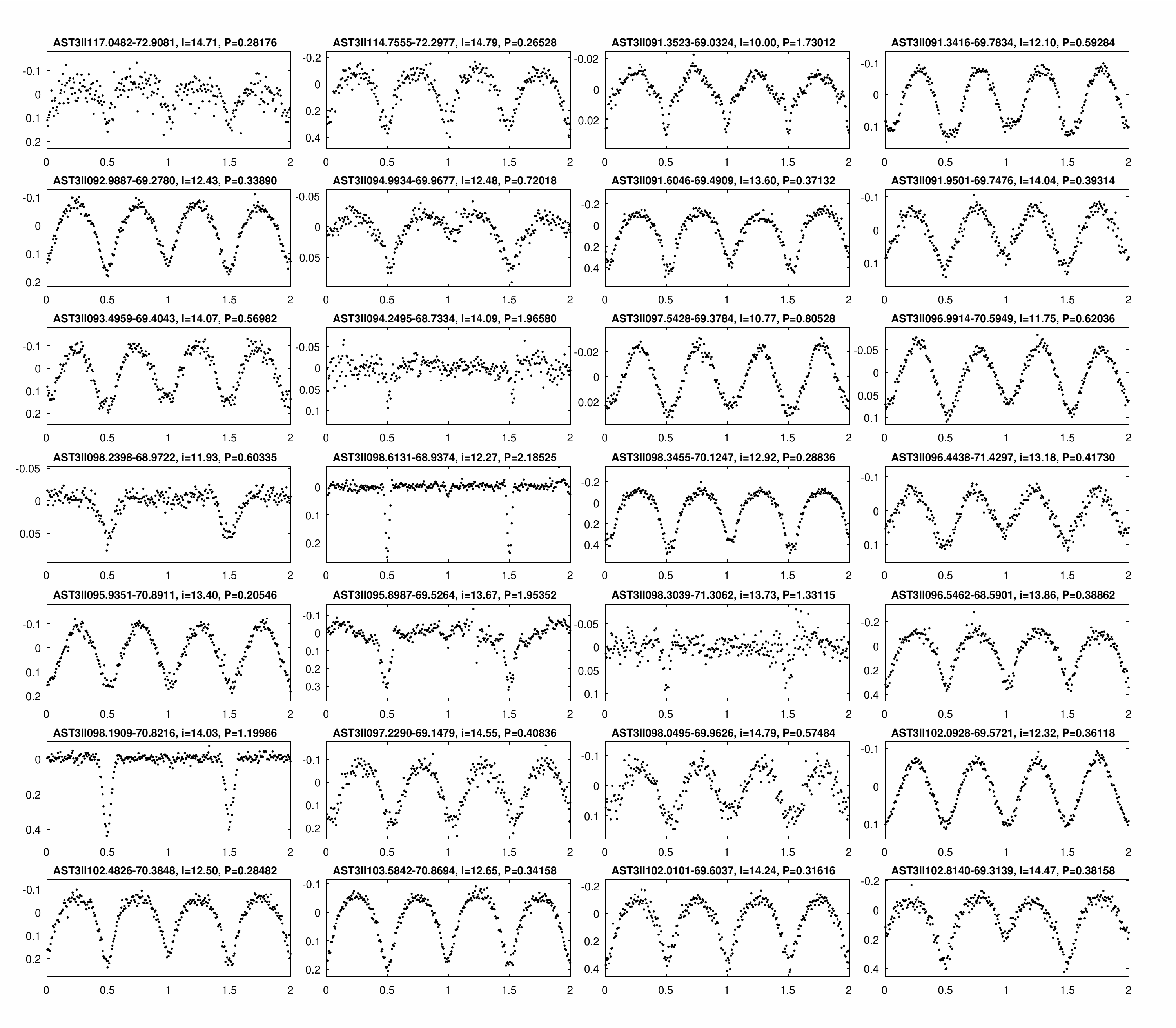}
\caption{Eclipsing binaries found within the data obtained in 2016 by AST3-II, continued...\label{fig:binaries3}}
\end{figure}

\begin{figure}
\centering
\includegraphics[width=\textwidth]{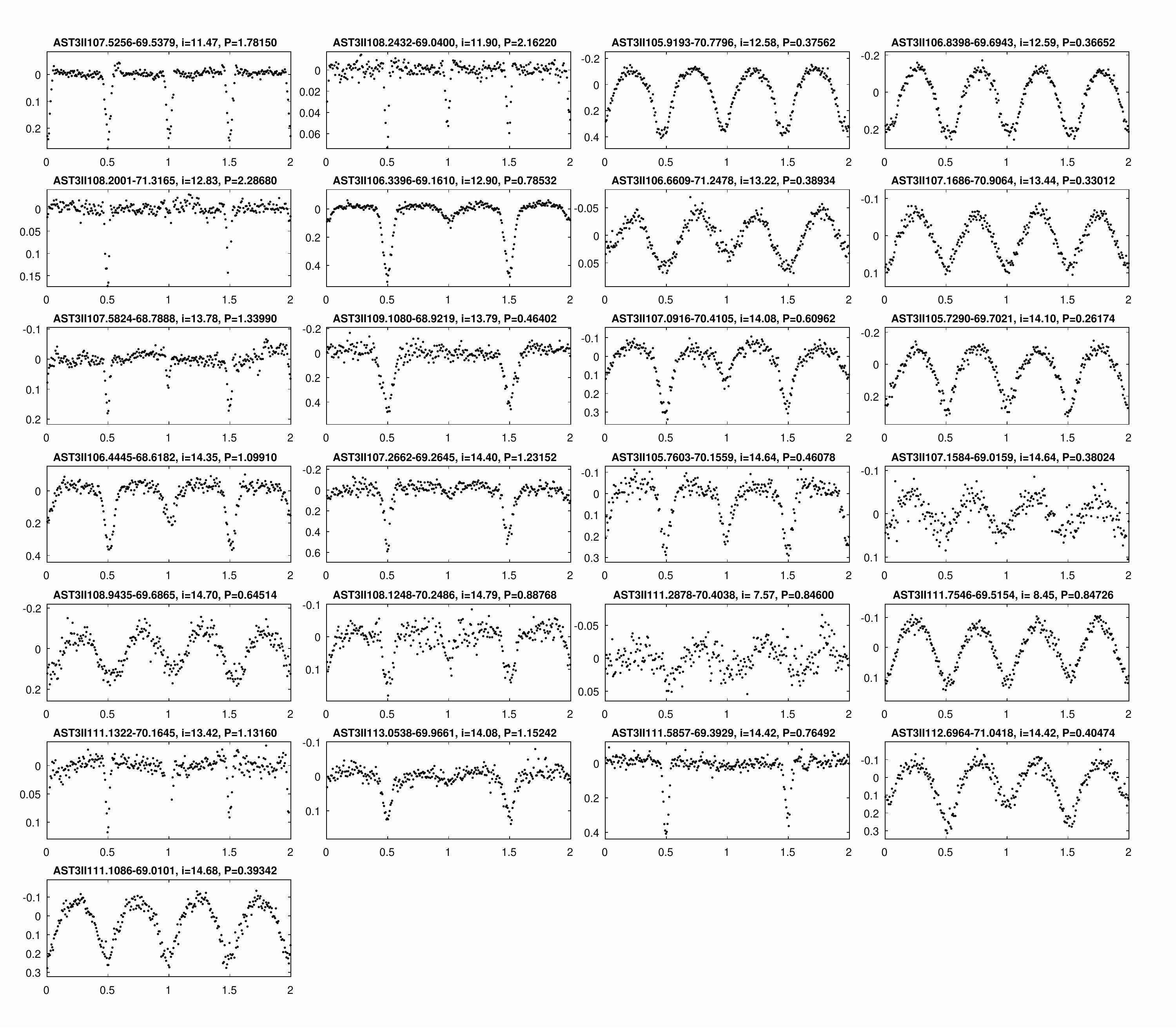}
\caption{Eclipsing binaries found within the data obtained in 2016 by AST3-II, continued...\label{fig:binaries4}}
\end{figure}

We scanned 10 fields within the Southern CVZs of \textit{TESS} in the polar nights of 2016. The raw image data acquired by the AST3-II telescope were reduced. Each catalog that is released along with this work is first cross-matched with the APASS point source catalog in the Sloan-\textit{i} band and is calibrated by the fitted zero-point magnitude (see Figure \ref{fig:zeropoint}). The matching radius is set to be $2^{\prime\prime}$. Although we defocused the telescope (FWHM $\sim 5^{\prime\prime}$ ) and adopted a photometry aperture up to $12^{\prime\prime}$, there is no significant blending issue. The reasons are, first we have avoid the crowd field, e.g. fields close to the LMC (Large Magellan Cloud) , second we adopted a short exposure time (10 seconds) and not to many faint stars were caught, and last we have a high spatial resolution (pixel-scale $\sim1^{\prime\prime}$~pixel$^{-1}$)---neighboring targets can be easily distinguished even if they fall in the same aperture. Lightcurves of stars brighter than 15-th magnitude are extracted, detrended (see Figure \ref{fig:rms}) and also released along with this work. From the these lightcurves, we have found 221 variables within the Southern CVZ of \textit{TESS}, including 117 binaries and 104 pulsating stars. Their phase-folded lightcurves are shown in Figures \ref{fig:binaries1}--\ref{fig:variables4}. Our final variable catalog is cross-matched with the latest AAVSO database and the \textit{Tess} Input Catalog (TIC, \citealp{Stassun17}) using a matching radius of $5^{\prime\prime}$. This large matching radius is in consideration of our defocused PSF. In practice, if multiple matches happened within a matching radius, we select the ones with nearest magnitude. Fortunately, the space distribution of known variables in AAVSO and pre-selected bright stars in \textit{TESS} is quite sparse. So no confusion in cross match was reported by our matching process. We list the IDs of matched targets from both the AAVSO database and the TIC catalog in Table \ref{tab:variables-table}. The target stellar properties such as stellar radius, mass, and variable star type are also listed. We note that there are 67 targets (identified with a CTL flag of 1 in Table \ref{tab:variables-table}) that found in the exoplanet Candidate Target List (CTL) for \textit{TESS}. Targets in the CTL list are preselected objects that have higher priorities to search for transiting exoplanets using short-cadence ($\sim$2 minutes) lightcurves. Our detection of variability in these targets indicates that additional care needs to be taken when searching for transit signals in these targets. There are 42 variables in common between our catalog and the AAVSO database. Panel a.~in Figure \ref{fig:cross-match} shows the comparison between the periods of common variables found by AST3-II and in the AAVSO database. There are only two mismatches in period: AST3II096.5462-68.5901 (ASAS J062611-6835.4) and AST3II106.3396-69.1610 (YZ Vol). We plot their phase curves in panels b1, b2 and c1, c2, according to the periods from AST3-II and the AAVSO database. The periods found from the AST3-II observations (panels b1 and c1) are clearly more convincing.

As shown in Figure \ref{fig:rms}, the valid magnitude range for our survey is from $\textbf{\textit{m}}_\textit{i} = 10$ to $\textbf{\textit{m}}_\textit{i}  = 15$. Most stars brighter than $\textbf{\textit{m}}_\textit{i} =10$ magnitude are likely to be saturated. However, on occasion the extinction caused by bad weather or frosting on the front window of the telescope can be as high as 4 magnitudes. During such times of poor photometric conditions only the brightest stars can be observed, which has the fortunate result that some parts of the lightcurves of very bright stars ($7.5 < \textbf{\textit{m}}_\textit{i} < 10$) are still usable. After filtering out the remaining saturated measurements, the photometric precision from these highly extincted images is sufficient to identify some variables with large periodic variations. For examples, target ``AST3II111.7546-69.5154'' has an magnitude of 8.45 in $\textit{i}$-band and a variation period of 0.84726 days. It has been labeled as an ``EW/KE'' variable \footnote{\textbf{EW, the W Ursae Majoris-type eclipsing variables. KE, the contact systems of early (O-A) spectral type. Their variations are both caused by two ellipsoidal components that are almost in contact and not easy to be classified.}} in the AAVSO data base with a variation period of 0.84743 days. Target ``AST3II092.1221-71.1200'' is of $9.09$ magnitude \textbf{in $\textit{i}$-band} and has a variation period of 1.71539 days. In the AAVSO catalog, its name is  ``ASAS J060829-7107.2'' which is labeled as an ``ACV'' variable \footnote{\textbf{ACV, the $\alpha^2$ Canum Venaticorum variables which are main-sequence stars with spectral types from B8 to A7 and displaying strong magnetic fields. Their brightness vary with a period from 0.5 to 160 days and an amplitudes within 0.1 mag.}} with a period of 1.71443 days.

Another thing worthy of notice in Figure \ref{fig:rms} is that the RMS of faint stars are not improved by binning from 12 minutes to 36 minutes. A possible reason is that the fainter is a star the less data points are in its lightcurve. And most of the valid measurements are usually crowded into those days with good observing conditions. For the other days with large extinction, lightcurves of faint stars are very sparse, say 1-2 points per hour on average. When we divide these time series into equally spaced intervals with 12 or 36 minutes, many intervals are actually empty or have 1 measurement only. As a result the RMS of the entire lightcurve may be dominated by these sparse parts and won't be improved by increasing binning time from 12 minutes to 36 minutes. Another reason is that some systematic errors may still bury in the photometry of stars fainter than $\textbf{\textit{m}}_\textit{i}=14.0$. This effect raises an uncertainty in the RMS at a level of several percent, which is not a serious problem for most variables stars, since the amplitudes of their brightness variations are usually larger than $10 \%$. For transiting exoplanet searching, we abandon stars fainter than 14.0-th magnitude.

\subsection{Variables Found by AST3-II}

\subsubsection{Eclipsing Binaries}
While detailed studies of the variable stars we have found are beyond the scope of this paper, we present a simple classification of the variables according to the shape and frequency analysis of their lightcurves. Since we have a lack of sample lightcurves that have a similar error model to the AST3-II observations, machine learning methods can not be easily employed. However, samples from this work will be used in a training library for future machine classifications. Our classification was done by two groups of people independently using the lightcurve morphology, supplemented by Fourier analysis of the brightness variations in cases where the shape was not clear.  Each variable was labeled only when the two groups achieved the same or similar result, after occasional robust discussion.

The first major category of variables we found are eclipsing binaries, and these can be further sub-divided into Algol-type eclipsing systems (EAs), $\beta$ Lyrae-type eclipsing systems (EBs), and W Ursae Majoris-type eclipsing variables (EWs).
The EA systems have different depths between the primary and secondary minima, and have clearly defined times for the beginning and the end of the eclipses; the most common EA systems are detached eclipsing binaries \citep{Catelan15}, although some of them are semi-detached binaries, as, in fact, is the prototype of this class, Algol \citep{Kolbas15}.
The EB systems always show a continuous variation in brightness and have an obvious deeper primary eclipse. The EW systems also show a continuous change in brightness, but the difference between the depths of the primary and secondary minima is no longer so obvious. Sometimes this difference is indistinguishable under poor photometric precision and
the target may be easily mis-classified as a pulsating star if we only consider the shape of its lightcurve. A frequency analysis of the lightcurve is needed in this situation. The EW systems consist of two components that are almost in contact, thus most of them have periods shorter than one day.

\begin{figure}
\centering
\includegraphics[width=\textwidth,trim={0 0 0 0},clip]{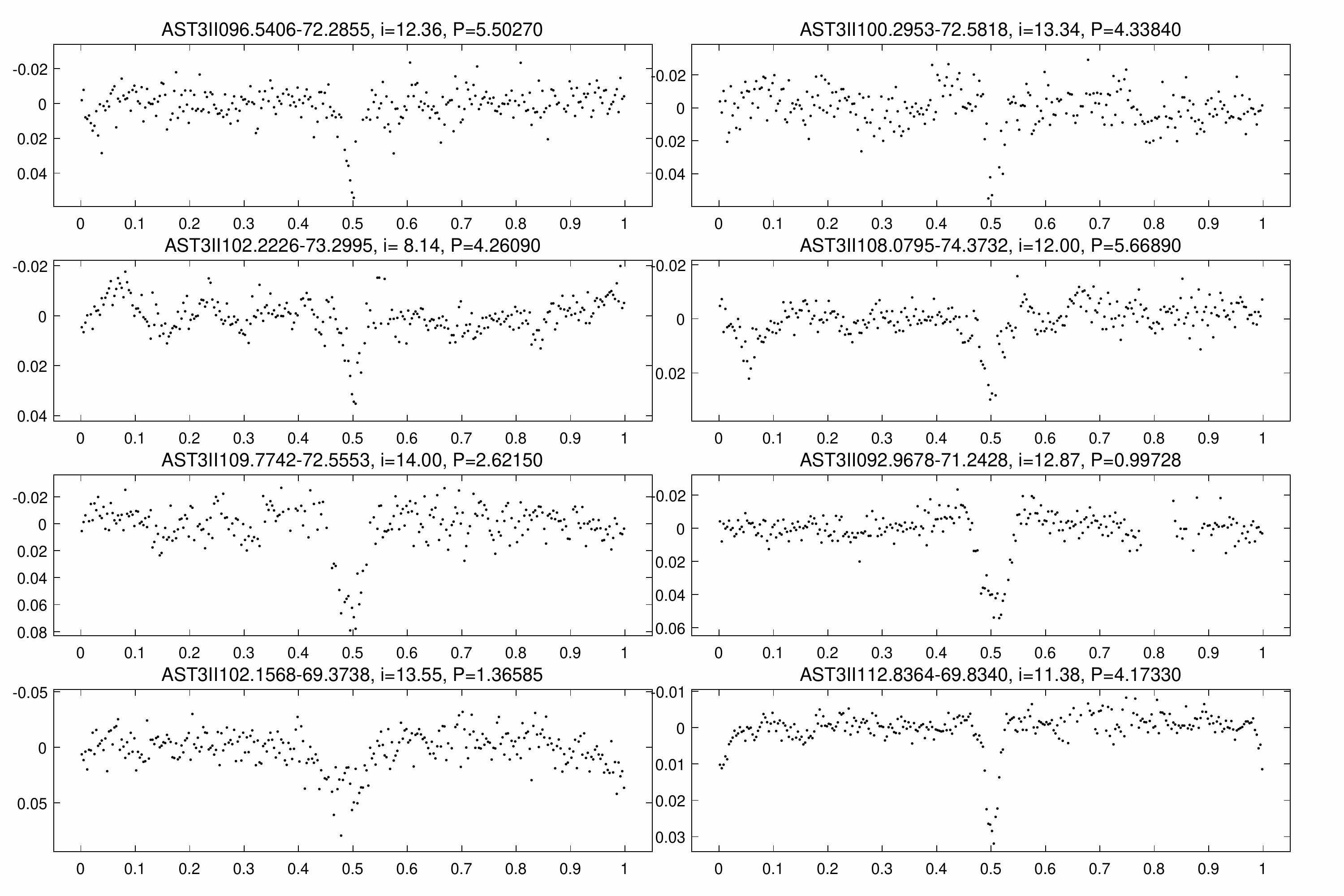}
\caption{Low depth eclipsing binaries. We show them separately from the ordinary eclipsing binaries since their depths are less than $5\%$ and they show insignificant secondary eclipses. These could be easily mistaken for transiting exoplanets if the photometric precision was insufficient to show the sharp ``V''-shape bottom. The x-axis and the y-axis of each panel are the phases [0,1] and the the vartiation in magnitude $\Delta \textbf{\textit{m}}_\textit{i}$, respectively.\label{fig:lowbinaries}}
\end{figure}

\begin{figure}
\centering
\includegraphics[width=\textwidth]{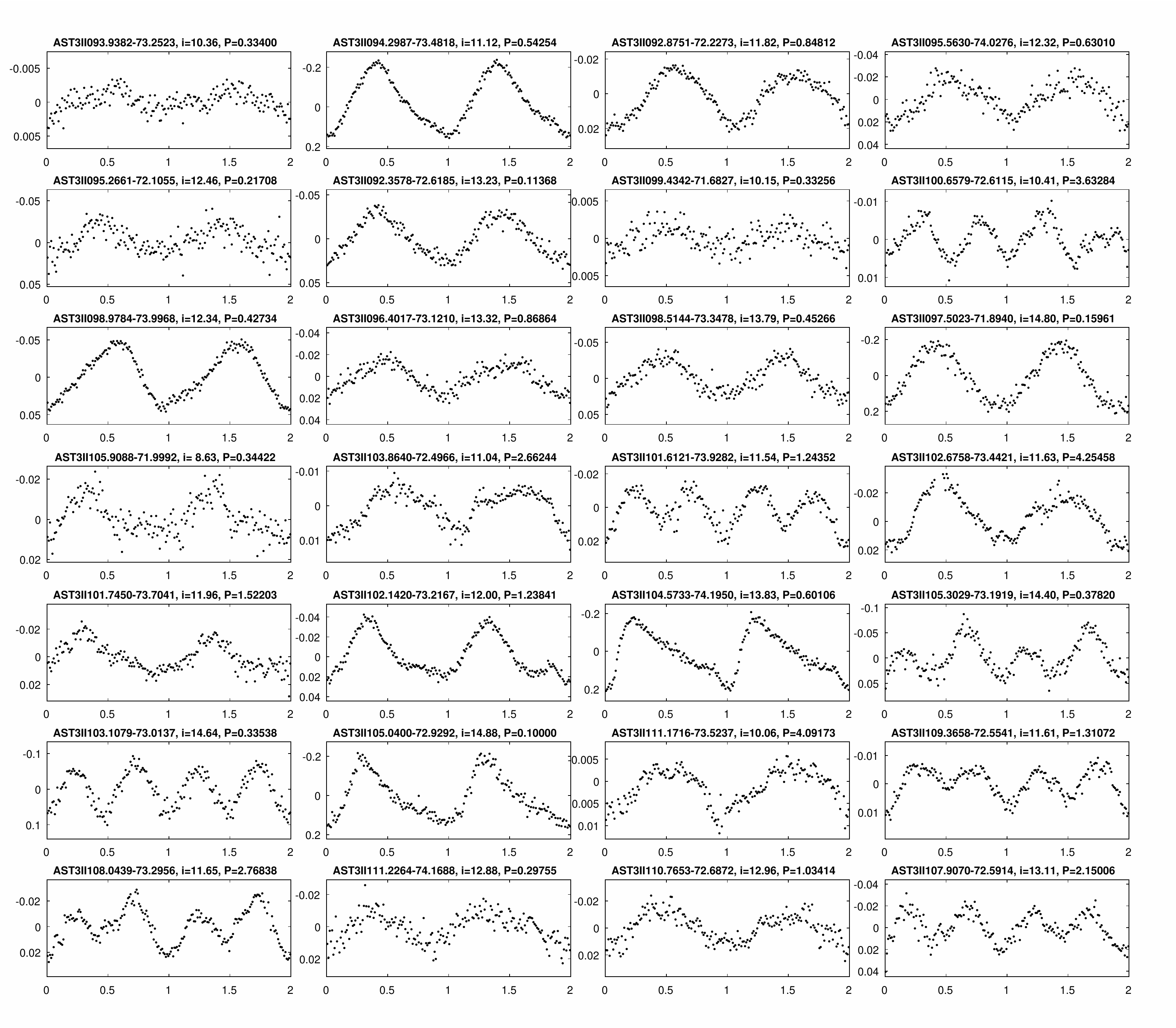}
\caption{Pulsating variables found by AST3-II in 2016. The label above each panel contains the $\textit{i}$-band median magnitude (which has been subtracted) for the curves, and the period in days. The x-axis and the y-axis of each panel are the phases [0,2] (Note that each lightcurve is folded to 2 times of its period, so the measurements around phase 1.5 is not identical to that around phase 0.5.) and the vartiation in magnitude $\Delta \textbf{\textit{m}}_\textit{i}$, respectively.\label{fig:variables1}}
\end{figure}

\begin{figure}
\centering
\includegraphics[width=\textwidth]{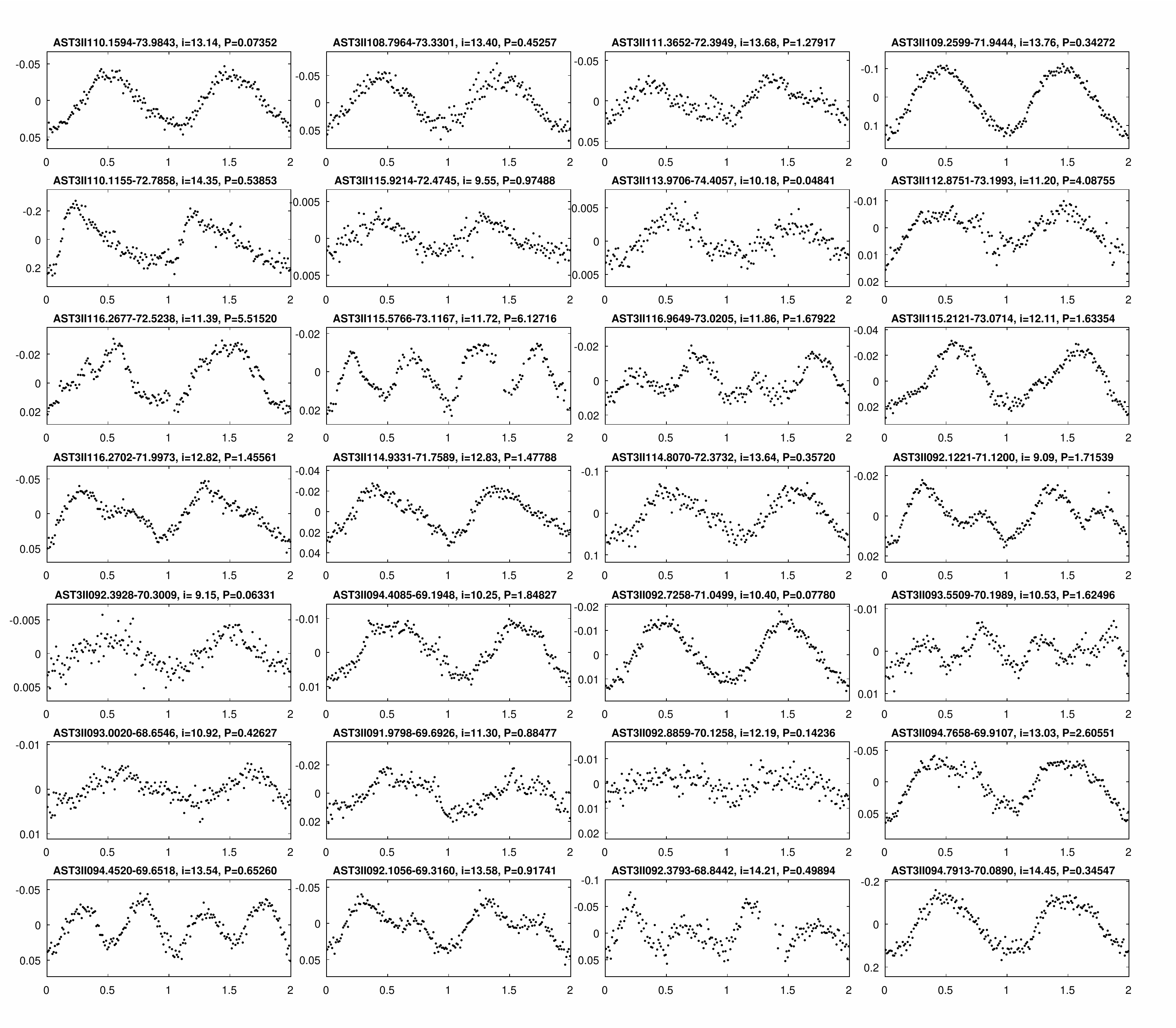}
\caption{Pulsating variables found by AST3-II in 2016, continued...\label{fig:variables2}}
\end{figure}

\begin{figure}
\centering
\includegraphics[width=\textwidth]{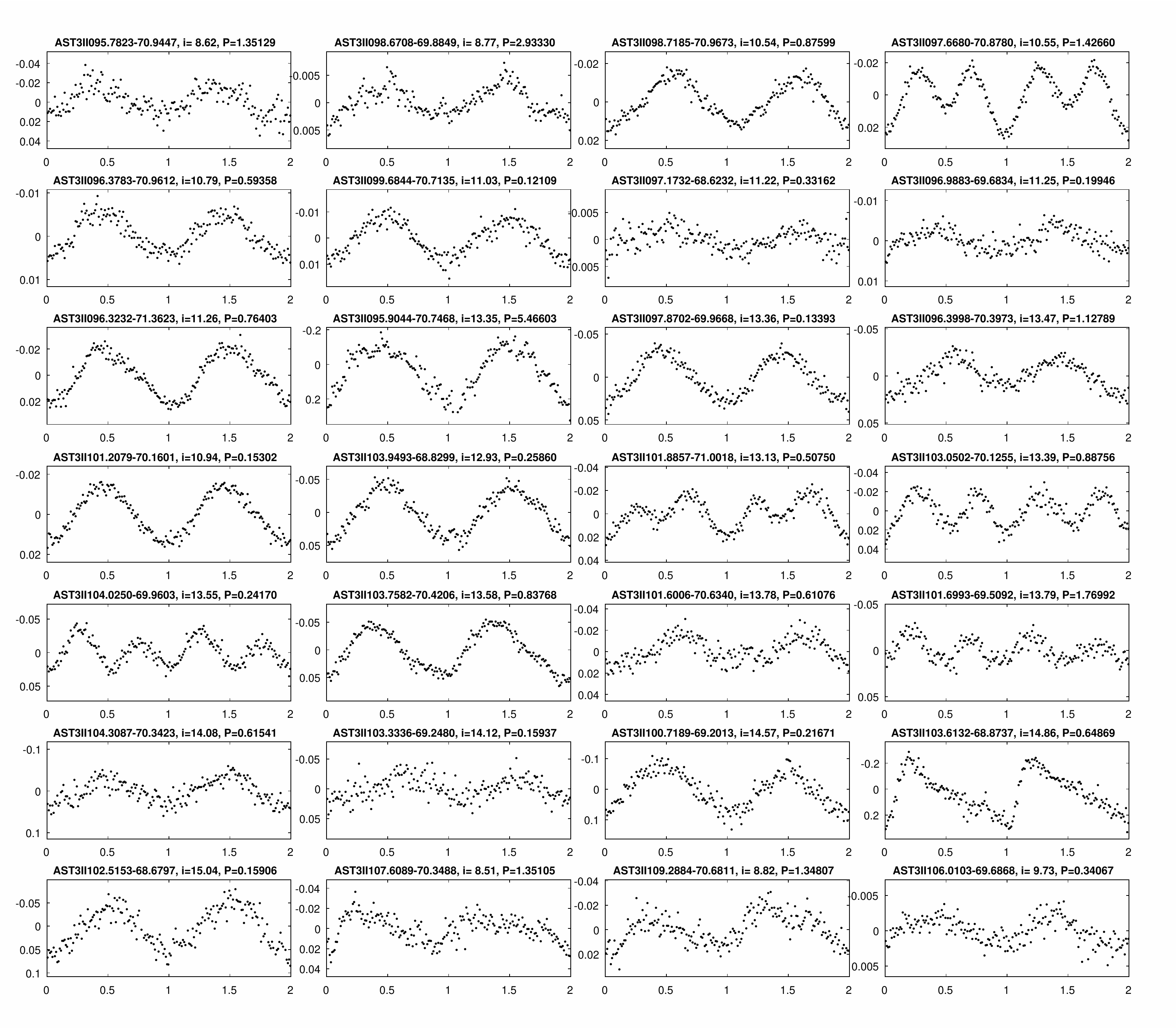}
\caption{Pulsating variables found by AST3-II in 2016, continued...\label{fig:variables3}}
\end{figure}

\begin{figure}
\centering
\includegraphics[width=\textwidth,trim={0  {0.4\textwidth} 0 0},clip]{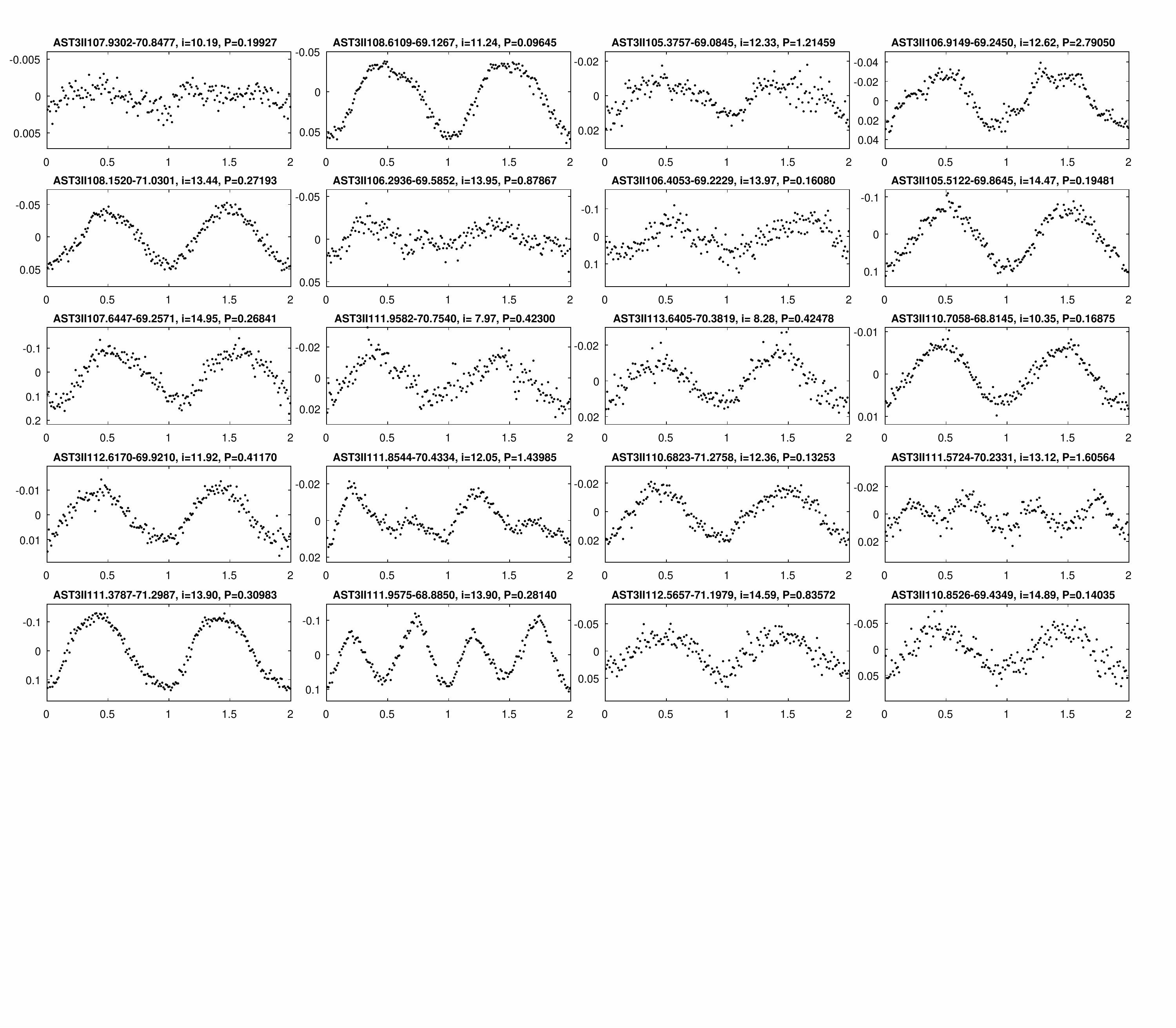}
\caption{Pulsating variables found by AST3-II in 2016, continued...\label{fig:variables4}}
\end{figure}

In total, we have detected 117 binaries, 86 of which are new detections from the AST3-II project in 2016. There are 69 EWs, 19 EBs, and 29 EAs in our sample. The classifications of 20 stars are not well determined and \textbf{each of them could be classified under several different variable classes}, including eclipsers, pulsators, and others. For these uncertain systems, we mark them by the symbol ``$?$'' or separate different types by a pipe symbol ``$|$'' in Table \ref{tab:variables-table}.  We also show 8 binaries with eclipse depth $\leq 5\%$ that are very similar in appearance to transit signals but have a sharp ``V''-shaped bottom, in Figure \ref{fig:lowbinaries}.

\subsubsection{Pulsating Variables}
The second category of variables in our survey are the pulsating variables: $\delta$ Scuti, $\gamma$ Doradus, RR Lyrae stars, Cepheids and so on \citep{Catelan15}. $\delta$ Scuti variables are late A- and early F-type stars located in the instability strip above the main sequence belt in the Hertzsprung-Russell diagram. Their typical pulsation periods are found to be in the range from 0.02 to 0.25 days \citep{Breger00}. $\gamma$ Doradus stars are located in a similar position in the instability strip as the $\delta$ Scuti stars, but their pulsating periods are longer---between 0.3 days and 3 days \citep{Cuypers09}. RR Lyrae stars are radially pulsating giant stars with spectral types from A to F with periods in the range from 0.2 to 1.0 days \citep{Smith04}. Cepheid variables obey a period-luminosity relation and can be divided into two subclasses---type I and type II \citep{Catelan15}---based on their masses, ages, and evolutionary states. The brightness variation of most type I Cepheid variables (also known as $\delta$ Cepheids) shows a rapid rise to maximum and a slow decline back to minimum, which is similar to the variation of an RR Lyrae star but with a longer period from 1 to 60 days \citep{Soszynski08}. Type II Cepheids generally show a relatively broad maximum and a symmetric minimum \citep{Schmidt04}, and have periods of $\sim 0.8--35$ days and light amplitudes from 0.3 to 1.2 mag in $V$ band.

As for the eclipsing binary stars, our classification of pulsating variables is based on the shape of their lightcurves. However, since most pulsating stars have multiple frequencies, all of the variables that are classified as pulsating stars were analyzed by Fourier decomposition. The results of a frequency analysis played a very important role in deciding their nature. In our sample, there are 104 pulsators, which are classified into 29 $\delta$ Scuti stars,  35 Cepheids, and 40 RR Lyraes. Uncertain systems are labeled by symbols ``$?$'' or ``$|$''.

%
%
%
%
%

\section{Summary}\label{sec:summary}
We present some of the latest results from the Chinese Exoplanet Searching Program from Antarctica. This program is based on the AST3-II telescope located at the Chinese Kunlun station at Dome A, Antarctica. This first data release contains a data-set obtained in the austral winter of 2016 with target fields within the Southern Continuous Viewing Zone of \textit{TESS}. The data-set is available to the community through the website of the School of Astronomy and Space Science, Nanjing University \footnote{\url{http://www.njutido.com/tido/data.html} or  \url{http://116.62.78.33/tido/data.html}} and the Chinese Astronomical Data Center \footnote{\url{http://casdc.china-vo.org/archive/ast3/II/dr1/}}. The released data consists of three parts: Calibrated Catalogs, Lightcurves for General Science Usage and a Catalog of new found Variables. In addition Science Images can be made available, via email requests to \email{huizhang@nju.edu.cn}. The Science Images include reduced FITS images with headers containing the observation information, including target name, world-coordinate system, observation date, and exposure time.

The Calibrated Catalogs contain all the sources found by our pipeline with a SNR above a threshold of 3-$\sigma$. Observation conditions and statistics of image qualities are also included in the header of each catalog (see Table \ref{tab:cataloghead}). The Lightcurves for General Science Usage include sources that are cross-matched with the sources in the APASS catalog with a limiting magnitude of $\textbf{\textit{m}}_\textbf{i} \le15$. There are two subsets of lightcurves: (1) original lightcurves that inherit all the information from the corresponding catalogs, and (2) polished lightcurves which are detrended and binned to a cadence of 12 minutes. The Catalog of Variables shows coordinates, magnitudes, periods and stellar properties cross-matched with the AAVSO database and the TIC catalog for all the variables we have found within the Southern CVZ of \textit{TESS} (see Table \ref{tab:variables-table}). There are 42 variables are matched in the variable catalog of AAVSO. Although the classification is not the major goal of this work, we have achieved a high consistency with AAVSO database --- only 3 of these 42 variables are not consistent: target ``AST3II092.1221-71.1200'' ( ``ACV''  or ``CEP''), ``AST3II096.3232-71.3623'' (``ROT'' or ``RR''), ``AST3II102.4829-71.7105'' (``YSO''  or ``EW''). From the shape of our lightcurves, we think our classification is more convincing. Note that we have included the contact eclipsing binary, ``EC'', the detached eclipsing binary, ``ED'', and the semi-detached eclipsing binary, ``ESD'', in the AAVSO database into ``EW'' ,``EA'' and ``EB'', respectively. \textbf{The two types in each pair, e.g. ``EC'' and ``EW'', are basically the same type in two different classification systems.} Only There are 179 targets that are newly found variables and 67 targets are also listed in the potential planet candidates list of \textit{TESS}. Since we have only selected variables with regular shapes and obvious periods, many irregular and/or long-term variables are not included in our variable catalog. These objects will be studied and released in future works. The minimum variation reliably detected by our survey is below 5 mmag, showing that we have the ability to find transiting exoplanets. A detailed catalog of transiting exoplanet candidates found by our survey is in preparation.

\acknowledgments
This work was supported by the Natural Science Foundation of China (NSFC grants 11673011, 11333002, 11273019), National Basic Research Program (973 Program) of China (Grant Nos. 2013CB834900, 2013CB834904). The authors deeply appreciate all the CHINAREs for their great effort in installing/maintaining CSTAR, CSTAR-II, AST3-I, AST3-II and PLATO-A. This study has also been supported by the Chinese Polar Environment Comprehensive Investigation \& Assessment Program (Grant No. CHINARE2016-02-03), the australian Antarctic Division, and the Australian National Collaborative Research Infrastructure Strategy administered by Astronomy Australia Limited. Zhang is also grateful to the High Performance Computing Center (HPCC) of Nanjing University for reducing the data used in this paper. This research was made possible through the use of the AAVSO Photometric All-Sky Survey (APASS), funded by the Robert Martin Ayers Sciences Fund.

\software{Sextractor \citep{Bertin96}, 
                Swarp \citep{Bertin02}, 
                VARTOOLS \citep{Hartman16}, 
                Astrometry.net \citep{Lang10},
                MATLAB}.

%




\appendix

\section{Appendix Data}
\begin{longrotatetable}
\begin{deluxetable}{lclcllcp{3.5cm}lp{2.0cm}llc}
\tablecaption{Variables from AST3-II observation in 2016\label{tab:variables-table}}
\tablehead{
\colhead{Target\tablenotemark{a}}  & \colhead{mag$_\textit{i}$\tablenotemark{b}} &
\colhead{ID$_{tic}$\tablenotemark{c}}  &  \colhead{mag$_{tic}$\tablenotemark{c}}  &
\colhead{R$_{tic}$ \tablenotemark{c}} &  \colhead{M$_{tic}$\tablenotemark{c}} &
\colhead{CTL\tablenotemark{d}} &
\colhead{AAVSO Notation\tablenotemark{e}} & \colhead{P$_{aas}$ \tablenotemark{e}} & \colhead{Type$_{aas}$\tablenotemark{e}} &
\colhead{P$_{ast3}$} &  \colhead{Type$_{ast3}$\tablenotemark{f}} &   \\
\colhead{-} & \colhead{$[mag]$} &
\colhead{-} & \colhead{$[mag]$} &
\colhead{$[R_{\odot}]$} & \colhead{$[M_{\odot}]$} &
\colhead{$flag$} &
\colhead{-} & \colhead{$[days]$} & \colhead{-} &
\colhead{ $[days]$} & \colhead{-} &
}
\startdata
AST3II111.2878-70.4038 &   7.57 &  300384180 &   6.63 &  11.0074 &  0.7757 &  0 &                             - &         - &$               - $&  0.84600 &$           EW $\\
AST3II111.9582-70.7540 &   7.97 &  300508526 &   7.31 &  16.8009 &  0.7123 &  0 &                             - &         - &$               - $&  0.42300 &$          RR $\\
AST3II102.2226-73.2995 &   8.14 &  177258735 &   7.53 &   1.5909 &  1.3934 &  1 &                             - &         - &$               - $&  4.26090 &$          EA $\\
AST3II113.6405-70.3819 &   8.28 &  453079732 &   7.66 &   2.4195 &  1.4143 &  1 &                             - &         - &$               - $&  0.42478 &$          RR $\\
AST3II111.7546-69.5154 &   8.45 &  300443831 &   8.25 &   3.0748 &  1.6413 &  1 &                        ZZ Vol &   0.84743 &$           EW|KE $&  0.84726 &$          EW $\\
AST3II107.6089-70.3488 &   8.51 &  300039874 &   8.07 &   9.7297 &  0.9137 &  0 &                             - &         - &$               - $&  1.35105 &$         CEP $\\
AST3II095.7823-70.9447 &   8.62 &  167089430 &   8.05 &   2.8333 &  2.1349 &  1 &                             - &         - &$               - $&  1.35129 &$         CEP $\\
AST3II105.9088-71.9992 &   8.63 &  271554516 &   8.43 &   2.0157 &  1.6519 &  1 &                             - &         - &$               - $&  0.34422 &$          RR $\\
AST3II098.6708-69.8849 &   8.77 &  167344043 &   8.50 &   1.2255 &  0.9718 &  1 &                             - &         - &$               - $&  2.93330 &$         CEP $\\
AST3II109.2884-70.6811 &   8.82 &  300160946 &   8.29 &   0.4974 &  0.4977 &  0 &                             - &         - &$               - $&  1.34807 &$         CEP $\\
AST3II092.1221-71.1200 &   9.09 &   41259805 &   8.85 &   1.7874 &  1.9437 &  1 &           ASAS J060829-7107.2 &   1.71443 &$             ACV $&  1.71539 &$         CEP $\\
AST3II092.3928-70.3009 &   9.15 &   41360272 &   8.83 &   5.8939 &  1.7469 &  0 &                             - &         - &$               - $&  0.06331 &$        DSCT $\\
AST3II115.9214-72.4745 &   9.55 &  272127517 &   9.33 &   1.8948 &  1.4190 &  1 &                             - &         - &$               - $&  0.97488 &$          RR $\\
AST3II106.0103-69.6868 &   9.73 &  299899924 &   9.40 &   1.8372 &  1.6901 &  1 &                             - &         - &$               - $&  0.34067 &$          RR $\\
AST3II091.3523-69.0324 &  10.00 &   41172665 &   9.76 &   2.9070 &  1.4465 &  1 &                             - &         - &$               - $&  1.73012 &$          EA $\\
AST3II111.1716-73.5237 &  10.06 &  271795905 &   9.76 &   1.2572 &  1.2002 &  1 &                             - &         - &$               - $&  4.09173 &$         CEP $\\
AST3II099.4342-71.6827 &  10.15 &  167416361 &   9.86 &   1.3734 &  1.2409 &  1 &                             - &         - &$               - $&  0.33256 &$        DSCT $\\
AST3II113.9706-74.4057 &  10.18 &  271971704 &   9.88 &   2.4153 &  1.6948 &  1 &                             - &         - &$               - $&  0.04841 &$        DSCT $\\
AST3II107.9302-70.8477 &  10.19 &  300086363 &   9.90 &   3.4197 &  1.1236 &  1 &                             - &         - &$               - $&  0.19927 &$     RR|DSCT $\\
AST3II094.4085-69.1948 &  10.25 &   41595212 &   9.98 &   1.6549 &  1.2327 &  1 &                             - &         - &$               - $&  1.84827 &$         CEP $\\
AST3II110.7058-68.8145 &  10.35 &  300327061 &  10.00 &   1.8652 &  1.4767 &  1 &                             - &         - &$               - $&  0.16875 &$          RR $\\
AST3II093.9382-73.2523 &  10.36 &  141868094 &  10.08 &   2.0054 &  1.4850 &  1 &                             - &         - &$               - $&  0.33400 &$        DSCT ?$\\
AST3II092.7258-71.0499 &  10.40 &   41362881 &  10.12 &   3.3323 &  1.5383 &  1 &                             - &         - &$               - $&  0.07780 &$        DSCT $\\
AST3II100.6579-72.6115 &  10.41 &  176872638 &  10.26 &        - &       - &  0 &                             - &         - &$               - $&  3.63284 &$         CEP $\\
AST3II093.5509-70.1989 &  10.53 &   41483281 &  10.24 &   1.3562 &  1.5535 &  1 &                             - &         - &$               - $&  1.62496 &$         CEP? $\\
AST3II098.7185-70.9673 &  10.54 &  167361929 &  10.23 &   2.0106 &  1.4631 &  1 &                             - &         - &$               - $&  0.87599 &$          RR $\\
AST3II097.6680-70.8780 &  10.55 &  167249549 &  10.22 &   3.4671 &  1.4169 &  1 &                             - &         - &$               - $&  1.42660 &$         CEP $\\
AST3II097.5428-69.3784 &  10.77 &  167248486 &  10.45 &   1.8288 &  1.4376 &  1 &                             - &         - &$               - $&  0.80528 &$          EW $\\
AST3II096.3783-70.9612 &  10.79 &  167163582 &  10.49 &        - &       - &  0 &                             - &         - &$               - $&  0.59358 &$          RR $\\
AST3II094.1155-74.4470 &  10.89 &  141871560 &  10.51 &   1.9905 &  1.5946 &  1 &                      NSV 2922 &   0.63045 &$              EW $&  0.63052 &$          EW $\\
AST3II093.0020-68.6546 &  10.92 &   41463672 &  10.68 &   1.8286 &  1.5935 &  1 &                             - &         - &$               - $&  0.42627 &$          RR $\\
AST3II101.2079-70.1601 &  10.94 &  176960346 &  10.66 &   1.5066 &  1.3332 &  1 &                             - &         - &$               - $&  0.15302 &$        DSCT $\\
AST3II093.7797-71.5701 &  10.96 &   41533230 &  10.66 &   1.3007 &  1.1688 &  1 &           ASAS J061507-7134.2 &   1.42051 &$  EC|DCEP-FO|$$ESD $&  1.42086 &$          EW $\\
AST3II099.6844-70.7135 &  11.03 &  167417105 &  10.70 &   2.6568 &  1.7240 &  1 &                             - &         - &$               - $&  0.12109 &$        DSCT $\\
AST3II103.8640-72.4966 &  11.04 &  177349463 &  10.76 &   1.0578 &  1.0657 &  1 &                             - &         - &$               - $&  2.66244 &$         CEP $\\
AST3II094.2987-73.4818 &  11.12 &  141870888 &  10.79 &        - &       - &  0 &                        RV Men &   0.54229 &$             RRC $&  0.54254 &$          RR $\\
AST3II091.6806-72.2152 &  11.17 &  141766191 &  10.83 &   1.4415 &  1.0253 &  1 &                             - &         - &$               - $&  0.37354 &$          EW $\\
AST3II112.8751-73.1993 &  11.20 &  271891181 &  10.85 &   1.0634 &  0.9906 &  1 &                             - &         - &$               - $&  4.08755 &$         CEP $\\
AST3II097.1732-68.6232 &  11.22 &  167207431 &  10.90 &   1.7974 &  1.2539 &  1 &                             - &         - &$               - $&  0.33162 &$          RR $\\
AST3II110.2915-71.8165 &  11.24 &  300291165 &  11.13 &        - &       - &  0 &           ASAS J072110-7149.0 &   0.35427 &$              EA $&  0.35434 &$       EB|EW $\\
AST3II108.6109-69.1267 &  11.24 &  300139147 &  10.91 &   1.7439 &  1.5913 &  1 &           ASAS J071427-6907.6 &   0.09647 &$            DSCT $&  0.09645 &$        DSCT $\\
AST3II096.9883-69.6834 &  11.25 &  167203167 &  10.94 &   1.3137 &  1.1658 &  1 &                             - &         - &$               - $&  0.19946 &$        DSCT $\\
AST3II096.3232-71.3623 &  11.26 &  167163906 &  10.91 &   0.6935 &  0.8363 &  1 &           ASAS J062517-7121.9 &   0.75890 &$             ROT $&  0.76403 &$          RR $\\
AST3II113.5745-74.2172 &  11.30 &  271904441 &  11.68 &   1.3449 &  1.2080 &  1 &                             - &         - &$               - $&  0.69178 &$          EW ?$\\
AST3II091.9798-69.6926 &  11.30 &   41256640 &  10.99 &   1.6020 &  1.4501 &  1 &                             - &         - &$               - $&  0.88477 &$          RR $\\
AST3II115.5460-72.9271 &  11.31 &  272087157 &  10.91 &        - &       - &  0 &                             - &         - &$               - $&  0.31996 &$          EW? $\\
AST3II112.8364-69.8340 &  11.38 &  300559128 &  10.99 &   1.3798 &  1.2397 &  1 &                             - &         - &$               - $&  4.17330 &$          EA $\\
AST3II102.2495-73.3640 &  11.39 &  177258700 &  10.95 &   2.3260 &  1.5876 &  1 &           ASAS J064900-7321.8 &   0.68796 &$              EC $&  0.68792 &$          EW $\\
AST3II116.2677-72.5238 &  11.39 &  272128498 &  10.98 &        - &       - &  0 &                             - &         - &$               - $&  5.51520 &$         CEP $\\
AST3II107.5256-69.5379 &  11.47 &  300034498 &  11.19 &   1.4353 &  1.2911 &  1 &           ASAS J071006-6932.3 &   1.78290 &$              ED $&  1.78150 &$          EA $\\
AST3II101.6121-73.9282 &  11.54 &  177254093 &  11.17 &   2.0275 &  1.4758 &  1 &                             - &         - &$               - $&  1.24352 &$         CEP $\\
AST3II109.3658-72.5541 &  11.61 &  271640842 &  11.34 &   1.3425 &  1.1772 &  1 &                             - &         - &$               - $&  1.31072 &$         CEP $\\
AST3II102.6758-73.4421 &  11.63 &  177283525 &  11.33 &   1.3975 &  1.0873 &  1 &                             - &         - &$               - $&  4.25458 &$         CEP $\\
AST3II108.0439-73.2956 &  11.65 &  391946675 &  11.27 &        - &       - &  0 &                             - &         - &$               - $&  2.76838 &$         CEP $\\
AST3II115.5766-73.1167 &  11.72 &  272087305 &  11.38 &   1.3772 &  1.1055 &  1 &                             - &         - &$               - $&  6.12716 &$         CEP $\\
AST3II096.9914-70.5949 &  11.75 &  167203947 &  11.48 &        - &       - &  0 &           ASAS J062758-7035.7 &   0.62027 &$      EC|RRC|$ $ESD $&  0.62036 &$          EW $\\
AST3II094.9034-72.0335 &  11.79 &  141914317 &  11.48 &   1.2774 &  1.0286 &  1 &           ASAS J061937-7202.0 &   0.43718 &$              EC $&  0.43722 &$          EW $\\
AST3II092.8751-72.2273 &  11.82 &  141807839 &  11.19 &        - &       - &  0 &                             - &         - &$               - $&  0.84812 &$          RR $\\
AST3II116.9649-73.0205 &  11.86 &  272190346 &  11.57 &   2.1749 &  2.1012 &  1 &                             - &         - &$               - $&  1.67922 &$         CEP $\\
AST3II108.2432-69.0400 &  11.90 &  300091984 &  11.57 &   1.3587 &  1.2205 &  1 &                             - &         - &$               - $&  2.16220 &$          EA $\\
AST3II112.6170-69.9210 &  11.92 &  300556532 &  11.54 &        - &       - &  0 &                             - &         - &$               - $&  0.41170 &$          RR $\\
AST3II098.2398-68.9722 &  11.93 &  167307524 &  11.86 &   1.2569 &  1.1312 &  1 &                             - &         - &$               - $&  0.60335 &$          EA $\\
AST3II101.7450-73.7041 &  11.96 &  177253966 &  11.60 &   2.4601 &  1.5650 &  1 &                             - &         - &$               - $&  1.52203 &$         CEP $\\
AST3II108.0795-74.3732 &  12.00 &  391947238 &  11.67 &   1.2661 &  1.1390 &  1 &                             - &         - &$               - $&  5.66890 &$          EA $\\
AST3II102.1420-73.2167 &  12.00 &  177258779 &  11.64 &        - &       - &  0 &                             - &         - &$               - $&  1.23841 &$         CEP $\\
AST3II105.4951-72.9830 &  12.01 &  388180826 &  11.86 &   1.5144 &  1.3659 &  1 &                             - &         - &$               - $&  0.37226 &$          EW $\\
AST3II101.8520-71.5969 &  12.03 &  177017182 &  11.73 &   2.0432 &  1.2507 &  1 &                             - &         - &$               - $&  0.32736 &$          EW $\\
AST3II091.1306-72.1367 &  12.03 &  141713380 &  11.72 &   1.0337 &  0.9785 &  1 &                             - &         - &$               - $&  0.38590 &$          EW|RR $\\
AST3II111.8544-70.4334 &  12.05 &  300449915 &  11.68 &   1.0956 &  1.0564 &  1 &                             - &         - &$               - $&  1.43985 &$         CEP $\\
AST3II102.4683-73.4007 &  12.07 &  177283493 &  11.77 &        - &       - &  0 &                             - &         - &$               - $&  0.75318 &$          EB $\\
AST3II091.3416-69.7834 &  12.10 &   41173515 &  11.77 &   1.6011 &  1.4493 &  1 &           ASAS J060521-6947.1 &   0.59305 &$          EC|RRC $&  0.59284 &$       EW|RR $\\
AST3II115.2121-73.0714 &  12.11 &  272085493 &  11.76 &        - &       - &  0 &                             - &         - &$               - $&  1.63354 &$         CEP $\\
AST3II092.8859-70.1258 &  12.19 &   41363877 &  11.83 &   1.1597 &  1.0548 &  1 &                             - &         - &$               - $&  0.14236 &$     DSCT|EA $\\
AST3II114.8573-73.7690 &  12.20 &  271999940 &  11.88 &   1.1989 &  1.0777 &  1 &           ASAS J073926-7346.2 &   0.56424 &$          ESD|EC $&  0.56410 &$          EB $\\
AST3II098.6131-68.9374 &  12.27 &  167339240 &  11.95 &   1.1524 &  1.0495 &  1 &                             - &         - &$               - $&  2.18525 &$          EA $\\
AST3II098.6280-72.0523 &  12.30 &  142105466 &  11.95 &   1.1155 &  1.0522 &  1 &           ASAS J063431-7203.1 &   0.43320 &$          EC|ESD $&  0.43318 &$          EW $\\
AST3II102.0928-69.5721 &  12.32 &  177018607 &  12.00 &   1.0583 &  0.9855 &  0 &                             - &         - &$               - $&  0.36118 &$          EW $\\
AST3II095.5630-74.0276 &  12.32 &  141944605 &  11.99 &   1.2912 &  1.1605 &  1 &                             - &         - &$               - $&  0.63010 &$        DSCT ?$\\
AST3II105.3757-69.0845 &  12.33 &  177238312 &  12.02 &   1.3690 &  1.2298 &  0 &                             - &         - &$               - $&  1.21459 &$         CEP $\\
AST3II098.9784-73.9968 &  12.34 &  142106818 &  11.94 &        - &       - &  0 &                             - &         - &$               - $&  0.42734 &$          RR $\\
AST3II096.5406-72.2855 &  12.36 &  142013932 &  12.02 &   1.2720 &  1.1440 &  0 &                             - &         - &$               - $&  5.50270 &$          EA $\\
AST3II110.6823-71.2758 &  12.36 &  300328626 &  12.08 &   1.5047 &  1.3567 &  0 &                             - &         - &$               - $&  0.13253 &$        DSCT $\\
AST3II092.9887-69.2780 &  12.43 &   41464424 &  12.05 &   0.9920 &  0.9401 &  1 &           ASAS J061157-6916.7 &   0.33890 &$     ESD|DSCT|$ $EC $&  0.33890 &$          EW $\\
AST3II108.2786-71.7907 &  12.44 &  300137432 &  12.31 &   0.9607 &  0.9175 &  1 &                        TY Vol &   0.32763 &$              EW $&  0.32762 &$          EW $\\
AST3II095.2661-72.1055 &  12.46 &  141914372 &  12.10 &        - &       - &  0 &                             - &         - &$               - $&  0.21708 &$          RR $\\
AST3II094.9934-69.9677 &  12.48 &  167005326 &  12.16 &   1.4051 &  1.2629 &  0 &                             - &         - &$               - $&  0.72018 &$          EB $\\
AST3II107.4982-72.1355 &  12.50 &  391927558 &  12.21 &   1.3305 &  1.1161 &  0 &           ASAS J070959-7208.2 &   0.54292 &$          ESD|EC $&  0.54316 &$          EB $\\
AST3II102.4826-70.3848 &  12.50 &  177032797 &  12.35 &   1.2760 &  0.9323 &  1 &                             - &         - &$               - $&  0.28482 &$          EW $\\
AST3II096.2029-72.0109 &  12.52 &  142013583 &  12.22 &   1.1731 &  1.0646 &  0 &                             - &         - &$               - $&  0.32200 &$          EW? $\\
AST3II105.9193-70.7796 &  12.58 &  299900499 &  12.29 &   1.0019 &  0.9471 &  1 &           ASAS J070341-7046.8 &   0.37554 &$         EC|DSCT $&  0.37562 &$          EW $\\
AST3II094.4370-73.0193 &  12.59 &  141870501 &  12.33 &   1.2483 &  1.1240 &  0 &           ASAS J061745-7301.2 &   0.53189 &$             ESD $&  0.53202 &$          EB $\\
AST3II106.8398-69.6943 &  12.59 &  300009837 &  12.42 &   1.4030 &  1.2610 &  0 &           ASAS J070722-6941.7 &   0.36651 &$              EC $&  0.36652 &$          EW $\\
AST3II106.9149-69.2450 &  12.62 &  300010099 &  12.19 &        - &       - &  0 &                             - &         - &$               - $&  2.79050 &$         CEP $\\
AST3II103.5842-70.8694 &  12.65 &  177113550 &  12.32 &   1.2062 &  1.0899 &  0 &                             - &         - &$               - $&  0.34158 &$          EW $\\
AST3II102.4829-71.7105 &  12.70 &  177033514 &  12.44 &   1.5974 &  1.4456 &  0 &  ASASSN-V J064955.83-714237.5 &         - &$             YSO $&  1.00686 &$          EW $\\
AST3II113.0724-72.9353 &  12.72 &  271892852 &  12.33 &        - &       - &  0 &                             - &         - &$               - $&  4.14840 &$      EA| $$RS $\\
AST3II093.7557-72.2885 &  12.73 &  141869057 &  12.38 &   1.0858 &  1.0038 &  0 &                             - &         - &$               - $&  0.43568 &$          EW $\\
AST3II108.4769-72.3906 &  12.74 &  271594955 &  12.33 &   1.2730 &  1.1448 &  0 &           ASAS J071355-7223.5 &   0.41165 &$              EC $&  0.41164 &$          EW $\\
AST3II116.9547-73.9117 &  12.77 &  272191334 &  12.35 &   1.4148 &  1.2720 &  0 &           ASAS J074749-7354.7 &   0.41929 &$              EC $&  0.41930 &$          EW $\\
AST3II116.2702-71.9973 &  12.82 &  272128027 &  12.47 &        - &       - &  0 &                             - &         - &$               - $&  1.45561 &$         CEP $\\
AST3II107.6780-73.6250 &  12.83 &  391926737 &  12.45 &   1.4546 &  1.3092 &  0 &           ASAS J071043-7337.5 &   0.59408 &$          ESD|EC $&  0.59408 &$          EB $\\
AST3II108.2001-71.3165 &  12.83 &  300137123 &  12.55 &   1.2374 &  1.1150 &  0 &                             - &         - &$               - $&  2.28680 &$          EA $\\
AST3II114.9331-71.7589 &  12.83 &  300709487 &  12.43 &        - &       - &  0 &                             - &         - &$               - $&  1.47788 &$         CEP $\\
AST3II097.6441-71.6785 &  12.86 &  167250056 &  12.50 &   1.1120 &  1.0214 &  0 &                             - &         - &$               - $&  0.85770 &$          EB $\\
AST3II092.9678-71.2428 &  12.87 &   41362672 &  12.50 &        - &       - &  0 &                             - &         - &$               - $&  0.99728 &$          EA $\\
AST3II111.2264-74.1688 &  12.88 &  271795440 &  12.54 &   0.8139 &  0.7894 &  1 &                             - &         - &$               - $&  0.29755 &$     DSCT|RR $\\
AST3II106.3396-69.1610 &  12.90 &  299939799 &  12.60 &   1.3331 &  1.1974 &  0 &                        YZ Vol &   1.57097 &$              EA $&  0.78532 &$          EB $\\
AST3II098.3455-70.1247 &  12.92 &  167338355 &  12.45 &   1.3545 &  0.8715 &  0 &           ASAS J063323-7007.5 &   0.28836 &$              EC $&  0.28836 &$          EW $\\
AST3II103.9493-68.8299 &  12.93 &  177116027 &  12.60 &   1.3316 &  1.1961 &  0 &                             - &         - &$               - $&  0.25860 &$          RR $\\
AST3II110.7653-72.6872 &  12.96 &  271724440 &  12.65 &        - &       - &  0 &                             - &         - &$               - $&  1.03414 &$         CEP $\\
AST3II094.7658-69.9107 &  13.03 &  167005378 &  12.62 &        - &       - &  0 &                             - &         - &$               - $&  2.60551 &$         CEP $\\
AST3II107.9070-72.5914 &  13.11 &  391946253 &  12.81 &   1.3958 &  1.2544 &  0 &                             - &         - &$               - $&  2.15006 &$         CEP $\\
AST3II111.5724-70.2331 &  13.12 &  300443364 &  12.74 &   1.5702 &  1.4194 &  0 &                             - &         - &$               - $&  1.60564 &$         CEP $\\
AST3II101.8857-71.0018 &  13.13 &  177016884 &  12.76 &        - &       - &  0 &                             - &         - &$               - $&  0.50750 &$          RR $\\
AST3II110.1594-73.9843 &  13.14 &  271697152 &  12.68 &   1.5338 &  1.3845 &  0 &                             - &         - &$               - $&  0.07352 &$        DSCT $\\
AST3II096.4438-71.4297 &  13.18 &  167163979 &  12.82 &   1.6282 &  1.4757 &  0 &                             - &         - &$               - $&  0.41730 &$          EW $\\
AST3II106.6609-71.2478 &  13.22 &  299945169 &  13.00 &   0.8687 &  0.8427 &  1 &                             - &         - &$               - $&  0.38934 &$          EW $\\
AST3II092.3578-72.6185 &  13.23 &  141806774 &  12.91 &   1.7082 &  1.5550 &  0 &                             - &         - &$               - $&  0.11368 &$        DSCT $\\
AST3II090.6260-73.7799 &  13.28 &  141711100 &  12.95 &   1.1278 &  1.0322 &  0 &           ASAS J060230-7346.8 &   0.43001 &$ EC|RRC|$ $DSCT|ESD $&  0.42996 &$          EW $\\
AST3II104.7510-73.5358 &  13.31 &  177387467 &  12.92 &   1.4841 &  1.3371 &  0 &                             - &         - &$               - $&  0.78224 &$          EW $\\
AST3II113.2799-73.8991 &  13.31 &  271893557 &  12.96 &   0.9121 &  0.8799 &  1 &                             - &         - &$               - $&  0.29744 &$          EW $\\
AST3II096.3806-73.2253 &  13.32 &  141979627 &  12.90 &   0.8276 &  0.8034 &  1 &                             - &         - &$               - $&  0.24426 &$          EW $\\
AST3II096.4017-73.1210 &  13.32 &  141979700 &  12.95 &        - &       - &  0 &                             - &         - &$               - $&  0.86864 &$          RR $\\
AST3II100.2953-72.5818 &  13.34 &  142149364 &  12.90 &        - &       - &  0 &                             - &         - &$               - $&  4.33840 &$          EA $\\
AST3II105.7928-73.9199 &  13.35 &  388181338 &  12.99 &   1.7702 &  1.6185 &  0 &                             - &         - &$               - $&  1.58656 &$          EB $\\
AST3II095.9044-70.7468 &  13.35 &  167089250 &  12.90 &        - &       - &  0 &  ASASSN-V J062337.04-704448.5 &   5.52500 &$             CWB $&  5.46603 &$         CWB $\\
AST3II097.8702-69.9668 &  13.36 &  167251575 &  13.01 &   1.7199 &  1.5667 &  0 &                             - &         - &$               - $&  0.13393 &$        DSCT $\\
AST3II114.7338-72.9593 &  13.37 &  271999317 &  12.89 &   1.2502 &  1.1256 &  0 &                             - &         - &$               - $&  0.32968 &$          EW $\\
AST3II096.1773-72.9143 &  13.38 &  141979847 &  12.97 &   1.1861 &  1.0743 &  0 &                             - &         - &$               - $&  0.35684 &$          EW $\\
AST3II103.0502-70.1255 &  13.39 &  177075511 &  13.11 &        - &       - &  0 &                             - &         - &$               - $&  0.88756 &$          RR $\\
AST3II095.9351-70.8911 &  13.40 &  167089377 &  13.12 &        - &       - &  0 &                             - &         - &$               - $&  0.20546 &$          EW $\\
AST3II108.7964-73.3301 &  13.40 &  271639568 &  13.03 &   1.2686 &  1.1411 &  0 &                             - &         - &$               - $&  0.45257 &$          RR $\\
AST3II111.1322-70.1645 &  13.42 &  300330083 &  13.00 &        - &       - &  0 &                             - &         - &$               - $&  1.13160 &$          EA $\\
AST3II107.1686-70.9064 &  13.44 &  300015474 &  13.14 &        - &       - &  0 &                             - &         - &$               - $&  0.33012 &$          EW $\\
AST3II108.1520-71.0301 &  13.44 &  300086267 &  13.10 &   1.7057 &  1.5524 &  0 &                             - &         - &$               - $&  0.27193 &$        DSCT $\\
AST3II096.3998-70.3973 &  13.47 &  167163087 &  13.13 &   1.3377 &  1.2015 &  0 &                             - &         - &$               - $&  1.12789 &$         CEP $\\
AST3II104.8638-74.1627 &  13.50 &  177387851 &  13.17 &        - &       - &  0 &                             - &         - &$               - $&  2.50724 &$          EA $\\
AST3II094.4520-69.6518 &  13.54 &   41595694 &  13.17 &   1.5192 &  1.3705 &  0 &                             - &         - &$               - $&  0.65260 &$          RR $\\
AST3II104.0250-69.9603 &  13.55 &  177161450 &  13.18 &        - &       - &  0 &                             - &         - &$               - $&  0.24170 &$        DSCT $\\
AST3II102.1568-69.3738 &  13.55 &  177018730 &  13.25 &   0.9939 &  0.9415 &  0 &                             - &         - &$               - $&  1.36585 &$          EA? $\\
AST3II092.5544-73.8344 &  13.57 &  141809397 &  13.19 &   1.1713 &  1.0632 &  0 &                             - &         - &$               - $&  0.33874 &$          EW $\\
AST3II092.1056-69.3160 &  13.58 &   41331305 &  13.15 &   0.6797 &  0.6541 &  0 &                             - &         - &$               - $&  0.91741 &$          RR $\\
AST3II103.7582-70.4206 &  13.58 &  177115026 &  13.30 &        - &       - &  0 &                             - &         - &$               - $&  0.83768 &$          RR $\\
AST3II100.0861-72.5257 &  13.59 &  142149387 &  13.26 &   1.1969 &  1.0826 &  0 &  ASASSN-V J064020.58-723133.0 &   0.49752 &$              EW $&  0.49736 &$          EW $\\
AST3II091.6046-69.4909 &  13.60 &   41227678 &  13.12 &   0.9864 &  0.9361 &  0 &           ASAS J060625-6929.4 &   0.37134 &$          EC|ESD $&  0.37132 &$          EW $\\
AST3II095.3040-72.3847 &  13.64 &  141945928 &  13.30 &   1.2085 &  1.0916 &  0 &  ASASSN-V J062112.94-722305.3 &   0.43400 &$              EW $&  0.43378 &$          EW $\\
AST3II114.8070-72.3732 &  13.64 &  271998864 &  13.26 &   1.6298 &  1.4772 &  0 &                             - &         - &$               - $&  0.35720 &$          RR $\\
AST3II095.8987-69.5264 &  13.67 &  167088180 &  13.38 &        - &       - &  0 &                             - &         - &$               - $&  1.95352 &$          EA $\\
AST3II111.3652-72.3949 &  13.68 &  271797324 &  13.28 &        - &       - &  0 &                             - &         - &$               - $&  1.27917 &$         CEP $\\
AST3II116.4310-73.9974 &  13.69 &          - &      - &        - &       - &  0 &                             - &         - &$               - $&  0.50372 &$          EW $\\
AST3II098.3039-71.3062 &  13.73 &  167309182 &  13.41 &   1.0812 &  1.0008 &  0 &                             - &         - &$               - $&  1.33115 &$          EA $\\
AST3II109.2599-71.9444 &  13.76 &  300160216 &  13.46 &   1.6193 &  1.4670 &  0 &  ASASSN-V J071702.37-715640.0 &   0.34266 &$             RRC $&  0.34272 &$          RR $\\
AST3II107.5824-68.7888 &  13.78 &  300035002 &  13.54 &        - &       - &  0 &                             - &         - &$               - $&  1.33990 &$          EA $\\
AST3II101.6006-70.6340 &  13.78 &  176980970 &  13.25 &        - &       - &  0 &                             - &         - &$               - $&  0.61076 &$          RR $\\
AST3II109.1080-68.9219 &  13.79 &  300162041 &  13.55 &        - &       - &  0 &                             - &         - &$               - $&  0.46402 &$          EA $\\
AST3II098.5144-73.3478 &  13.79 &  142104577 &  13.46 &        - &       - &  0 &                             - &         - &$               - $&  0.45266 &$          RR $\\
AST3II101.6993-69.5092 &  13.79 &  176986279 &  13.43 &        - &       - &  0 &                             - &         - &$               - $&  1.76992 &$          CEP $\\
AST3II104.5733-74.1950 &  13.83 &  177355123 &  13.47 &   1.3275 &  1.1924 &  0 &           ASAS J065818-7411.7 &   0.60111 &$            RRAB $&  0.60106 &$          RR $\\
AST3II096.5462-68.5901 &  13.86 &  167125813 &  13.46 &   1.2416 &  1.1185 &  0 &           ASAS J062611-6835.4 &   0.27978 &$          ESD|EC $&  0.38862 &$          EW $\\
AST3II111.3787-71.2987 &  13.90 &  300384759 &  13.60 &   1.3078 &  1.1750 &  0 &  ASASSN-V J072530.96-711755.7 &   0.30986 &$             RRC $&  0.30983 &$          RR $\\
AST3II111.9575-68.8850 &  13.90 &  300448831 &  13.56 &        - &       - &  0 &                             - &         - &$               - $&  0.28140 &$        DSCT $\\
AST3II106.2936-69.5852 &  13.95 &  299939522 &  13.46 &        - &       - &  0 &                             - &         - &$               - $&  0.87867 &$          RR $\\
AST3II104.9029-73.9904 &  13.96 &  177387750 &  13.69 &   1.0375 &  0.9716 &  0 &                             - &         - &$               - $&  0.40066 &$          EW $\\
AST3II097.2789-73.7791 &  13.97 &  142050170 &  13.58 &   0.9329 &  0.8965 &  0 &                             - &         - &$               - $&  0.28862 &$          EW $\\
AST3II106.4053-69.2229 &  13.97 &  299943998 &  13.60 &        - &       - &  0 &                             - &         - &$               - $&  0.16080 &$        DSCT $\\
AST3II109.7742-72.5553 &  14.00 &  271695366 &  13.69 &   1.3067 &  1.1741 &  0 &                             - &         - &$               - $&  2.62150 &$          EA $\\
AST3II099.4564-73.5644 &  14.02 &  142142482 &  13.77 &   1.2053 &  1.0892 &  0 &                             - &         - &$               - $&  0.37834 &$          EW $\\
AST3II098.1909-70.8216 &  14.03 &  167308872 &  13.66 &   1.1810 &  1.0704 &  0 &                             - &         - &$               - $&  1.19986 &$          EA $\\
AST3II091.9501-69.7476 &  14.04 &   41256694 &  13.74 &   1.1345 &  1.0369 &  0 &                             - &         - &$               - $&  0.39314 &$          EW $\\
AST3II093.4959-69.4043 &  14.07 &   41484182 &  13.62 &   1.4363 &  1.2920 &  0 &                             - &         - &$               - $&  0.56982 &$          EW $\\
AST3II117.2315-72.1701 &  14.08 &  272189663 &  13.65 &   1.0356 &  0.9703 &  0 &                             - &         - &$               - $&  0.25154 &$          EW $\\
AST3II107.0916-70.4105 &  14.08 &  300015200 &  13.77 &        - &       - &  0 &                             - &         - &$               - $&  0.60962 &$          EB $\\
AST3II113.0538-69.9661 &  14.08 &  300600720 &  13.65 &   1.3886 &  1.2477 &  0 &                             - &         - &$               - $&  1.15242 &$       EB|EA $\\
AST3II104.3087-70.3423 &  14.08 &  177163611 &  13.67 &        - &       - &  0 &                             - &         - &$               - $&  0.61541 &$          RR $\\
AST3II113.2682-72.2934 &  14.09 &  271892411 &  13.78 &   1.2579 &  1.1320 &  0 &                             - &         - &$               - $&  1.11378 &$       EA|EB $\\
AST3II113.8272-74.2497 &  14.09 &  271971611 &  13.94 &   1.2426 &  1.1193 &  0 &                             - &         - &$               - $&  0.34376 &$          EW $\\
AST3II094.2495-68.7334 &  14.09 &   41594692 &  13.80 &   1.4040 &  1.2620 &  0 &                             - &         - &$               - $&  1.96580 &$          EA $\\
AST3II103.9694-71.9303 &  14.10 &  177114179 &  13.87 &        - &       - &  0 &                             - &         - &$               - $&  0.23812 &$          EW $\\
AST3II105.7290-69.7021 &  14.10 &  284196017 &  13.73 &        - &       - &  0 &                             - &         - &$               - $&  0.26174 &$          EW $\\
AST3II103.3336-69.2480 &  14.12 &  177078161 &  13.79 &        - &       - &  0 &                             - &         - &$               - $&  0.15937 &$        DSCT $\\
AST3II092.0825-72.1979 &  14.16 &  141766811 &  13.74 &        - &       - &  0 &                             - &         - &$               - $&  3.24340 &$          EA $\\
AST3II092.3793-68.8442 &  14.21 &   41339662 &  13.90 &   1.1181 &  1.0256 &  0 &                             - &         - &$               - $&  0.49894 &$     RR|DSCT $\\
AST3II094.8862-73.8941 &  14.22 &  141940658 &  13.90 &   1.2313 &  1.1100 &  0 &                             - &         - &$               - $&  0.40800 &$          EW $\\
AST3II102.0101-69.6037 &  14.24 &  177018590 &  13.78 &   1.0681 &  0.9921 &  0 &  ASASSN-V J064802.44-693613.4 &   0.31616 &$              EW $&  0.31616 &$          EW $\\
AST3II096.6214-74.2658 &  14.30 &  142015305 &  13.92 &        - &       - &  0 &                             - &         - &$               - $&  1.15166 &$          EA $\\
AST3II104.8267-73.9244 &  14.33 &  177387712 &  14.17 &        - &       - &  0 &                             - &         - &$               - $&  0.38164 &$          EB $\\
AST3II106.4445-68.6182 &  14.35 &  299943609 &  14.14 &   1.4526 &  1.3073 &  0 &                             - &         - &$               - $&  1.09910 &$          EB $\\
AST3II110.1155-72.7858 &  14.35 &  271696340 &  14.24 &   1.2130 &  1.0952 &  0 &  ASASSN-V J072027.77-724709.6 &   0.53787 &$            RRAB $&  0.53853 &$          RR $\\
AST3II110.5742-71.7080 &  14.39 &  300328886 &  13.99 &   1.1655 &  1.0590 &  0 &                             - &         - &$               - $&  0.45036 &$          EW $\\
AST3II107.2662-69.2645 &  14.40 &  300014528 &  14.00 &   1.2244 &  1.1043 &  0 &                             - &         - &$               - $&  1.23152 &$          EB $\\
AST3II105.3029-73.1919 &  14.40 &  370236528 &  14.11 &        - &       - &  0 &                             - &         - &$               - $&  0.37820 &$        DSCT $\\
AST3II103.8244-72.1353 &  14.41 &  177349687 &  14.35 &   1.0011 &  0.9465 &  0 &                             - &         - &$               - $&  0.33874 &$          EW $\\
AST3II111.5857-69.3929 &  14.42 &  300443903 &  13.89 &        - &       - &  0 &                             - &         - &$               - $&  0.76492 &$          EA $\\
AST3II112.6964-71.0418 &  14.42 &  300558347 &  14.10 &   1.4076 &  1.2653 &  0 &                             - &         - &$               - $&  0.40474 &$       EB|EW $\\
AST3II094.7913-70.0890 &  14.45 &  167005196 &  14.10 &        - &       - &  0 &                             - &         - &$               - $&  0.34547 &$          RR $\\
AST3II102.8140-69.3139 &  14.47 &  177035211 &  14.14 &   1.2512 &  1.1264 &  0 &                             - &         - &$               - $&  0.38158 &$          EB $\\
AST3II105.5122-69.8645 &  14.47 &  284196104 &  14.20 &   1.6581 &  1.5050 &  0 &                             - &         - &$               - $&  0.19481 &$        DSCT $\\
AST3II114.2903-71.7404 &  14.49 &  300655606 &  14.16 &   1.4035 &  1.2615 &  0 &                             - &         - &$               - $&  0.37428 &$          EW $\\
AST3II097.2290-69.1479 &  14.55 &  167206981 &  14.29 &   1.1424 &  1.0424 &  0 &                             - &         - &$               - $&  0.40836 &$          EW $\\
AST3II100.7189-69.2013 &  14.57 &  176936445 &  14.24 &   0.8807 &  0.8534 &  0 &                             - &         - &$               - $&  0.21671 &$        DSCT $\\
AST3II112.5657-71.1979 &  14.59 &  300557359 &  14.11 &        - &       - &  0 &                             - &         - &$               - $&  0.83572 &$          RR $\\
AST3II103.9372-73.1591 &  14.61 &  177350624 &  14.22 &   0.7959 &  0.7707 &  0 &          SSS-J065544.7-730933 &   0.29535 &$              EW $&  0.29536 &$          EW $\\
AST3II116.7978-74.0328 &  14.62 &  272187821 &  14.26 &        - &       - &  0 &          SSS-J074711.5-740158 &   0.38556 &$              EW $&  0.38544 &$          EW $\\
AST3II105.7603-70.1559 &  14.64 &  284196250 &  14.47 &        - &       - &  0 &                             - &         - &$               - $&  0.46078 &$       EB|EA $\\
AST3II107.1584-69.0159 &  14.64 &  300014365 &  14.28 &   1.1028 &  1.0152 &  0 &                             - &         - &$               - $&  0.38024 &$          EW $\\
AST3II103.1079-73.0137 &  14.64 &  177306954 &  14.37 &   1.0519 &  0.9813 &  0 &                             - &         - &$               - $&  0.33538 &$    DSCT|EW $\\
AST3II111.1086-69.0101 &  14.68 &  300378377 &  14.30 &   1.1978 &  1.0833 &  0 &                             - &         - &$               - $&  0.39342 &$          EW $\\
AST3II096.3027-73.3588 &  14.70 &  141979538 &  14.42 &        - &       - &  0 &                             - &         - &$               - $&  0.37346 &$       EB|EW $\\
AST3II108.9435-69.6865 &  14.70 &  300158892 &  14.35 &   1.5320 &  1.3827 &  0 &                             - &         - &$               - $&  0.64514 &$          EW $\\
AST3II117.0482-72.9081 &  14.71 &  272190253 &  14.36 &   1.1212 &  1.0277 &  0 &                             - &         - &$               - $&  0.28176 &$          EW $\\
AST3II108.7658-72.9983 &  14.72 &  271639765 &  14.44 &   1.1647 &  1.0584 &  0 &                             - &         - &$               - $&  2.27788 &$          EA $\\
AST3II111.1314-74.2812 &  14.76 &  271795366 &  14.43 &   1.0413 &  0.9741 &  0 &                             - &         - &$               - $&  0.35092 &$          EW $\\
AST3II114.7555-72.2977 &  14.79 &  271998796 &  14.65 &        - &       - &  0 &                             - &         - &$               - $&  0.26528 &$          EW $\\
AST3II098.0495-69.9626 &  14.79 &  167308253 &  14.38 &   1.4496 &  1.3044 &  0 &                             - &         - &$               - $&  0.57484 &$          EW $\\
AST3II108.1248-70.2486 &  14.79 &  300086700 &  14.51 &   1.6266 &  1.4741 &  0 &                             - &         - &$               - $&  0.88768 &$       EA|EB $\\
AST3II097.5023-71.8940 &  14.80 &  167250176 &  14.59 &        - &       - &  0 &                             - &         - &$               - $&  0.15961 &$        DSCT $\\
AST3II103.6132-68.8737 &  14.86 &  177112378 &  14.47 &        - &       - &  0 &  ASASSN-V J065427.19-685225.2 &   0.64953 &$            RRAB $&  0.64869 &$          RR $\\
AST3II105.0400-72.9292 &  14.88 &  177387115 &  14.39 &   1.3128 &  1.1794 &  0 &  ASASSN-V J070009.61-725545.1 &   0.09995 &$            HADS $&  0.10000 &$        DSCT $\\
AST3II110.8526-69.4349 &  14.89 &  300327486 &  14.46 &        - &       - &  0 &                             - &         - &$               - $&  0.14035 &$        DSCT $\\
AST3II107.6447-69.2571 &  14.95 &  300039207 &  14.58 &   1.7947 &  1.6443 &  0 &                             - &         - &$               - $&  0.26841 &$          RR $\\
AST3II102.5153-68.6797 &  15.04 &  177022890 &  14.65 &        - &       - &  0 &                             - &         - &$               - $&  0.15906 &$        DSCT $\\
\enddata
\tablenotetext{a}{IDs of AST3-II targets in a format of `AST3II+Ra+Dec'.}
\tablenotetext{b}{$\textit{i}$-band magnitudes from the APASS catalog.}
\tablenotetext{c}{IDs, \textit{TESS} magnitudes, stellar radii and stellar masses from the \textit{TESS} Input Catalog \citep{Stassun17}.}
\tablenotetext{d}{CTL flags, 1 means this target is also selected in the Candidate Target List of \textit{TESS} \citep{Stassun17}.}
\tablenotetext{e}{Names, periods and type designations from the AAVSO database. Explanation for each type designation is placed in footnote f.}
\tablenotetext{f}{Variable star type designations following VSX. $ACV$: $\alpha^2$ Canum Venaticorum variables; $CEP$: Cepheids; $CWB$: W Virginis variables with periods shorter than 8 days. Also known as BL Herculis variables; $DSCT$: Variables of the $\delta$ Scuti type (those with amplitudes larger than 0.15 magnitude and assymetric light curves are designated $HADS$);  $EA (ED)$: 	Algol eclipsing systems; $EB (ESD)$: $\beta$ Lyrae-type eclipsing systems; $EW (EC)$: W Ursae Majoris-type eclipsing variables; $ROT$:Spotted stars that weren't classified into a particular class; $RR (RRAB/RRC) $: Variables of the RR Lyrae type;  $RS$: RS Canum Venaticorum-type binary systems. $YSO$: Young Stellar Object of unspecified variable type;
\url{https://www.aavso.org/vsx/index.php?view=about.vartypes}}
\end{deluxetable}
\end{longrotatetable}

\end{CJK*}
\end{document}